\documentclass[12pt,preprint]{aastex}

\usepackage{epsf}
\usepackage{psfig}

\shorttitle{NUCLEAR STAR CLUSTERS IN SPIRAL GALAXIES}
\shortauthors{Rossa et al.}

\newcommand{\lta}{\lesssim}
\newcommand{\gta}{\gtrsim}

\newcommand{\kms}{\>{\rm km}\,{\rm s}^{-1}}

\newcommand{\Mpc}{\>{\rm Mpc}}
\newcommand{\pc}{\>{\rm pc}}
\newcommand{\Msun}{\>{\rm M_{\odot}}}
\newcommand{\Lsun}{\>{\rm L_{\odot}}}

\begin{document}

\title{HST/STIS SPECTRA OF NUCLEAR STAR CLUSTERS IN SPIRAL GALAXIES:
DEPENDENCE OF AGE AND MASS ON HUBBLE TYPE\altaffilmark{1}}

\author{J\"{o}rn Rossa\altaffilmark{2}, 
Roeland P. van der Marel\altaffilmark{2},
Torsten B\"{o}ker\altaffilmark{3},
Joris Gerssen\altaffilmark{4},
Luis C. Ho\altaffilmark{5},
Hans-Walter Rix\altaffilmark{6},
Joseph C. Shields\altaffilmark{7}, and
Carl-Jakob Walcher\altaffilmark{6}}

\altaffiltext{1}{Based on observations made with the NASA/ESA Hubble
Space Telescope, obtained [from the Data Archive] at the Space Telescope
Science Institute, which is operated by the Association of Universities for
Research in Astronomy, Inc., under NASA contract NAS 5-26555. These
observations are associated with proposals \#9070 and \#9783.}

\altaffiltext{2}{Space Telescope Science Institute, 3700 San Martin Drive, 
Baltimore, MD 21218; jrossa@stsci.edu, marel@stsci.edu}

\altaffiltext{3}{Astrophysics Division, RSSD, European Space Research and 
Technology Centre (ESTEC), NL-2200 AG Noordwijk, The Netherlands; 
tboeker@rssd.esa.int}

\altaffiltext{4}{Department of Physics, University of Durham, Rochester 
Building, Science Laboratories, South Road, Durham DH1 3LE, United Kingdom;
joris.gerssen@durham.ac.uk}

\altaffiltext{5}{The Observatories of the Carnegie Institution of Washington, 
813 Santa Barbara Street, Pasadena, CA 91101-1292; lho@ociw.edu}

\altaffiltext{6}{Max-Planck-Institut f\"ur Astronomie, K\"onigstuhl 17, 
D-69117 Heidelberg, Germany; rix@mpia.de; walcher@mpia.de} 

\altaffiltext{7}{Department of Physics and Astronomy, Clippinger Research 
Laboratories, Ohio University, 251B, Athens, OH 45701-2979;
shields@phy.ohiou.edu}

%%%%%%%%%%%%%%%
% Abstract
%%%%%%%%%%%%%%%

\begin{abstract}
We study the nuclear star clusters in spiral galaxies of various
Hubble types using spectra obtained with STIS on-board {\it HST}. We
observed the nuclear clusters in 40 galaxies, selected from two
previous HST/WFPC2 imaging surveys. At a spatial resolution of $\sim
0\farcs2$ the spectra provide a better separation of cluster light
from underlying galaxy light than is possible with ground-based
spectra. Approximately half of the spectra have a sufficiently high 
signal-to-noise ratio for detailed stellar population analysis. For
the other half we only measure the continuum slope, as quantified by
the $B-V$ color. To infer the star formation history, metallicity and
dust extinction, we fit weighted superpositions of single-age stellar
population templates to the high signal-to-noise spectra. We use the
results to determine the luminosity-weighted age, mass-to-light ratio,
and masses of the clusters. The models provide excellent fits to the
data and generally require a mixture of populations of different
ages. Approximately half of the sample clusters contain a population
younger than 1\,Gyr. The luminosity-weighted age ranges from 10\,Myrs
to 10\,Gyrs. The stellar populations of NCs are generally best fit as a
mixture of populations of different ages. This indicates that NCs did not 
form in a single event, but instead they had additional star formation long 
after the oldest stars formed. On average, the sample clusters in late-type 
spirals have a younger luminosity-weighted mean age than those in early-type 
spirals ($\overline{\langle \log \tau \rangle_L} = 8.37 \pm 0.25$ vs.~$9.23 
\pm 0.21$). The average mass-weighted ages are older by $\sim0.7$ dex,
indicating that there often is an underlying older population that
does not contribute much light, but does contain most of the mass. The
average cluster masses are smaller in late-type spirals than in
early-type spirals ($\overline{\log M} = 6.25 \pm 0.21$ vs.~$7.63
\pm 0.24$), and exceed the masses typical of globular clusters. The
cluster mass correlates loosely with total galaxy luminosity. It
correlates more strongly with both the Hubble type of the host galaxy and
the luminosity of its bulge. The latter correlation has the same slope
as the well-known correlation between supermassive black hole mass and
bulge luminosity. The properties of both nuclear clusters and black
holes in the centers of spiral galaxies are therefore intimately
connected to the properties of the host galaxy, and in particular its
bulge component. Plausible formation scenarios will have to account
for this. We discuss various possible selection biases in our results,
but conclude that none of them can explain the differences seen
between clusters in early- and late-type spirals. The inability to
infer spectroscopically the populations of faint clusters does
introduce a bias towards younger ages, but not necessarily towards
higher masses.
\end{abstract}

%%%%%%%%%%%%%%%
% Keywords
%%%%%%%%%%%%%%%

\keywords{galaxies: evolution ---  galaxies: nuclei --- galaxies: spiral --- 
galaxies: star clusters ---  galaxies: stellar content}

\clearpage

%%%%%%%%%%%%%%%
% Beginning of main text
%%%%%%%%%%%%%%%

\section{INTRODUCTION}
\label{s:intro}

The centers of spiral galaxies have been studied for decades using
ground-based observations. However, with the advent of high-spatial
resolution observations from the {\it Hubble Space Telescope} (HST) it
became clear that there was still much to be learned. A particularly
exciting finding has been the fact that most spiral galaxies have a
nuclear star cluster (NC) in their very center. The agreement between
the positions of these NCs and the isophotal centers of the galaxies
is generally better than one arcsecond and is consistent with zero to
within the error bars \citep{boe02}. These NCs have now been studied
in detail for large samples using broad-band imaging with HST's Wide
Field and Planetary Camera \citep[WFPC;][]{phi96}, Wide Field and
Planetary Camera~2 \citep[WFPC2;][]{car97,car98,mat99,boe02,boe04} and
Near-Infrared Imaging Camera and Multi Object Spectrometer
\citep[NICMOS;][]{car01,car02}. These projects have shown that at
least 50\% of early-type spirals and at least 75\% of late-type
spirals host a NC. These NCs could hold important clues to the formation
and evolution of spiral galaxies in general. And, in particular, they
may be closely associated with the secular evolution processes that
form pseudo-bulges and destroy bars
\citep[e.g.,][]{ata92a,ata92b,nor96,car99,kor04}. The imaging studies
have shown that the NC luminosities typically range from
$10^{5}$--$10^{8} \Lsun$ and that their effective radii typically
range from 2--$10\pc$. About 50\% of the NCs has an effective radius
in the range of 2.4--5.0\,pc \citep{boe04}. The dependence of these
properties on Hubble type, galaxy luminosity, bulge type, bar class,
nuclear environment, and other quantities has also been studied in
some detail. The next logical step is to build a more detailed
understanding of physical properties such as the NC mass, age, and
stellar populations. These quantities are difficult to derive from
multi-band imaging, due to various well-known degeneracies between
age, metallicity, dust, and population mixtures. Spectroscopic
observations are therefore called for to make further progress.

Up to a few years ago only NCs in a few individual nearby galaxies had
been studied spectroscopically in some detail, including M33
\citep{kor93,lon02}, IC\,342 \citep{boe99} and NGC\,4449 \citep{boe01}.
More recently data have become available for some larger samples.
\citet{wal05,wal06} observed a sample of nine NCs in late-type
spiral galaxies using the Ultraviolet and Visual Echelle Spectrograph
(UVES) on the Very Large Telescope (VLT). The data were used to
determine the cluster masses from measurements of the velocity
dispersion. In addition, the stellar populations were inferred from
models for the spectral absorption line features. Advantages of these
VLT data included both high signal-to-noise ratio ($S/N$) and high
spectral resolution. On the other hand, the modest spatial resolution
of these and other ground-based observations are a disadvantage, since
it makes it difficult to disentangle the NC light from underlying
bulge and disk light. This can still be done acceptably for late-type
spirals, which do not have prominent bulges. However, for early-type
spirals this issue quickly becomes prohibitive. This problem can be
alleviated significantly by using the Space Telescope Imaging
Spectrograph (STIS) on HST, but at the expense of reduced $S/N$ and
spectral resolution. This approach was used by \citet{sar05}, who
studied the centers of 23 early-type spirals. Their sample was
selected to contain only galaxies for which the central regions are
dominated by emission lines. These galaxies comprise a mix of
\hbox{H\,{\sc ii}} galaxies, LINERs and Seyfert galaxies. Their
results therefore provide insight into the central regions of spiral
galaxies and bulges in general, but not necessarily NCs in particular.

Common findings that have emerged from the spectroscopic studies are
that the NCs span a wide range of characteristic ages, from 10\,Myrs
to 10\,Gyrs, and that NCs typically contain a mixture of populations
of different ages. What has remained less clear is how and to what
extent cluster ages and masses depend on the Hubble type of the
galaxy. Such a dependence would not be surprising, given that many
other characteristic properties of galaxy nuclei vary strongly as a
function of Hubble type, including e.g.: bulge/disk ratio; bulge
brightness profile shape \citep[de Vaucouleurs $R^{1/4}$ law or
exponential; e.g.,][]{cou96,jon96}; AGN fraction
\citep[e.g.,][]{ho97}; and black hole mass 
\citep[e.g.,][]{tre02,marc03,har04}.
Any dependence of NC properties on Hubble type would provide new
constraints on theories for both NC formation and spiral galaxy
evolution.

Motivated by the aforementioned considerations we present here the
results of a comprehensive study of the properties of NCs as a
function of Hubble type. We selected galaxies with known NCs from
previous HST imaging surveys. The galaxies range from Hubble type Sa
to Sm. They were observed with HST/STIS to ensure sufficient spatial
resolution to allow a meaningful separation of NC light from
underlying bulge and disk light, even for the earliest Hubble types.
As in \citet{sar05} and \citet{wal06} we fit the spectra with
weighted superpositions of single-age stellar population templates
from \citet{bru03}. This yields for each NC the star formation
history, metallicity, dust extinction, luminosity-weighted age,
mass-to-light ratio, and mass.

The paper is laid out as follows. Section~\ref{s:sample} describes the
selection and characteristics of the sample. Section~\ref{s:obsreduc}
discusses the acquisition and reduction of the HST/STIS spectra.
Section~\ref{s:synthesis} discusses the spectral population fitting
procedure used for analyzing the spectra, as well as the robustness of
its results. Section~\ref{s:broadband} discusses the methodology used
for inferring total NC luminosities and broad-band colors from the
spectra. Section~\ref{s:discussion} presents all the results, while
Section~\ref{s:caveats} discusses various systematic effects that
could potentially influence the results, including contamination from
AGN emission, uncertainties in the subtraction of underlying bulge and
disk light, and sample biases. Section~\ref{s:correlations} analyzes
how the NC properties correlate with the global properties of the host
galaxy, and in particular its Hubble type. Section~\ref{s:summary}
presents a summary of the main conclusions.

\section{SAMPLE SELECTION}
\label{s:sample}

For the current spectroscopic survey we selected galaxies from the two
imaging studies of NCs in spiral galaxies with HST/WFPC2. We chose
early-type spirals from the sample of \citet{car98} and late-type
spirals from the sample of \citet{boe02,boe04}.

\citet{car98} imaged a sample of 40 spiral galaxies. The parent sample 
was selected from the UGC Catalog \citep{nil73} and the ESO-LV catalog
\citep{lau89} with the following primary selection criteria: (1) Hubble type 
between Sa and Scd (Hubble T-type between 1 and 6); (2) heliocentric
velocity less than $2500\,\kms$; (3) inclination angle, estimated from
the apparent axial ratio, smaller than $75^\circ$; and (4) angular
diameter larger than 1\arcmin. The 40 galaxies presented in
\citet{car98} form the random subset that were observed between 1996
March and 1997 February. NCs were detected in 23 of the 40
galaxies. For spectroscopic follow-up with HST/STIS we selected only
the 16 galaxies in which a NC had been detected with a magnitude
brighter than $21.0$ in the WFPC2 F606W filter (similar to the
$V$-band). Other NCs were considered too faint to obtain a spectrum
with a reasonable $S/N$ with STIS in the available exposure time. We
did not include the galaxy ESO\,205$-$G7, although it has a central
source of magnitude $19.1\pm0.2$. Ground-based spectra that we
obtained of this object with the UVES spectrograph on the ESO/VLT in
2001 (unpublished) reveal a broad H$\alpha$ line with a FWZI of
200\,{\AA}, indicating that it is an AGN.

\citet{boe02} imaged a sample of 77 spiral galaxies. The parent sample
was selected from the RC3 catalog of bright galaxies \citep{dev91}
with the following primary selection criteria: (1) Hubble type between
Scd and Sm (Hubble T-type between 6 and 9); (2) heliocentric velocity
less than $2000\,\kms$; and (3) inclination angle, estimated from the
apparent axial ratio, smaller than $51^\circ$. The 77 galaxies
presented in \citet{boe02} form a subset that was randomly selected
from a ``snapshot'' pool to fill gaps in the HST observing
schedule. NCs were detected in 59 of the 77 galaxies. For
spectroscopic follow-up with HST/STIS we selected only the 40 galaxies
in which a NC had been detected with a magnitude brighter than $I =
20.5$. Other NCs were considered too faint to obtain a spectrum with a
reasonable $S/N$ with STIS in the available exposure time. Our
spectroscopic program for the late-type spirals was also a snapshot
program, so not all 40 galaxies were actually observed with STIS. In
the end, data were successfully obtained for a randomly selected
subset of 24 galaxies\footnote{The galaxy ESO\,138$-10$ was also
observed in the context of program GO\#9070, but this galaxy is not
included in the sample for the present paper. Due to a pointing error
no spectrum was obtained of the nucleus.}. One additional galaxy from
the 40 (namely NGC\,4411B) was observed together with the early-type
galaxies (to fill the observing time vacated by ESO\,205$-$G7).

Some important properties and quantities of the sample galaxies are
listed in Tables~\ref{t:obs1} and~\ref{t:obs2}. The 15 early-type
spirals are listed in Table~\ref{t:obs1} and the 25 late-type
spirals are listed in Table~\ref{t:obs2}. The listed properties
include position coordinates, galaxy type, heliocentric velocity, and
the apparent and absolute magnitude of the NC. Our
spectroscopic sample contains some bias in the sense that we are not
including the faintest NCs detected in imaging studies. We discuss the 
impact of this bias on our scientific results in Section~\ref{ss:biases}.

\section{OBSERVATIONS AND DATA REDUCTION}
\label{s:obsreduc}

\subsection{Observational Strategy}
\label{ss:obsstrat}

We obtained longslit spectra with STIS on-board the {\it HST}. All
observations used the CCD of the instrument, which is a SITe, thinned,
backside illuminated 1024$\times$1024 pixel array, with a pixel size
of $21\,\mu$m$\times21\,\mu$m. A detailed description is given in the
STIS instrument handbook \citep{kim03}. The early-type spirals were
observed in the context of HST GO program \#9783 and the late-type
spirals in the context of HST Snapshot program \#9070 (PI on both
programs: B\"oker). Except for the adopted exposure times, both
programs used a similar observing strategy.

The observation sequence for each target started with a standard STIS
point-source acquisition, in which a $5 \times 5$ arcsecond image is
first obtained.  For the acquisition image, the telescope was pointed
at the location of the galaxy center as determined from the existing
WFPC2 images, and the image exposure time was chosen to provide $S/N$
sufficient for accurate centroiding using the brightest pixels. After
adjusting the pointing based on the result, we used the grating G430L
to obtain low resolution spectra, with the aim to study the stellar
populations of the NCs. The resulting spectra cover the wavelength
range from $2888.6$--$5703.2$\,{\AA} with a pixel scale of
$2.73$\,{\AA} in the dispersion direction. The pixel scale in the
spatial direction is 0\farcs05071\,pix$^{-1}$. We used the
$52\arcsec\times0\farcs2$ slit. The FWHM of the resulting line-spread
function is $1.4$ pixels for a point source and $4.0$ pixels for a
constant surface brightness extended source \citep{kim03}. The NCs are
generally barely resolved spatially with HST \citep{boe04} and fall
between these two extremes. For a $2.5$ pixel FWHM at 4000\,{\AA} the
spectral resolution is $R = 586$. The spectroscopic exposure time was
divided over two exposures, with a 5 pixel offset along the slit in
order to correct for hot pixels and cosmic ray events during data
reduction.  For GO program \#9783 each of the two exposures was itself
composed of two individual sub-exposures. The journal of observations,
including the integration times and the $S/N$ per pixel values, is listed 
in Table~\ref{t:exp}. The observations of the late-type spirals generally 
had shorter exposure times than those of the early-type spirals (by a factor 
$\sim 2.4$). This is due to the fact that the former were obtained in the
context of a snapshot program.

\subsection{Data reduction}
\label{ss:datared}

To start the data reduction we downloaded the latest instrument
calibration files from the HST Data Archive. For each target we then
applied the basic pipeline process to each of the two exposures. This
includes overscan subtraction and trimming, bias correction, dark
correction, and flat-fielding. We then aligned the two exposures by
shifting one of the two by 5 pixels in the spatial
direction. Subsequently, the task {\em ocrreject} in the
IRAF\footnote{IRAF is distributed by the National Optical Astronomy
Observatory, which is operated by the Association of Universities
for Research in Astronomy, Inc. (AURA) under cooperative agreement
with the National Science Foundation.} STSDAS package was used to
combine the images with rejection of both cosmic rays and hot
pixels. After this step there remained both some positive and negative
outliers in the combined data (the former are due to improperly
removed cosmic rays and hot pixels; the latter are due to pixels that
are hotter in the dark frame than in the science data). To correct
these outliers we applied the task {\em cosmicrays}, first on the
image itself and then on a negative version of itself. Following this,
we performed a two-dimensional rectification and a flux
calibration. For the wavelength calibration we used arc-lamp spectra
obtained in the same HST orbit. The spectra were resampled
logarithmically to a scale of $138\,\kms$ per pixel to facilitate the
spectral fitting discussed below.

For each galaxy we extracted and co-added the central four rows of the
two-dimensional spectrum. This yields a one-dimensional spectrum for a
region of $0\farcs20 \times 0\farcs20$ around the galaxy center. This
translates to 17\,pc$\times$17\,pc at the mean distance to our
galaxies. The extracted spectra contain primarily light from the
bright NC, but there is also some contribution from the underlying
galaxy (bulge and disk). To estimate this contribution we also
extracted and co-added the rows on the two-dimensional spectrum at
distances between $0\farcs15$ and $0\farcs30$ from the center. We
subtracted the result (after proper scaling to the same number of
rows) from the nuclear spectra. This also subtracts a small non-zero
background count-rate that we found on most of the reduced
two-dimensional spectra. The resulting one-dimensional nuclear spectra
are the ones that we have used for the analysis in the remainder of
this paper. It should be noted that even before subtraction the galaxy
light makes only a relatively modest contribution to the nuclear
spectrum. For the late-type spirals the median galaxy light
contribution to the nuclear spectrum is 12\%, and for the early-type
spirals it is 29\%. After the subtraction any remaining galaxy light
contamination is probably much smaller than these percentages,
although not quite zero. This is discussed further in
Section~\ref{ss:subtraction}. Most of the NCs have effective radii
below $0\farcs1$, but there are a few with larger sizes
\citep{boe04}. So in some cases our approach also subtracts a small
amount of NC light. This is discussed further in
Section~\ref{s:broadband}.

\section{SPECTRAL POPULATION FITTING}
\label{s:synthesis}

\subsection{Modeling Approach}
\label{ss:approach}

There are several possible approaches for using our spectra to
constrain the age, metallicity, and dust extinction for the NCs in our
sample. We will use here the fact that a population with arbitrary
star formation history can be written as a sum of single-age stellar
populations. Due to continued advances in stellar population synthesis
modeling, the spectra of single-age stellar populations are now
reasonably well known at a spectral resolution of a few Angstroms
\citep[e.g.,][]{bru03}. By fitting a weighted superposition of the 
spectra of single-age stellar populations to an observed spectrum one
can infer the population characteristics and star formation history of
the object. We will refer to this approach as ``spectral population
fitting''. \citet{sar05} and \citet{wal06} previously used this
approach to study the centers of spiral galaxies and their
NCs. An alternative approach would be to extract and model individual 
line strength indices. This is advantageous because different absorption
lines are sensitive to different properties of the stellar population. 
For example, some lines are mostly sensitive to age, while others are 
mostly sensitive to metallicity.  However, we do not use
absorption line indices here.  Because of the limited $S/N$ of our
spectra we can measure individual absorption lines only with limited
accuracy. By contrast, the technique of spectral population fitting
has the advantage that all pixels in the spectrum are modeled at the
same time, so that the effect of noise is mitigated considerably. It
also allows for a more straightforward interpretation of spectra with
mixed populations of different age and/or metallicity. \citet{wal06}
used both line strength indices and spectral population fitting to
analyze their ground-based VLT data of NCs, which have higher $S/N$
and spectral resolution than our data, but lower spatial
resolution. Broadly speaking, it was found that the two methods gave
consistent results.

We developed new software that fits an observed galaxy spectrum as a
linear sum of population templates. The fit also allows for
differences in redshift and line width between the galaxy and template
spectra. The code is based on the Gauss-Hermite Pixel Fitting software
for analyzing galaxy kinematics \citep{vdm94}. It is similar to the
code used by \citet{wal06}, which is based on the code for analyzing
galaxy kinematics developed by \citet{rix92}. The best-fit is defined
as the one that minimizes the $\chi^2$ difference between the observed
and model spectra. It is found as the matrix solution of a
non-negative least squares (NNLS) problem. This enforces the fact that
none of the templates can have negative weight in the solution. To
find the solution we used the fast and efficient algorithm of
\citet{law74}. Mathematical proofs exist that this algorithm will
always find a global minimum (instead of a spurious local minimum). We
verified the accuracy of our code by analyzing a small sub-sample of
our survey with both our own code and that of \citet{wal06} and found
excellent agreement between the results.

As templates we used the models of single-age stellar populations
prepared by \citet{bru03}. From their collection we used the model
spectra based on the Padua 1994 evolutionary tracks, utilizing a
\citet{cha03} initial mass function (IMF) with lower and upper mass
cut-offs of $m_{\rm{low}}$ = $0.1\,\Msun$ and $m_{\rm{up}}$ =
$100\,\Msun$. We used model spectra for four different metallicities
$Z =$ 0.004, 0.008, 0.02 and 0.05 and 14 different ages $\log \tau
=$6.00, 6.48, 6.78, 7.00, 7.48, 7.78, 8.00, 8.48, 8.78, 9.00, 9.48,
9.78, 10.00 and 10.30.\footnote{The age $\tau$ is expressed in years
throughout this paper, unless noted otherwise. The metallicity value
$Z=0.02$ is traditionally referred to as the ``solar metallicity''
($Z_{\odot}$). For historical consistency we will use this terminology
here. However, it should be noted that it has recently been argued
that the metallicity of our Sun is actually lower, namely $Z=0.012$
\citep{asp05}.} The models have solar abundance ratios. To address the
effect of dust we made versions of each template spectrum with
different amounts of extinction $A_V$, using the extinction law of
\citet{car89}. The templates have a spectral resolution of $3${\AA},
which is better than the resolution of the STIS data. All templates
were rebinned to the same logarithmic scale with $138\kms$ per pixel
used for the galaxy spectra. We performed all fits over the wavelength
range from 3540{\AA} to 5680{\AA}, for which galaxy data as well as
model predictions are available. This region covers the prominent
Balmer lines H$\beta$, H$\gamma$, and H$\delta$ plus other prominent
diagnostic absorption lines such as the Mg\,Ib, Ca and Fe lines.

\subsection{Results}
\label{ss:results}

We applied the method of spectral population fitting to all of the NC
spectra. We found that roughly half of the spectra did not have
sufficient $S/N$ for a meaningful analysis of the absorption line
properties. So we decided to retain for the spectral population
fitting only those NCs for which the spectra had $S/N \gta 5$ per
pixel (see Table~\ref{t:exp}). This criterion was chosen because at lower 
$S/N$ it proved impossible to even recover the known galaxy redshift 
(listed by NED, the NASA Extragalactic Database) from the absorption lines 
in the spectrum. Visual inspection of the low $S/N$ spectra also confirmed
that little could be learned about their absorption lines. In practice
the adopted $S/N$ limit corresponds to a magnitude limit of $m_B
\leq$ 20.35\,mag for the NCs observed in program GO\#9783 (see
Section~\ref{s:sample}) and $m_B \leq$ 19.55\,mag for the NCs observed
in program GO\#9070. The cutoff is fainter for the early-type spirals
observed in GO\#9783 than for the late-type spirals observed in
GO\#9070 because the former were observed with longer exposure
times. In total there are ten early- and nine late-type galaxies
brighter than the adopted cutoffs. They are listed in
Table~\ref{t:age}. For the NCs that are fainter than the adopted
magnitude cutoffs we use the STIS spectra only for a study of the
continuum shape, and not of the absorption lines. This is discussed in
Section~\ref{s:broadband} below.

To model the spectra we performed fits of different fixed metallicity
$Z$ and extinction $A_V$. We used a grid with the four available
metallicity values and extinctions increasing from zero upwards in
steps $0.1$ mag. For each fixed $(Z,A_V)$ combination we used our code
to find the weighted mix of populations of different ages that fits
the observed spectrum with the minimum $\chi^2$, which we denote
$\chi^2_{\rm min}(Z,A_V)$. Subsequently, we identified the
best-fitting extinction and metallicity by finding the combination for
which $\chi^2_{\rm min}(Z,A_V)$ is lowest. This yields for each galaxy
the best-fit model that we use in the remainder of the paper.

In principle we could have fit each NC spectrum with a weighted
sum of {\it all} the available templates, including templates of
different age, metallicity and extinction. However, this would
introduce considerable degeneracy amongst the templates, leading to a
non-unique fit that would be hard to interpret. Since we did not do
this, it should be kept in mind that the best-fitting metallicity $Z$
and extinction $A_V$ should be interpreted as ``average'' quantities
only. Our strategy provides no insight into possible variations in
metallicity and extinction within each NC. One might argue though
that such variations are not large. Either way, the adopted
approach is sufficient for the purposes of the present paper, and is
similar to that adopted by \citet{wal06}.

The observed spectra are shown in black in Figure~\ref{f:spectraone},
with the best-fit models overplotted in red. The NCs are shown in
order of increasing age. The fits are generally good, as confirmed by
the values of the reduced $\chi^2$ listed in Table~\ref{t:age}. A few
galaxies show emission lines (e.g., NGC 3277). These were masked in
the fit. The best-fit metallicity and extinction for each galaxy are
listed in Table~\ref{t:age}. We also list NC properties that can be
derived from the age-mix that best fits the data, namely: the
luminosity-weighted mean value of $\log \tau$, which we call $\langle
\log \tau \rangle_L$; the mass-weighted mean value of $\log \tau$,
$\langle \log \tau \rangle_M$; and the $B$-band mass-to-light ratio in
solar units, $M/L_B$. We note also that the best-fit models
successfully recovered the known redshift for all galaxies, and that
the inferred line broadening was found to be consistent with
expectation based on the instrumental resolution of the STIS setup
(see Section~\ref{ss:obsstrat}).
 
\subsection{Robustness of Population Fits}
\label{ss:robustness}

With our aforementioned method we find the best-fitting metallicity
and extinction for each NC. The uncertainties on these quantities can
be obtained in a straightforward manner using the $\Delta \chi^2$
statistic. This yields 1$\sigma$ uncertainties that are generally
similar to or smaller than the spacing of our $(Z,A_V)$ grid. The
inferred metallicity and extinction should therefore be fairly
accurate, as judged by these formal uncertainties. On the other hand,
there could be additional systematic uncertainties that are more
difficult to quantify (e.g., due to uncertainties in the modeling of
stellar evolution). So to be conservative we adopt throughout this
paper the spacing of our grids as estimates of the uncertainties in
$Z$ and $A_V$, i.e., $\Delta \log Z = 0.4$ and $\Delta A_V = 0.1$.

In the present context we are mostly interested in constraining the
ages of the NCs. As it turns out, the inferred ages are not very
sensitive to either $Z$ or $A_V$. This is illustrated by
Figures~\ref{f:ageext} and~\ref{f:agemetal}. Figure~\ref{f:ageext}
shows for all NCs the best-fit value of the luminosity-weighted mean
age $\langle \log \tau \rangle_L$ as a function of extinction. For
this plot we determined for each NC the metallicity and age mix that
give the lowest $\chi^2$ for fixed values of $A_V$ in the range $0.0$
to $0.6$ (this range contains the best-fit extinction for most of the
galaxies in the sample; see Table~\ref{t:age}). In general, the
implied $\langle \log \tau \rangle_L$ decreases monotonically with
increasing $A_V$.  This is because with increased extinction, the
population must be intrinsically bluer, and therefore younger, to
provide the same continuum slope. However, even a change of $0.6$ mag
in the assumed $A_V$ changes the best fit age only by 0.2--0.5
dex. This is much smaller than the spread in ages amongst the
different NCs. Similarly, Figure~\ref{f:agemetal} shows $\langle \log
\tau \rangle_L$ as a function of metallicity, with $A_V$ and the age
mix at each metallicity chosen to minimize $\chi^2$. For young
populations with $\langle \log \tau \rangle_L \lta 8.5$, the inferred
age is almost independent of $Z$. For older populations the age
decreases when the metallicity increases. This is because
higher-metallicity populations tend to be redder, so that a younger
age (i.e., bluer continuum) is needed to provide the same continuum
slope. However, even a change of $1.1$ dex in $\log Z$ only changes
the inferred age by $\sim 0.3$ dex. This is also much smaller than the
spread in ages amongst the different NCs. So the overall age
distribution of the NC population is insensitive to possible
systematic uncertainties in the inferred extinctions and
metallicities. Note also in this context that Figures~\ref{f:ageext}
and~\ref{f:agemetal} conservatively assume that we have no knowledge
of the actual $Z$ and $A_V$. In reality, the values of these
quantities are actually well constrained by our data.

Our template library is based on models with solar abundance ratios.
Of course, the NCs under study may not have exactly these same
abundance ratios. This might introduce small systematic biases in our
results. However, the influence on the inferred ages is probably
small. Figure~\ref{f:agemetal} shows that, for the purposes of the
present study, the inferred ages are relatively insensitive to
variations in overall metallicity. Abundance ratio variations are
probably a second-order effect as compared to variations in $Z$. Our
results are therefore not likely to be sensitive to them either. This
argument is supported by the good agreement between the $M/L$ ratios
inferred from our stellar population study and those from dynamical
studies, as discussed in Section~\ref{ss:masses}.

Another possible concern might be that by allowing an arbitrary mix of
ages (i.e., an arbitrary star formation history) we might somehow have
provided the fit too much freedom. This could in principle decrease
the accuracy of the inferred ages. To address this issue we also
performed single-age fits to all NC spectra. We determined the
$\chi^2$ of the fit to the data for each individual template of fixed
$Z$, $A_V$ and age, and then adopted the template with the lowest
$\chi^2$ as the best fit. The ages $\log \tau$ thus inferred agree
very well with the luminosity-weighted ages inferred from the
composite-age fits. For the sample as a whole we find that $\log \tau
- \langle \log \tau \rangle_L$ has an average of $0.11 \pm 0.08$, with
an RMS scatter of 0.36. The scatter is similar to the spacing in age
between the different templates in our library, which is $\sim 0.3$
dex.

Another approach for assessing the accuracy of the inferred ages is to
perform Monte-Carlo simulations. We did this in \citet{boe03}. In that
paper we constructed artificial galaxy spectra as a superposition of
randomly chosen templates from the library with randomly chosen
weights. Artificial Poisson noise was then added to the spectra with
$S/N$ values representative of the range encountered in our
observations. The artificial spectra were analyzed in similar fashion
as our real data, using the same library of templates (i.e., same
metallicity and extinction) used to build the artificial galaxy
spectra. Statistics were performed on the relation between the
luminosity-weighted age $\langle \log \tau \rangle_L$ of the input
spectra and the values for the same quantity inferred with the
code. The 1$\sigma$ spread in the residuals was typically found to be
$\lta 0.3$ dex for values of $\log \tau \gta 8.0$, which is the age
for most of the NCs in our sample. The spread decreases for older
populations and is about a factor of two smaller at $\log \tau \approx
9.0$. The spread is larger for younger populations, but a very young
population is never mistaken for an old one ($\log
\tau \gta 9.0$). Also, young NCs (such as the one identified in NGC\,2139) 
tend to be bright and produce spectra of high $S/N$, which decreases
the uncertainty.

There are five late-type spirals in our sample that we also studied in
\citet{wal06}. For four of these galaxies (NGCs 300, 428, 1493 and
2139) the STIS data had sufficient $S/N$ for spectral population
fitting. A cross-comparison of our results for these galaxies with the
results in \citet{wal06} provides a useful consistency check. The
latter paper used data from the VLT, which provides higher spectral
resolution but lower spatial resolution and a smaller wavelength
range. Despite these differences, the luminosity-weighted ages
inferred from the VLT and STIS data agree with a mean difference of
$0.03 \pm 0.12$ dex and a RMS scatter of $0.24$ dex. So the agreement
is excellent, and provides no reason to mistrust the results from
either study. There are no early-type spirals for which a direct
comparison to our STIS results is possible. \citet{sar05} performed a
study with HST/STIS of the centers of early-type spiral galaxies with
emission lines, but there are no galaxies in common between our
samples to make a direct comparison.

\section{BROAD-BAND MAGNITUDES AND $B-V$ COLOR}
\label{s:broadband}

Approximately half of our spectra have insufficient $S/N$ for spectral
population fitting. However, even for the faintest NCs the spectra
have sufficient $S/N$ for the calculation of integrated properties. In
particular, we used the spectra to estimate the apparent magnitudes of
the NCs in the Johnson $B$ and $V$ passbands. These bands cover the
wavelength range of our spectra and are of particular interest because
they are used most frequently in stellar population studies.

The STIS pipeline calibrates spectra to absolute fluxes in erg
s$^{-1}$ cm$^{-2}$ {\AA}$^{-1}$ using reference files that are based
on observations of spectrophotometric standard stars. We used the {\em
synphot} package in IRAF/STSDAS to calculate integrals over the
spectra, weighted with the transmission curves of the $B$ and $V$
passbands, respectively. The results were converted to magnitudes by
comparison to results obtained in a similar manner for the star Vega,
which has magnitude zero in all Johnson bands to within $\sim 0.03$\,mag 
\citep{bes98,boh04}. The random uncertainties in the magnitudes follow 
from propagation of the flux uncertainties calculated by the STIS
pipeline.  Our spectra do not quite cover the entire wavelength range
of the $V$-band. In order to calculate $V$-band magnitudes we
extended the spectra to redder wavelengths using a linear
extrapolation. Any potential systematic errors thus introduced are not
expected to be large. A constant flux error of 10\% over the
extrapolated region would change the calculated $V$-band magnitudes by only
$0.03$. More generally, the relative accuracy of the inferred
magnitudes (i.e. the systematic errors in the implied colors $B-V$) is 
expected to be $\lta 0.05$ mag.

The magnitudes obtained by integration over the spectra measure only
the light in our STIS aperture of $0\farcs20 \times 0\farcs20$. To
estimate the total magnitudes of the NCs we therefore need to apply a
correction for NC light that falls outside the aperture. We did this
in two steps. First, we added a constant correction of $2.5 \log(0.71)
= -0.37$ mag, to account for the fact that even for a point source
only $71$\% of the light falls in an aperture of $0\farcs20 \times
0\farcs20$ \citep{kim03}. And second, we added a correction that
depends on the extent of the NC. This correction is $2.5 \log (f_{\rm
e}/f_{\rm p})$, where $f_{\rm e}$ and $f_{\rm p}$ are the fractions of
the light of an extended and a point source, respectively, that fall
within a $0\farcs20 \times 0\farcs20$ aperture. The fractions were
calculated using a simple Gaussian model for the light distribution,
with the FWHM as the only free parameter, using eq.~(A3) of \citet{vdm97}. 
For $f_{\rm e}$ we used the FWHM measured from the intensity profile 
along the slit and for $f_{\rm p}$ we used the actual FWHM of the STIS PSF. 
The use of a simple Gaussian model for the core of the light distribution 
is adequate, given that the correction of $-0.37$ mag has already accounted 
for the scattering of light into the extended wings of the PSF. The final NC 
magnitudes, including the aperture corrections, are listed in 
Table~\ref{t:color}. The table also lists the implied $B-V$ color.

As a consistency check, we compared our $V$-band magnitudes $m_V$ with
the instrumental magnitudes in the WFPC2/F606W filter system given by
\citet{car98} for the NCs in early-type spirals. We find that on
average $m_V - m_{\rm F606W} = 0.04$, with an RMS scatter of $0.40$
mag. For comparison, single-age solar-metallicity stellar populations
of $10^8$ and $10^{10}$ yrs have $m_V - m_{\rm F606W} = 0.12$ and
$0.32$, respectively (given the $V-I$ from \citet{bru03} and the
transformations given by \citet{hol95}). We also combined our
$V$-band magnitudes with the $I$-band magnitudes given by
\citet{boe02} for the NCs in late-type spirals to estimate $V-I$
colors. These colors can be used as a consistency check for those NCs
for which an independent color estimate is available. This is the case
for the nine NCs for which spectral population fitting was
performed in Section~\ref{s:synthesis}. For these NCs we calculated
the predicted $V-I$ color based on the best-fitting stellar population
mix and dust extinction, as given in Table~\ref{t:age}. We find that
on average $(V-I) - (V-I)_{\rm pred} = 0.18$, with an RMS scatter of
$0.53$ mag. These comparisons suggest that our magnitudes are probably
accurate in an absolute sense to $\sim 0.2$ mag, when averaged over
the sample. However, for individual NCs the 1$\sigma$ accuracy is
probably no better than $\sim 0.5$ mag. These absolute accuracies are
reasonable, given the uncertainties associated with galaxy light
subtraction and aperture corrections. Of course, photometry of images
generally provides a more accurate method for estimating NC
magnitudes. Nonetheless, our results are useful because they augment
existing results in the literature with results in bluer passbands.

\section{DISCUSSION OF SPECTRAL POPULATION FITTING RESULTS}
\label{s:discussion}

\subsection{Metallicities and Extinctions}
\label{ss:metalext}

The main focus of our investigation is the determination of the ages
and masses of the NCs. A study of the NC metallicities and extinctions
is not the primary interest of our paper. Accurate studies of
metallicities would probably benefit from an analysis of individual
line strength indices in spectra of higher $S/N$, and accurate studies
of extinction would probably benefit from analysis of spectra that
cover a larger wavelength range. Nonetheless, our results obtained
from the spectral population fitting (see Table~\ref{t:age}) do have
fairly small formal random uncertainties. While there might be
systematic uncertainties, there are no specific omissions in our
analysis that would lead us to mistrust the results. So it is of some
interest to examine our findings for the metallicities and extinctions
listed in Table~\ref{t:age}.

Figure~\ref{f:Zhisto} shows histograms of the metallicity
distributions. The bottom panel is for the complete sample of NCs for
which we performed spectral population fitting. The top two panels
show the results for the early- and late-type spirals separately.  The
average inferred metallicity is ${\bar Z} = 0.030 \pm 0.005$ for the
early-type spirals, ${\bar Z} = 0.018 \pm 0.004$ for the late-types,
and ${\bar Z} = 0.024 \pm 0.006$ for the complete sample. The quoted
errors are formal errors, calculated as the RMS scatter among the
individual datapoints divided by $\sqrt{N}$. Given the crudeness of
our sampling in metallicity, these errors are probably optimistic. So
on the whole, the primary conclusions are that the metallicities are
close to solar and that the metallicities tend to be somewhat higher
in early- than in late-type spirals. These results are quite
consistent with what is known about the metallicities of spiral
galaxies, and the dependence on Hubble type, from studies of nebular
gas \citep[e.g.,][]{zar94}.

Figure~\ref{f:avhisto} is similar to Figure~\ref{f:Zhisto}, and shows
histograms of the dust extinction distributions. The average inferred
extinction is ${\bar A}_V = 0.32 \pm 0.13$ for the early-type
galaxies, ${\bar A}_V = 0.39 \pm 0.09$ for the late-types, and ${\bar
A}_V = 0.35 \pm 0.08$ for the complete sample. The average for the
early-types is influenced somewhat disproportionately by
NGC\,3455. Its inferred extinction of $A_V = 1.4$ is considerably
larger than the values found for all other NCs. Without this galaxy
the early-types would have ${\bar A}_V = 0.20 \pm 0.06$. Either way,
the average extinction is marginally higher for the NCs in late-type
galaxies than for the NCs in early-type spirals. This is as expected,
given that late-type spirals are more gas rich, show enhanced star
formation and radiate more FIR emission than early-type spirals
\citep[e.g.,][]{ken83,dev87}. In general, the results from spectral
population fitting become more uncertain as the amount of dust
extinction increases. However, for the sample here we do not generally
find the dust extinctions to be particularly high. So there is little
reason to be concerned that the inferred ages are significantly
impacted by any remaining uncertainties in the exact amount of
extinction, or the detailed shape of the extinction law. As a
consistency check on our analysis, we also compared the inferred
extinctions to the known Galactic foreground value for each galaxy. In
all cases we found the inferred value to be larger than the Galactic
foreground value (see Tables~\ref{t:obs1} and~\ref{t:obs2}) to within
the $\sim 0.1$ accuracy dictated by our sampling of $A_V$. This is
what would be expected for a successful fit, given that the total
extinction should never be smaller than the known foreground
extinction.

It is of interest to compare our results to those of \citet{sar05} and
\citet{wal06}. The HST study of \citet{sar05} addressed the stellar
populations in the centers of 23 galactic bulges in early-type spirals
for which the central regions are dominated by emission lines. These
galaxies comprise a mix of \hbox{H\,{\sc ii}} galaxies, LINERs and
Seyfert galaxies. \citet{sar05} found that good fits could be obtained
with models of either $Z=0.02$ or $0.05$, consistent with the results
for early-type spirals in our study. Their average extinction of
${\bar A}_V = 0.37 \pm 0.05$ also agrees well with our value of ${\bar
A}_V = 0.32 \pm 0.13$. The VLT study of \citet{wal06} addressed the
NCs in a sample of nine late-type spirals. The average metallicity for
the \citet{wal06} sample is ${\bar Z} = 0.015 \pm 0.004$. This is
consistent with the value ${\bar Z} = 0.018 \pm 0.004$ for the late-type
galaxies in our sample. The average $I$-band extinction found by 
\citet{wal06} corresponds to ${\bar A}_V = 0.56 \pm 0.18$, as compared to 
${\bar A}_V = 0.39 \pm 0.09$ for the late-type spirals in our sample. Our 
conclusion that the dust extinctions are not particularly high is also 
consistent with the findings of \citet{car01}, who studied the $V-H$ and 
$J-H$ colors of NCs in spiral galaxies using HST imaging. So overall, our 
results agree well with those of other authors.

\subsection{Cluster Ages}
\label{ss:ages}

We have derived luminosity-weighted ages for the 19 NCs for which we
performed spectral population fitting (Table~\ref{t:age}).
Figure~\ref{f:agehisto} shows histograms of the derived ages for the
early-type, late-type and total sample. The average
luminosity-weighted ages are $\langle \log \tau \rangle_L = 9.23 \pm
0.21$ (early-type), $8.37 \pm 0.25$ (late-type) and $8.82 \pm 0.18$
(combined sample).

The inferred average luminosity-weighted ages are smaller than the
typical ages of galactic disks, for which the start of star formation
was typically at least 5--10 Gyrs ago
\citep[e.g.,][]{kno99,bin00,aba03}. This indicates that star formation
has continued in the centers of spiral galaxies after the initial
formation of the galaxy. We find the ages of the NCs in late-type
spirals to be smaller than those in early-type spirals. This is
illustrated further by Figure~\ref{f:hubblevsage}, which shows a clear
relation between $\langle\log\,\tau\rangle_L$ and the Hubble T-type of
the host galaxy (the parameters of the best linear fit and the
Spearman rank-order correlation coefficient are given in
Table~\ref{t:fitcoeff}). This result is probably not unexpected, given
that late-type spirals are more gas-rich \citep[e.g.,][]{you91}. So
there might have been more gas flows towards the center from which to
form stars. Also, our finding for the nuclear clusters is consistent
with the fact that the underlying disks of late-type spirals formed
more of their stars recently than early-type spirals
\citep[e.g.,][]{ken98}. The three youngest clusters in our sample are
hosted in late-type spirals, and the four oldest clusters are hosted
in early-type spirals.

It is again of interest to compare our results to those of
\citet{sar05} and \citet{wal06}. The average luminosity-weighted age
for nuclear clusters in early-type spirals reported here, $9.23 \pm
0.21$, is similar to the value $9.50 \pm 0.13$ inferred from the
results of \citet{sar05}. The small (non-significant) difference could
be due to the fact that the \citet{sar05} sample has relatively more
galaxies of very early types (Sa or even S0) than our study
does. Also, the galaxies in their sample do not necessarily host
nuclear star clusters and are much more biased towards AGN
activity. So in view of these differences, the agreement between our
results is surprisingly good. In the same spirit, the average
luminosity-weighted age for nuclear clusters in late-type spirals
reported here, $8.37 \pm 0.25$, is similar to the value $8.15 \pm
0.19$ inferred from the results of \citet{wal06}. Their VLT sample
overlaps in part with our late-type sample (there are four galaxies in
common with the STIS subsample for which we performed spectral
population fitting: NGCs 300, 428, 1493 and 2139). The range of Hubble
types in both samples is very similar, as is the range of cluster
luminosities (the average absolute magnitude of the clusters in the
\citet{wal06} sample is $\langle M_I \rangle = -12.75$, as compared
to $\langle M_I \rangle = -12.64$ for the NCs in late-type spirals for
which we performed spectral population fitting). So the agreement of
the present results with those inferred from the \citet{wal06} VLT
spectra is more or less what was expected. The VLT spectra had much
higher spectral resolution and $S/N$ than our data. So the agreement
does indicate that there is no reason for concern that the limited
spectral resolution and $S/N$ of our STIS spectra might have somehow
biased our results.

We note that to make a fair comparison of the results of \citet{wal06}
and \citet{sar05} to ours we had to convert their results to the same
definitions that we have used here. Our quantity $\langle \log \tau
\rangle_L$ is the average logarithm of the age of the population,
weighted by luminosity. By contrast, \citet{wal06} and
\citet{sar05} quote $\log \langle \tau \rangle_L$, which is the 
logarithm of the average age of the population, weighted by
luminosity. We prefer the former statistic, because it more clearly
highlights the contributions and presence of young populations. For
example, a cluster for which 50\% of the light comes from a $10^6$ yr
old population and the other 50\% comes from a $10^{10}$ yr old
population has $\langle \log \tau \rangle_L = 8.0$ and $\log \langle
\tau \rangle_L = 9.70$. Note also that while we and \citet{sar05} weight by
luminosity in the spectral range covered by the STIS spectra,
\citet{wal06} weight by luminosity in the $I$-band. We corrected for this 
as well in making the aforementioned comparisons.
     
In addition to the luminosity-weighted ages we also calculated
mass-weighted ages $\langle \log \tau \rangle_M$ for our own STIS
spectral population fits (see Table~\ref{t:age}). The sample averages
are $\langle \log \tau \rangle_M = 9.89 \pm 0.14$ (early-type), $9.07
\pm 0.21$ (late-type) and $9.50 \pm 0.16$ (combined sample). These
mass-weighted averages are considerably older than the
luminosity-weighted averages, typically by $\sim0.7$ dex. This is
because young populations are much brighter than old populations. So
while young populations contribute a significant fraction of the
light, their contribution to the total NC mass is more limited.  There
often is an underlying older population that does not contribute much
light, but does contain most of the mass.

Although it is possible to calculate average ages in various ways, it
should be kept in mind that no single number can capture the
complexity of mixed populations that is present in the NCs. Like
\citet{sar05} and \citet{wal06} before us we find that mixed
populations generally give significantly better fits to the observed
spectra than single-age fits. The median difference in $\chi^2$
between the best mixed-age and single-age fits for the sample galaxies
is $\Delta \chi^2 = 92$.  There are a total of 16 free parameters in
the mixed-age fits (metallicity, extinction, and the 14 weights of
single-age templates). The standard theory of confidence level
estimation using $\Delta \chi^2$ statistics \citep[e.g.,][]{pre92}
dictates that the 1-$\sigma$ confidence limit around the best
mixed-age fit corresponds to a contour of $\Delta \chi^2 = 18$ in the
16-dimensional parameter space. Conversely, a value $\Delta \chi^2 =
92$ is statistically significant at more than the 99.999\% level.  The
present results provide the first demonstratation that mixed-age
populations are present also in the NCs of early-type spirals.
\citet{wal06} studied only late-type spirals. \citet{sar05} studied
early-type spirals, but their galaxies did not generally or
necessarily host NCs. With our new results it is now clear that all
NCs, independent of host Hubble type, did not form in a single
event. Instead they had additional star formation even long after the
oldest stars formed. Table~\ref{t:contr} lists the derived luminosity
fractions for the 19 NCs for which we performed spectral population
fitting.  Instead of listing the contributions of each individual
single-age template, we have binned the weights into three (somewhat
arbitrarily chosen) age groups. The groups correspond respectively to
``young populations'' ($\log \tau \leq 7.8$), ``intermediate-age''
populations ($7.8 < \log \tau \leq 9.2$), and ``old'' populations
($\log \tau \geq 9.2$). In four of the ten NCs in early-type spirals
more than half of the light comes from an old population. By contrast,
in the NCs in late-type spirals the light is always dominated by an
intermediate-age or young population.

\subsection{Cluster Masses}
\label{ss:masses}

The spectral population fits to the STIS spectra yield the
mass-to-light ratio in the $B$-band, $M/L_B$ (Table~\ref{t:age}).
These values range from $0.02$ to roughly $9.0$. The average $M/L_B$
values are $3.64 \pm 1.03$ (early-type sample) and $0.52 \pm 0.19$
(late-type sample). The smaller value for the NCs in late-type spirals
is due to the fact that on average they have younger stellar
populations, as discussed above.

The NC mass $M$ is obtained upon multiplication of the $B$-band
luminosity $L_B$ and the $M/L_B$ value of the best composite spectral
population fit (Table~\ref{t:age}). The former can be calculated from
the apparent magnitude $m_B$ (Table~\ref{t:color}), the extinction
$A_B$ (which follows from the $V$-band extinction $A_V$ listed in
Table~\ref{t:age}), and the distance $D$ in Mpc. The distance can be
calculated as $v_{\rm sys}/H_0$, where $v_{\rm sys}$ is the systemic
velocity of the galaxy corrected for Virgocentric infall as listed in
Tables~\ref{t:obs1} and~\ref{t:obs2}. We use here a Hubble constant
$H_0 = 70\,\rm{km\,s^{-1}\,Mpc^{-1}}$. This yields for the NC mass
that
\begin{equation}
\label{mass}
  \log M = \log(M/L_B) - 0.4\,m_B + 
     2\log v_{\rm{sys}} + 0.530\,A_V + 8.502  .
\end{equation} 
Here $M$ and $M/L_B$ are both given in solar units. The constant
$0.530$ is equal to $0.4 A_B / A_V $, where $A_B/A_V = 1.324$ is based
on the extinction law of \citet{rie85}. The constant $8.502$ is equal
to $0.4 (25 - 5 \log H_0 + M_{B,\odot})$, where $M_{B,\odot}$ is the
$B$-band solar absolute magnitude \citep{bin98}. We used
equation~(\ref{mass}) to calculate the masses for the 19 NCs for which
we did spectral population fits. The results are listed in
Table~\ref{t:age}.

Figure~\ref{f:masses} shows histograms of the derived masses for the
early-type, late-type and total sample. The average masses are
$\overline{\log M} = 7.63 \pm 0.24$ (early-type), $6.25 \pm 0.21$
(late-type) and $6.98 \pm 0.22$ (combined sample). So on average the
NCs in our sample are much more massive than globular clusters (GCs)
in the Milky Way, which have a characteristic mass of $\log M = 5.18$
\citep[e.g.,][]{har91}. On the other hand, some of the most massive
GCs in the Local Group, such as $\omega$\,Cen in the Milky Way
\citep{mey02} and G1 in M\,31 \citep{mey01}, and in other nearby
galaxies such as Cen\,A \citep{mar04} have masses that overlap with
the low-mass end of the NC mass distribution. Such massive globular
clusters are also abundant in mergers and starburst
galaxies. Dynamical masses for the most massive young star clusters in
the Antennae (NGC\,4038/39) range from $5.81 \leq \log M \leq 6.67$
\citep{men02}, with the mass distribution showing a power-law increase
towards lower masses ($\psi (M) \propto M^{-2}$ over the range $4.0
\leq \log M \leq 6.0$ \citep{zha99,whi99}). More recently, for the GC
W3 in the merger remnant NGC\,7252 an even higher dynamical mass of
$\log M = 7.9\pm0.1$ was derived \citep{mara04}. The most massive star
clusters in M\,82 have typical masses of $5.54 \leq \log M \leq 6.18$
\citep{mcc03}. Although all these objects span a range of masses, they
do all obey a common relation in terms of their ``Fundamental Plane''
properties \citep{wal05}. These similarities extend also to related
objects such as the nuclei of dwarf elliptical galaxies \citep{geh02}
and the so-called ``ultra-compact dwarf'' galaxies \citep{has05}. This
indicates that the formation of all these objects could be governed by
similar physical processes.

One striking aspect of Figure~\ref{f:masses} is that the mass
distribution of NCs in early-type spiral galaxies is centered around
much higher masses than that in late-type spirals. This is illustrated
further by Figure~\ref{f:hubblevsmass}, which shows a clear relation
between cluster mass $M$ and the Hubble T-type of the host galaxy (the
parameters of the best linear fit and the Spearman rank-order
correlation coefficient are given in Table~\ref{t:fitcoeff}). As many
as six out of ten NCs in early-type spirals for which we performed
spectral population fits have masses $\log M \gta 8.0$ (the apparent
bimodality in the mass distribution for the early-type sample seen in
the top panel of Figure~\ref{f:masses} is most likely due to low
number statistics and not worth attaching much significance to).  For
comparison, this exceeds the masses of many dwarf galaxies
\citep[e.g.,][]{bin88}. This result for early-type spirals connects
with an earlier finding of \citet{car02}. They found that NCs in
early-type spirals span a much larger range in luminosity than the NCs
found in late-type spirals, with luminosities extending up to $\sim
10^{8.4} L_\sun$ in the $H$-band. Our work shows that the luminosity
of these very bright NCs is not generally the result of a very young
population with a very low $M/L$. Instead, most of these very bright
NCs have predominantly old populations. Hence, they are not just
luminous, but also very massive.

Our mass estimates agree reasonably well with the few other published
estimates for NCs in spiral galaxies. \citet{wal05} determined the
masses of NCs in nine late-type spiral galaxies from a combination of
velocity dispersion measurements and dynamical modeling. Four of the
galaxies (NGCs 300, 428, 1493 and 2139) are common to the STIS
subsample for which we performed spectral population fitting. For
these galaxies the average residual between our two studies (our
values minus theirs) is $\Delta \log M = -0.21 \pm 0.13$\,dex. This
good agreement is consistent with the findings of \citet{wal06}, who
also found that mass estimates from dynamical modeling and spectral
population fitting are more or less consistent. This agreement
implies, among other things, that any systematic errors due to the
assumed solar abundance ratios in the spectral population fitting
cannot be large. The average NC mass of all nine late-type spiral
galaxies in the \citet{wal05} sample is $\overline{\log M} = 6.51 \pm
0.18$, as compared to $\overline{\log M} = 6.25 \pm 0.21$ for the
late-type spirals in our sample. Dynamical mass measurements also
exist for the NCs in the nearby late-type spirals M33 \citep{kor93}
and IC~342 \citep{boe99}. The inferred values of $\log M \lta 5.85$
for M33 and $\log M \sim 6.78$ for IC~342 fit within the range of our
NC mass determinations for late-type spirals.

We have not attempted to infer formal uncertainties on the $M/L$
values and masses implied by the spectral population fits. One reason
for this is that the large number of free parameters and the
possibility of correlations between the parameters make this a
difficult problem. Also, it is likely that the true errors are
dominated by systematics, rather than the $S/N$ of the data. Either
way, the average uncertainty can be estimated from the comparison of
our results to the dynamical measurements of
\citet{wal05}. Based on this we estimate that the average error in
both $M/L$ and $M$ does not exceed $\sim 0.3$ dex. However, it should be
noted that the errors need not be the same for the different galaxies in
the sample. When the light of the NC is dominated primarily by a
single population, either young or old, the inferred $M/L$ and $M$
should be fairly robust. But if instead the light is dominated by a
young population while at the same time the mass is dominated by an
old population, then the $M/L$ and $M$ are more uncertain. When these
quantities are determined primarily by a population that forms only a
small percentage of the total light then the result is obviously more
susceptible to systematic uncertainties of various kinds. Similarly,
the values of $\langle \log \tau \rangle_M$ discussed in
Section~\ref{ss:ages} are then also more uncertain. An additional
uncertainty comes from our coarse sampling of the ages in our
template library. For example, we include spectra of $\log \tau =
10.00$ and $10.30$, but a population that is as old as the Universe
($\sim 13.5 \times 10^9$ yr, so that $\log \tau = 10.13$) can only be
approximated as a weighted mixture of the two. For the NCs in
early-type spirals we find that one has a luminosity-weighted age
$\langle \log \tau \rangle_L$ that actually exceeds the canonical age
of the Universe in a $\Lambda$CDM cosmology, and there are five NCs
that have a mass-weighted age $\langle \log \tau \rangle_M$ for which
this is true. So it is possible that the $M/L$ and $M$ values for
these NCs have been slightly overestimated.

\section{ASSESSMENT OF SYSTEMATIC EFFECTS}
\label{s:caveats}     

\subsection{AGN Contributions}
\label{ss:AGN}

The galaxies that we selected for STIS spectroscopy were all chosen
because they were found to have a compact nuclear source, as
identified from HST imaging. In principle such a source could be
either a star cluster or non-thermal emission from an AGN. However,
one can discriminate between these possibilities using the size of the
source. A star cluster is generally spatially resolved with HST in
nearby galaxies, whereas AGN emission is unresolved.\footnote{It is
usually assumed that the detection of a {\it resolved} nuclear source
implies that the source is not an AGN. While this is generally a valid
argument, it must be applied with some care. Continuum emission from
an AGN is indeed generally not spatially resolved. However, the
emission-line region in many low-luminosity AGN can often be resolved
in HST images \citep[e.g.,][]{pog00}. While a filter such as F606W (as
used by, e.g., \citet{car98}) transmits mostly continuum, it does have
a small emission-line contribution from H$\alpha$+[\hbox{N\,{\sc ii}}]
as well. Therefore, even if the nuclear source is an AGN, it is not
entirely ruled out that it could appear slightly extended in F606W.}
\citet{boe04} did a detailed study of the sizes of the nuclear sources
in late-type spirals. They found that all sources in their sample were
spatially resolved, indicating that they are star clusters. This is
not unexpected, given that we know from emission line studies that AGN
are very rare in the latest Hubble types \citep{ho97}. \citet{car98}
and \citet{car02} measured the sizes and optical vs.~near-IR colors of
nuclear sources in mostly early-type spirals. As in the late-type
spirals, the sources were almost always found to be extended. The
colors were found to be in the range $V-H = -0.5$ to $3.0$. This is
similar to the range covered by stellar population synthesis
predictions for star clusters with different ages and metallicities
\citep[e.g.,][]{bru03}. So in both late- and early-type spirals, the
nuclear sources detected from HST imaging appear in large majority to
be star clusters, and not AGNs. \citet{car98} did find some nuclear
sources for which the spatial extent was not inconsistent with a point
source. However, none of these made it into our spectroscopic STIS
sample.

The fact that the nuclear sources seen in HST imaging are generally
star clusters does not by itself preclude the possibility that an AGN
might contribute some emission as well. Based on information in NED,
only three of the 40 galaxies in our sample were previously classified
as an AGN (NGC 5377) or a LINER (NGCs 4540 and 6384). Only one of
these galaxies, NGC 5377, is part of the high $S/N$ STIS sample for
which we have performed spectral population fits. Indeed, NGC 5377
shows emission lines in our STIS spectra. Only two other galaxies show
emission lines as well, namely NGC 3177 and NGC 3277 (the latter has
the strongest emission lines of the three galaxies, including
[\ion{O}{2}] and [\ion{O}{3}]). And in these two galaxies the emission
lines could be due to \hbox{H\,{\sc ii}} regions, and need not be
indicative of an AGN. Nonetheless, the true fraction of AGN in our
sample could be larger, especially for early-type spirals. Based on
ground-based emission line spectra 10\% of Sb-Sbc galaxies are
classified as Type I AGN and 34\% as LINERs or transition objects
\citep{ho97}. Of course, emission lines themselves are masked in our
stellar population fitting. So an AGN would only affect our analysis
through its non-thermal continuum. This would have two effects: it
would change the overall continuum slope of the spectrum and it would
decrease the equivalent width of the absorption lines. These effects
are very similar to (and in our modeling are probably degenerate with)
the effects of changes in extinction in metallicity. However, the
sensitivity of the inferred population ages to uncertainties in
extinction and metallicity is not particularly large for the purposes
of our study. This is dicussed in detail in
Section~\ref{ss:robustness} and quantified in Figures~\ref{f:ageext}
and \ref{f:agemetal}. Therefore, the sensitivity of the inferred
population ages to a possible non-thermal AGN contribution is probably
not particularly large either.

In principle it is possible that an unknown number of galaxies in our
sample is affected by the presence of an AGN continuum. In those
galaxies we would still expect our age and $M/L$ determinations of the
NC to be relatively unbiased. However, the total luminosity and mass
estimates would be too high by an amount that depends on exactly what
fraction of the light in the STIS spectrum is from an AGN. Since AGNs
are more common in early-type spirals than in late-type spirals, this
would affect early-type spirals more than late-type spirals. NCs in
early-type spirals are in fact found to have higher luminosities
\citep{car02} and masses (Section~\ref{ss:masses}) than those in
late-type galaxies. However, AGN continua can probably not
explain this effect. An AGN continuum can be very difficult to
distinguish from a young stellar population dominated by O
stars. Therefore, the amount of AGN continuum in the STIS spectra is
approximately constrained to be no more than the percentage of the
light in our stellar population fits that is attributed to a very
young stellar population. For the early-type spirals, where we are
most concerned about potential AGN contributions, these percentages
are generally zero or very low (cf.~Table~\ref{t:contr}). Therefore,
potential AGN contributions are not expected to affect our overall
conclusions.

\subsection{Bulge and Disk Light Subtraction}
\label{ss:subtraction}     

Another source of uncertainty in our analysis is the accuracy with
which the galaxy (bulge and disk) light underneath the NC can be
subtracted. The high spatial resolution of HST allows us to do this
subtraction much more accurately than what is possible in a
ground-based study. Nonetheless, the subtraction is not perfect and it
is important to address how this might have affected the results. As
described in Section~\ref{ss:datared}, we subtracted the galaxy light
by subtracting the spectrum observed at $\sim 0\farcs2$ from the NC. This
assumes implicitly that the bulge and disk have approximately constant
surface brightness in the central $0\farcs2$. This is a simplification,
because bulges tend to have cusped brightness profiles that increase
all the way towards the center. This is not likely to affect the
results for late-type spirals, which have faint or negligible
bulges. So galaxy contamination to the nuclear spectrum is small, and
exactly how it is modeled and corrected has little effect on the final
results. However, early-type spirals have prominent bulges, and for
them the galaxy light subtraction is more important. So it is
important to address how potential inaccuracies in the galaxy light
subtraction might have affected our results for early-type spirals.

One important test of our galaxy light subtraction procedure is to
compare the luminosities derived from our spectra with those derived
from HST/WFPC2 imaging by \citet{car98}. Our spectra only provide a
one-dimensional light profile that has been integrated over a $0\farcs2$
wide slit (which reduces the contrast between the cluster light and
the underlying galaxy light), sampled at $0\farcs05$ resolution. By
contrast, WFPC2 imaging provides a two-dimensional image that is
sampled at $0\farcs05$ resolution in both orthogonal
directions. Therefore, the WFPC2 images allow more accurate and
sophisticated techniques for galaxy light subtraction than was
possible with our STIS spectroscopy, including allowance for a
possible central surface brightness cusp in the bulge light.
Nonetheless, the luminosities inferred from our spectra show little
bias as compared to those inferred from the imaging, and appear
accurate in an average sense to $\sim 0.2$ mag (see
Section~\ref{s:broadband}). Therefore, the comparison with the WFPC2
imaging results by itself provides no reason to mistrust the accuracy
of the galaxy light subtraction in our STIS spectra.

What remains as a caveat is the possibility that both our study {\it
and} the \citet{car98} study might have overestimated the NC cluster
luminosity in early-type spirals, by underestimating the bulge
contribution to the central few pixels. Our own STIS dataset does not
really allow us to test this possibility in much quantitative detail. 
However, \citet{car02} in their appendix~B described detailed and extensive 
tests of the accuracy of their decompositions of HST images into NC and 
bulge components. They found no evidence that their inferred NC luminosities 
would be systematically biased. Depending on the exact properties of the NC 
and the surrounding bulge, they did find that it is possible to infer NC
luminosities that are too faint or too bright.  However, for NCs with
radii that are similar to those typically observed, the errors are
generally no worse than $\pm 1$ mag. When averaged over a realistic
range of NC and bulge properties, the average bias in NC magnitudes
was found to be close to zero.

We have found here that NCs in early-type spirals are more massive
than those in late-type spirals. The difference in average mass is
$1.4$ dex (see Section~\ref{ss:masses}). To explain these findings as
a result of inaccuracies in the galaxy light subtraction, one would
need to assume that NC luminosities in early-type spirals have been
overestimated by an average of $\sim 3.5$ mag. In view of the results
of \citet{car02}, this seems highly unlikely. Nonetheless, it should
be acknowledged that the identification and characterization of a NC
embedded in a bulge with a steep luminosity profile is a difficult and
potentially non-unique problem. Therefore, additional investigations
into the accuracy of such decompositions remain desirable.

Possible inaccuracies in the galaxy light subtraction would not only
affect the determination of the NC luminosity, as just discussed, but
they might also affect our age and $M/L$ determinations for the NC.
If bulge light has inadvertently contributed to the spectra that we
have analyzed, and if bulges have older ages and higher $M/L$ than
NCs, then we will have overestimated both the $M/L$ values and masses
of the NCs in early-type spirals. This would go in the same direction
as the correlation observed in Figure~\ref{f:hubblevsmass}, and we
therefore need to address the possible importance of this effect.

To proceed we need some understanding of the stellar population
gradients in the central $\sim 20 \pc$ of the early-type spirals in
our sample (for the galaxy light subtraction we use the observed
spectrum at $\sim 0\farcs2$ from the galaxy center, which corresponds to
$\sim 20 \pc$ at the mean distance of our sample.) Unfortunately,
little is known about this. The stellar population properties of
bulges have been well-studied \citep{wys97}. However, most of this
work refers to much larger ($\sim$ kpc) scales, and to ``classical''
bulges with $R^{1/4}$ light profiles. By contrast, NCs are
preferentially found in ``pseudo''-bulges with exponential light
profiles \citep{car98}, the properties of which are not nearly as well
understood \citep{kor04}. Our own STIS data cannot be used to address
radial population gradients, because the $S/N$ of the spectra drops
too rapidly away from the center. Also, color gradients from HST
imaging in different bands is not generally available for the galaxies
in our sample.

Despite the difficulties in constraining stellar population gradients
on the relevant spatial scales in our sample galaxies, there do exist
some indirect arguments and observations that can provide
insight. First, \citet{car01} imaged a sample of (mostly early-type)
spirals with HST in both optical and near-IR wavelengths. They showed
that the $V-H$ colors of the NCs in their sample were always similar
to and consistent with the $V-H$ color of the surrounding
pseudo-bulge. This suggests that any stellar population gradients in
the very central regions of spiral galaxies may not be large.  Second,
as a test, we analyzed our STIS spectra both with and without
subtraction of the galaxy light observed at $0\farcs2$. The inferred
ages were found to be very similar for the two sets of spectra: the
age difference for the NCs has a mean of $\Delta \log \tau =
0.22\pm0.06$ dex for early-type spirals and $0.07\pm0.01$ dex for
late-type spirals. If there had been large population gradients
between the light in the galaxy center and at $0\farcs2$, then one
might have expected the inferred age to be more sensitive to the
galaxy light subtraction.  And third, for late-type spirals we find
that our age determinations are in excellent agreement with those of
\citet{wal06}, even though the latter were obtained with five times
worse spatial resolution and mixed in much more disk and bulge light
into the spectrum that was analyzed.

None of the arguments that we have presented prove conclusively that
the luminosities and $M/L$ values of the NCs, particularly in
early-type spirals, might not have been overestimated. On the other
hand, none of the available data and tests that we and others have
done suggest that this has in fact been the case either. Given the
available evidence, it seems unlikely that errors resulting from
galaxy-light subtraction could by themselves explain the correlation
of cluster mass with Hubble type shown in
Figure~\ref{f:hubblevsmass}. \citet{car02} found that NCs in
early-type spirals tend to be brighter on average than those in
late-type spirals. Therefore, NCs in early-type spirals must be more
massive than those in late-type spirals unless either: (1)
\citet{car02} systematically overestimated the NC luminosities in 
early-type spirals; or (2) NCs in early-type spirals are younger and
have significantly lower $M/L$ than those in late-type spirals. The
first possibility is not something that we can verify independently
with our own STIS data, but does seem to have been tested extensively
(and ruled out) by \citet{car02}. The second possibility is exactly
the opposite of what we conclude from our STIS data, namely that NCs
in early-type spirals are older than those in late-type spirals (see
Section~\ref{ss:ages}). It seems unlikely that systematic errors in
$M/L$ due to inaccuracies in the galaxy-light subtraction could be
large enough to explain this.

So in summary, the difficulty of performing an accurate subtraction of
underlying galaxy bulge and disk light certainly does cause some
uncertainty in our results. However, estimates of the size of this
uncertainty do not appear large enough to be able to explain the
finding of Section~\ref{ss:masses} that early-type spirals are more
massive than those in late-type spirals.

\subsection{Sample Biases}
\label{ss:biases}     

\subsubsection{Selection for Spectroscopy and Population Fitting}
\label{sss:specselec}

The results for the ages and masses of NCs discussed in
Sections~\ref{ss:ages} and~\ref{ss:masses} pertain to those galaxies
for which our STIS spectra had sufficient $S/N$ for spectral
population fitting. This biases systematically against faint
NCs. Also, an explicit brightness cutoff was also applied in the
selection of the NC sample for the STIS observations themselves (see
Section~\ref{s:sample}). The luminosity and $M/L$ of a stellar
population depend systematically on age, in the sense that young
populations are more luminous and have lower $M/L$. It is therefore
likely that the paucity of faint clusters in our sample has introduced
biases in the inferred age and mass distributions of the NCs. It is
important to correct for these biases to obtain results that are
representative for NCs as a class.

Figure~\ref{f:maghistcar} shows histograms of the distribution of
absolute NC magnitudes $M_{\rm F606W}$ for early-type spirals. The top
panel shows the distribution for all NCs detected by \citet{car98}. The 
middle panel shows the histogram for the NCs which we have observed with 
HST/STIS. The bottom panel shows the histogram for those NCs for which we 
performed spectral population fitting. The average values of the 
distributions are: $\overline{M_{\rm F606W}} = -12.12$ (WFPC2 sample; 
\citet{car98}), $-12.79$ (STIS sample), and $-13.19$ (spectral population 
fitting sample). Figure~\ref{f:maghistboe} is similar to
Figure~\ref{f:maghistcar}, but shows the distributions for late-type
spirals studied by \citet{boe02} as a function of absolute NC
magnitude $M_I$. The average values of these distributions are:
$\overline{M_I} = -11.68$ (WFPC2 sample; \citet{boe02}), $-12.29$
(STIS sample), and $-12.64$ (spectral population fitting sample). Overall,
our sample spans an interesting range of Hubble types and
luminosities. However, Figures~\ref{f:maghistcar}
and~\ref{f:maghistboe} show that the NCs for which we have performed
spectral population fits are on average $\sim 1.0$ mag brighter
than the average NC in a spiral galaxy, and they also sample a more
limited range of brightness than would be found in a random sample of
NCs.

To translate the luminosity bias in our sample into a mass bias we
need to know the average extinction, age and $M/L$ for the faint
clusters. However, without high $S/N$ spectra and spectral population
fits these quantities are unknown. If age, $M/L$ and extinction do not
correlate with NC luminosity, then a bias of $-1.0$ mag in brightness
simply corresponds to a bias of $+0.4$ in $\log M$. However, this
assumption need not be correct. It is therefore of interest to try to
constrain observationally whether age, $M/L$ and extinction correlate
with NC luminosity. We explored this issue explicitly for those NCs
for which we performed spectral population fits, but found no obvious
trends. However, this is not a very stringent test of possible
correlations. The NCs for which spectral population fits were
performed cover a more limited range of luminosities than all NCs in
spiral galaxies (see Figures~\ref{f:maghistcar}
and~\ref{f:maghistboe}). Therefore, subtle correlations might easily
have been masked by the intrinsic scatter in the quantities of
interest.

The fact that we have measured $B-V$ colors for {\it all} galaxies in
our STIS sample (see Table~\ref{t:color}), and not just the brightest
ones, allows us to study potential correlations over a larger range of
NC magnitudes. Figure~\ref{f:bminusv} shows histograms of $B-V$ for
the NCs in our STIS sample for both early- and late-type
spirals. Separate histograms are shown for the NCs that were bright
enough for spectral population fitting (the NCs listed in
Table~\ref{t:age}) and those that were too faint (the NCs that are
listed in Table~\ref{t:color} but not in Table~\ref{t:age}). For the
bright sub-sample the sample averages are $\overline{B-V} = 0.77 \pm
0.08$ for early-type spirals and $\overline{B-V} = 0.45 \pm 0.09$ for
late-type spirals. The fact that NCs in early-type spirals tend to
have redder colors than those in late-type spirals was already hinted
at in the HST study of $V-H$ colors by \citet{car02} (see their
Figure~6). This result is revealed more clearly by our sample because
it contains many more late-type spirals. Our spectral population
fitting has shown that this color difference is due primarily to a
difference in age: NCs in early-type spirals are generally older than
those in late-type spirals, while the dust extinction is similar.

An interesting aspect of Figure~\ref{f:bminusv} is the difference
between the bright and faint sub-samples. For both early- and
late-type spirals the distribution for the faint sub-sample is shifted
towards redder colors as compared to the bright sub-sample. The shape
of the distributions for the faint sub-samples also differs somewhat
between the early- and late-type spirals. The faint NCs in late-type
galaxies peak at about $B-V \sim 0.6$--0.9, with a small tail towards
bluer colors. By contrast, the faint NCs in early-type spirals show a
broader distribution. However, this is driven primarily by the
unusually red colors for the NCs in NGC\,5188 and NGC\,6384. This
could be primarily a signature of dust. An inspection of the HST/WFPC2
image in \citet{car98} for NGC\,5188 reveals a very complex nuclear
morphology including several dust lanes close to the nucleus,
indicative of heavy extinction. To some lesser extent this also
applies to NGC\,6384.

For early-type spirals the sample average for the faint sub-sample
(for which no spectral population fits were performed) is
$\overline{B-V} = 1.21 \pm 0.20$, compared to $\overline{B-V} = 0.77
\pm 0.08$ for the bright sub-sample. The difference in color can be
explained in several ways. For example, one can attribute the entire
color difference to the effects of dust. In that case the faint
sub-sample must on average have $\Delta A_V = 1.36 \pm 0.68$ mag more
extinction.  That would also explain why the NCs in the faint
sub-sample appear on average $1.2$ mag fainter than those in the
bright sub-sample. In this scenario, there is no need to assume that
our NC mass and age estimates in Sections~\ref{ss:ages}
and~\ref{ss:masses} are biased in any way. Alternatively, one can
attribute the differences in color between the sub-samples to
differences in age, because older populations are redder. However,
even a population as old as a Hubble time is not as red as $B-V=1.21$;
for $Z=0.02$ one has $B-V = 0.98$ and for $Z=0.05$ one has $B-V =
1.11$ \citep{bru03}. So at least part of the color differences must be
due to dust (as already mentioned, this is certainly plausible for
NGC\,5188 and NGC\,6384). Either way, in this scenario the average
$\langle \log \tau \rangle_L = 9.23$ that we derived for NCs in
early-type spirals from spectral population fits is probably a lower
limit. The NCs that were too faint to study could be as old as a
Hubble time. A population as old as a Hubble time is $\sim 2.0$ mag
less luminous in the $V$-band than a population of the same mass with
$\log \tau = 9.23$. For comparison, the NCs in the sample of
\citet{car98} are on average only $1.0$ mag fainter than those for
which we have performed spectral population fits. So the difference in
brightness could be due entirely to a difference in age. There need
not be a difference in average mass amongst the bright and faint
samples.

For late-type spirals the sample average color for the faint
sub-sample is $\overline{B-V} = 0.69 \pm 0.05$, compared to
$\overline{B-V} = 0.45 \pm 0.09$ for the bright sub-sample. Again,
this difference in color can be attributed either to a difference in
dust extinction ($\Delta A_V = 0.74 \pm 0.31$) or a difference in age
($\Delta \log \tau = 0.41 \pm 0.12$; based on Bruzual-Charlot models
of solar metallicity, assuming an age $\langle \log \tau \rangle_L =
8.37$ for the brighter sub-sample, cf.~Section~\ref{ss:ages}). The
difference in age would correspond to a difference in brightness of
$0.55$ mag in the $I$-band, for a population of fixed mass. Both
scenarios would explain why the NCs in our faint sub-sample are on
average $0.6$ mag fainter than those in our bright sub-sample, without
any need to invoke a difference in NC mass between the sub-samples.
The average $I$-band absolute magnitude for our faint sub-sample is
still $0.4$ mag brighter than the average for all NCs in the
\citet{boe02} sample. This, too, could be due to differences in dust
absorption ($\Delta A_V = 0.85$), age ($\Delta \log \tau = 0.30$),
mass ($\Delta \log M = 0.16$), or any combination thereof. While some
difference in mass cannot be ruled out, a difference of $\Delta \log M
= 0.16$ would be no larger than the random uncertainties in our
spectral population fitting sample averages (which are $\Delta \log M
\approx 0.2$).

Figure~\ref{f:bminusv} shows differences in $B-V$ color between the
bright and faint NCs in our sample, and as discussed, this could be
due to differences in either dust extinction or age. While in general
it is difficult to discriminate between these alternative
explanations, differences in age could be the more plausible
explanation. For the NCs that we have been able to study in detail,
the amount of dust extinction was generally found to be quite small
(Section~\ref{ss:metalext}). Figure~\ref{f:bvage} shows the
correlation between $\langle \log \tau \rangle_L$ and $B-V$ for the
NCs for which we performed spectral population fits
(Table~\ref{t:age}).  It is clear that at least for this sub-sample
the $B-V$ color is almost uniquely an indicator of age, with older NCs
having redder colors.  Variations in dust extinction between the NCs
add only a small amount of scatter to an otherwise very tight relation
(the parameters of the best linear fit and the Spearman rank-order
correlation coefficient are given in Table~\ref{t:fitcoeff}).

As a result of our analysis of the $B-V$ colors of the NCs we conclude
that the inferred average ages discussed in Section~\ref{ss:ages} are
probably biased in the sense that they are younger than the true
average for all NCs in spiral galaxies. Depending on whether or not
there is a systematic difference in dust extinction between faint and
bright NCs, the bias might be anywhere in the range $\Delta \log \tau
= 0.0$--0.6 dex. For many early-type spirals this implies that the NC
population need not be any younger than the galaxy itself. This is not
generally true for NCs in late-type spirals. Even when a possible
age-bias is accounted for, it remains true that most NCs in late-type
spirals have average population ages significantly younger than a
Hubble time. More generally, while there are some biases in our
spectral population fitting results, none of these can account for the
differences that we have found between NCs in early-type and late-type
spirals. NCs in early-type spirals are older and more massive than
those in late-type spirals.

By contrast to our results for the cluster {\it ages}, we have found
no evidence that the average NC {\it masses}, discussed in
Section~\ref{ss:masses}, are systematically biased due to our
inability to spectroscopically analyze the populations of faint
clusters.

\subsubsection{Completeness of Imaging Surveys}
\label{sss:imcomplete}

Another potential selection effect inherent to our work is that the
original HST imaging surveys of \citet{car98} and \citet{boe02} may
have been biased against the detection of faint clusters.

\citet{boe02} found NCs in $\sim 75$\% of the galaxies in
their HST imaging study of late-type spiral galaxies. The absolute
magnitude distribution for these clusters showed a cutoff at $M_I
\approx -9$. It was argued that this cutoff is real, based on the fact that
clusters as faint as $M_I \approx -8$ could easily have been detected
in all galaxies. Late-type spirals do not have prominent bulges making
the detection of NCs relatively straightforward. On the other hand, no
firm upper limits were set on the luminosity of potential clusters in
the centers of those galaxies where no cluster was detected. So it
remains possible that some of these galaxies do host a NC, but that it
simply was too faint to be detected.

\citet{car98} found NCs in $\sim 50$\% of the galaxies in their HST
imaging study of mostly early- and intermediate-type spiral
galaxies. This should definitely be regarded as a lower limit. Due to
their brighter bulges, NCs are more difficult to detect in early-type
spirals than in late-type spirals. In several galaxies it was only
possible for \citet{car98} to set an upper limit to the magnitude of
any potential NC. In those cases, where the bulge had a very steep
$R^{1/4}$ profile, these upper limits were not particularly
stringent. So it is possible that some of these galaxies do host a NC,
but that it simply was too faint to be detected. As a corollary of
this, there is actually no firm statistical ground to conclude that
the prevalence of NCs is lower in early-type spirals than in late-type
spirals.

The possibility that some NCs might have gone undetected in the
original imaging surveys is of some concern. On the other hand, there
are two arguments that mitigate the potential impact of this on the
interpretation of our results. First, the detection rate of NCs has
been quite high in both early- and late-type spirals. Therefore, our
results are probably representative at least for the majority of the
NCs. So while our sample average NC properties could be biased towards
higher luminosities and masses, the bias is not expected to be very
large. Second, for those galaxies in which a NC has gone undetected in
the imaging surveys it is not necessarily true that the NC must be
atypically faint. Those galaxies might simply have had atypically
bright or concentrated bulges, or unusually complicated or dusty
morphologies. Therefore, our results may well be typical also for the
NCs that reside in galaxies in which they were not detected.

One of the more interesting findings of our work is the correlation of
NC mass with Hubble type. One potential selection effect in this
correlation is that low-mass clusters may have preferentially gone
undetected in early-type spirals, as compared to the late-type
spirals. In other words, NCs could reside in the bottom-left of
Figure~\ref{f:hubblevsmass} that were not represented in our
spectroscopic survey. But even so, there remains a clear distinction
between the NC masses in early and late Hubble types.  Many NCs in
early-type spirals are quite massive, $\log M_{\rm NC}\,[\Msun] \gta
7$, whereas such massive clusters are almost non-existent in late-type
spirals. This is not a selection effect. There is no reason why
massive NCs would not be detected in late-type spirals; they should in
fact be the easiest to detect. Therefore, the absence of points in the
upper-right part of Figure~\ref{f:hubblevsmass} is real.

\section{RELATION BETWEEN CLUSTER MASS AND GLOBAL GALAXY PROPERTIES}
\label{s:correlations}

An important finding of the WFPC2 study of \citet{car98} was that the
NC luminosity (i.e. absolute magnitude) correlates with the host
galaxy luminosity for early-type spirals. \citet{boe04} subsequently
found a similar correlation for NCs in late-type spirals. Both
studies find that more luminous galaxies tend to have more luminous
NCs. However, the relations between these quantities are not the same
for NCs in early-type and late-type spiral galaxies. They are offset
and have a somewhat different slope. At fixed galaxy luminosity, NCs
in early-type spirals tend to be more luminous than those in
late-type spirals. This dependence on Hubble type was seen also in the
study of \citet{car02}, who concluded that NCs in early-type spirals
can be significantly more luminous than those in late-type spirals.

One of the advantages allowed by the present spectral population
fitting study is that it yields the mass of the NCs, and not just
their luminosity. The mass of a cluster is at some level a more
fundamental quantity than its luminosity, because the latter can be
unduly influenced by variations in stellar population age and dust
extinction amongst galaxies. We therefore use the inferred cluster
masses to revisit the previously established correlations amongst NC
properties and galaxy properties.

Figure~\ref{f:masstotallum} shows the relation between NC mass $M$
(from Table~\ref{t:age}) and total blue galaxy luminosity $L_B$ for
the 19 galaxies for which we performed spectral population fits. The
quantity $L_B$ was calculated from the apparent magnitude $B_{T,0}$
(extinction- and K-corrected) listed in the RC3 \citep{dev91} and the
distances given in Tables~\ref{t:obs1} and~\ref{t:obs2}. There is a
correlation in the expected sense, with more luminous galaxies
typically hosting more massive clusters. However, the correlation is
not very strong, and is significant only with 92\% confidence (see
Table~\ref{t:fitcoeff}). One reason why the relation may appear weaker
than those presented by \citet{car98} and \citet{boe04} is that those
authors considered samples that covered a smaller range of Hubble
types. Since NCs masses correlate strongly with Hubble type (see
Figure~\ref{f:hubblevsmass}), any sample that mixes a large range of
Hubble types will present more scatter.

Since NC luminosities and masses appear to depend strongly on both
galaxy luminosity and Hubble type, it is useful to ask whether there is
not one single global galaxy property with which NC mass correlates
more strongly. If so, then that might be the more fundamental quantity
that drives these correlations. The situation is reminiscent of that
for the masses of black holes in galaxy centers. Those tend to be
larger in bright galaxies than faint galaxies (e.g., M87 vs. M32) and
tend to be larger in galaxies with big bulges than in those with small
bulges (e.g., the Sombrero galaxy NGC\,4594 vs.~the Milky Way). It was
realized that these facts indicate that while black hole mass
correlates only weakly with global galaxy properties such as Hubble
type or luminosity, it correlates much more strongly with quantities
that describe the bulge. In particular, tight correlations exist with
bulge velocity dispersion \citep[e.g.,][]{tre02}, mass
\citep[e.g.,][]{har04} and luminosity \citep[e.g.,][]{marc03}.

For most of the galaxies in our sample we do not have access to either
bulge velocity dispersions or detailed disk-bulge
decompositions. However, a rough estimate of the bulge-to-total light
ratio for each galaxy can be obtained from its Hubble $T$-type (which
we rounded to the nearest integer) using the correlations presented by
\citet{sim86}. Of course, it should be kept in mind that these 
correlations have significant scatter. Moreover, the correlations are
based on the assumption that the bulge can be described by an
$R^{1/4}$ surface brightness profile, which is suspect for very late
Hubble types \citep{kor04}. Nonetheless, the bulge-to-total light
ratios suggested by the correlations of \citet{sim86} provides an
useful low-order estimate. Upon multiplication by the total galaxy
luminosity, $L_B$, this yields an estimate for the bulge luminosity
$L_{{\rm bul,}B}$. Figure~\ref{f:massbulgelum} shows the NC mass for
the spectral population fitting sample as a function of the inferred
$L_{{\rm bul,}B}$.

There is a clear trend of increasing NC mass $M$ as a function of
increasing bulge luminosity. The correlation is statistically
significant at 99.99\% confidence (see Table~\ref{t:fitcoeff}). The
RMS scatter in $M$ around the best fit is 0.72 dex. This is larger
than the observational errors of $0.3$ dex in our mass estimates (see
Section~\ref{ss:masses}), so there does appear to be a certain amount
of intrinsic scatter in the correlation. The scatter in the
correlation of cluster mass with bulge luminosity is similar to that
for the correlation with Hubble type ($0.61$ dex,
cf.~Figure~\ref{f:hubblevsmass} and Table~\ref{t:fitcoeff}). Both
correlations are tighter than that with total galaxy luminosity
(Figure~\ref{f:masstotallum}). The correlation with bulge luminosity
is a combination of the correlations with Hubble type and with total
galaxy luminosity; it takes into account both the general fact that
early-type spirals have higher bulge-to-total luminosity ratios than
late-type spirals, and the specific fact the early-type spirals in our
sample are slightly more luminous than the late-type spirals in our
sample (Figure~\ref{f:masstotallum}).

Some care must be taken in the interpretation of the correlation
between NC mass $M$ and $L_{{\rm bul,}B}$. First of all, there might
have been a selection effect against NCs in the bottom right of
Figure~\ref{f:massbulgelum}, for the reasons discussed in
Section~\ref{sss:imcomplete}. Second, the correlation between NC mass
$M$ and $L_{{\rm bul,}B}$ is not truly a one-to-one
correlation. Probably not all spiral galaxies host NCs. \citet{boe02}
found NCs in $\sim 75$\% of the galaxies in their HST imaging study of
late-type spiral galaxies and \citet{car98} found NCs in $\sim 50$\%
of the galaxies in their HST imaging study of mostly early- and
intermediate-type spiral galaxies. Therefore, there exist galaxies
that have bulges, but no NCs. Conversely, some of the late-type spiral
galaxies that host NCs do not appear to have a bulge. \citet{sim86}
perform disk-bulge decompositions for galaxies with Hubble types as
late as Sd ($T = 7$), and our estimates of $L_{{\rm bul,}B}$ use their
results. However, detailed analysis of HST images of galaxies with
such late Hubble types indicates that some of these galaxies 
simply do not host bulges at all \citep{boesta03}. This is even more true
for galaxies of Hubble type Sm. For the two galaxies in
Table~\ref{t:age} with this Hubble-type we assigned a bulge-to-total
light ratio based on extrapolation of the fitting function of
\citet{sim86}.  However, the resulting values of $L_{{\rm bul,}B}$
should probably be considered upper limits (as indicated by arrows in
Figure~\ref{f:massbulgelum}), since these galaxies likely do not have
a bulge at all.

The correlation of NC mass with bulge luminosity suggests that the
global properties of the host galaxy play an important role in the
formation process of NCs. There is evidence that the properties of
both NCs and bulges depend sensitively on secular evolution
processes in galaxy disks \citep{kor04}. The stellar populations of
NCs are generally best fit as a mixture of populations of different
ages (see Section~\ref{ss:ages}). This indicates that NCs did not form
in a single event, but instead they had additional star formation long
after the oldest stars formed. This adds to other accumulating
evidence that secular evolution plays an important role in shaping the
central regions of spiral galaxies. Nonetheless, the physical
processes that form bulges and NCs are overall still understood
relatively poorly. While the results presented here provide useful new
additional constraints, it is not currently clear what causes the
observed differences in NC properties as a function of Hubble type.

The present day NC mass is a function of many variables, including:
the mass of the NC at its initial formation; the average mass that was
added during each subsequent burst of star formation; the amount of
time between bursts; and the number of bursts experienced since formation 
(which may or may not depend on the age of the galaxy). All of these
variables might plausibly depend on the Hubble type. Scenarios with
different combinations of these variables may all be able to explain
the available data equally well. On the other hand, it is useful to
note that some simple scenarios do appear to be ruled out by our
data. For example, one might have postulated that the histories of NCs
across the Hubble sequence are more of less identical except for the
fact that some NCs have more frequent or more massive bursts of star
formation than others. This would explain why some NCs are more
massive than others. But one would then expect that high-mass
clusters have average population ages that are equal to or younger
than those in low-mass clusters. However, this is not
observed. Figure~\ref{f:massvsage} shows the relation between $\log M$
and $\langle \log \tau \rangle_L$ (from Table~\ref{t:age}) implied by
our spectral population fitting results. Since both quantities
individually correlate with Hubble type they also correlate with one
another, in the sense that the NCs with the highest masses tend to
have the oldest populations. Therefore, the differences in NC masses
as a function of Hubble type must involve more parameters than just
variations in the mass and frequency of star formation bursts. For
example, it is possible that NCs in early-type spirals simply form
with higher masses than those in late-type spirals to begin with.

The relation between NCs and the overall globular cluster systems of
spiral galaxies remains at present unclear, but there could be
connections. One formation mechanism for NCs is through the inspiral
of globular clusters into the galactic center through dynamical
friction. This process typically takes anywhere between a Gyr and a
Hubble time \citep{mil04}, which is not inconsistent with the inferred
population ages of many of the NCs in our sample. The GCs of the Milky
Way are almost uniformly more metal poor than the NCs analyzed in our
sample. So in our Milky Way it is unlikely that one could form a NC
from GCs. However, the GC system of the Milky Way is not
representative of all other galaxies. Several galaxies are known with
populations of metal rich GCs such as the LMC \citep{wes97} and
NGC\,5128 \citep{yi04}. Metallicity considerations by themselves
therefore seem insufficient to validate this scenario. Also, it has
previously been found that individual (globular) clusters in galaxies
such as M33 follow a correlation between cluster mass $M$ and
population age $\langle \log \tau \rangle_L$ that is not dissimilar to
the correlation for NCs in Figure~\ref{f:massvsage} \citep{cha99}. It
is not obvious whether these results are related, because individual
star clusters are generally less massive ($10^{3-5} \Msun$) than NCs
and reside in a less dense environment. Nonetheless, it could indicate
a commonality in the physical process that governs the formation of
both types of clusters.  Such a commonality is further supported by
the finding that both types of cluster fall on the same fundamental
plane relations \citep{wal05}.

Another important quantity that correlates strongly with the bulge
luminosity is the mass $M_{\rm BH}$ of supermassive black holes (BHs)
in galactic nuclei. \citet{marc03} collected accurate black hole mass
determinations from a variety of sources and determined the best-fit
linear correlations of $\log M_{\rm BH}$ vs.~$L_{\rm bul}$ in various
bands. While the correlation has its lowest scatter in the near-IR
$K$-band, there is a clear correlation in the $B$-band as well. This
correlation is overplotted as a dashed line in
Figure~\ref{f:massbulgelum}. Its scatter is quoted by \citet{marc03}
as $0.48$ dex. The slope of the relation is identical to within the
statistical uncertainties to the slope of the relation between $\log
M_{\rm NC}$ and $L_{{\rm bul},B}$. The latter correlation has a
somewhat larger scatter though ($0.72$ dex,
cf.~Table~\ref{t:fitcoeff}). As mentioned previously, it should be
kept in mind that not all spiral galaxies have NCs, while some have
NCs but not bulges. On the other hand, these same things could be true
for BHs as well. With well-known exceptions such as the Milky Way and
NGC\,4258, accurate BH mass determinations for spiral galaxies remain
rather scarce \citep[see e.g.,][]{marco03,atk05}. So we do not really
know whether there exist some galaxies with bulges that do {\it not}
host BHs. And we do know that some supermassive BHs (as revealed by
AGN activity) exist in very late-type galaxies such as the Sd spiral
NGC\,4395 \citep{fil03,pet05} and the dE galaxy Pox\,52 \citep{bar04}
that probably do not have bulges. Also, the globular cluster G1 which
could have an intermediate-mass BH \citep[e.g.,][]{geb02,geb05} has
been suggested to have been the NC of a (now disrupted) dwarf
galaxy. Either way, it is not clear if there is a causal connection
between the NCs and BHs that both seem to be common in (spiral)
galaxies. Such a connection is not unimaginable, since star clusters
have often been suggested as the source of the seed BHs that
subsequently grow supermassive through merging or accretion
\citep[e.g.,][]{por02,vdm03}. Comparison of the lines in
Figure~\ref{f:massbulgelum} suggests that at fixed bulge luminosity
$M_{\rm NC} \approx 3.3 \times M_{\rm BH}$. However, this does not
imply that NCs and BHs necessarily exist in the same galaxies. At
present there are no individual galaxies for which both a NC mass
determination and a secure BH mass determination exist. There are some
galaxies with NCs that have BH mass upper limits
\citep[e.g.,][]{boe99,geb01,mer01}. The most interesting limit exists
for M33, which has $M_{\rm BH} / M_{\rm NC} \leq 10^{-2}$. The safest
conclusion from the data available at the present time is that both
the properties of NCs and BHs in the centers of spiral galaxies seem
intimately connected to the properties of the host galaxy, and in
particular its bulge component.

\section{SUMMARY AND CONCLUSIONS}
\label{s:summary}

We obtained longslit spectra with HST/STIS of a sample of 40 NCs in
spiral galaxies of both early and late Hubble types. The sample was
selected from the HST/WFPC2 imaging studies of \citet{car98} and
\citet{boe02}, respectively. At a spatial resolution of $\sim 0\farcs2$ the 
spectra provide a much better separation of the NC from underlying bulge 
and disk light than is possible with ground-based spectra.

For the 19 NCs with highest $S/N$ we performed spectral fitting to
analyze their stellar populations. A weighted superposition of
single-age stellar population models from the spectral library of
\citet{bru03} was fitted in a $\chi^2$ sense to the observed STIS
spectra over the wavelength range from 3540\,{\AA} to 5680\,{\AA}. The
models provide excellent fits to the data. They yield the star
formation history of each NC, the $B$-band mass-to-light ratio $M/L_B$,
and the luminosity-weighted and mass-weighted mean population ages,
$\langle \log \tau \rangle_L$ and $\langle \log \tau \rangle_M$. We
find the average uncertainty in the inferred age to be $\Delta
\langle \log \tau \rangle_L \lta 0.3$ dex. 

We used the spectra for all 40 targets to determine the NC luminosity
$L_B$ in the $B$ band (taking into account corrections for aperture
losses), as well as the $B-V$ color. The inferred luminosities are
broadly consistent with those obtained from the HST/WFPC2 imaging
studies. For the galaxies with spectral population fits we combined
$L_B$ with the inferred values of $M/L_B$ to estimate the NC mass
$M_{\rm NC}$. We find the average uncertainty in the inferred mass to
be $\Delta M_{\rm NC} \lta 0.3$ dex. The inferred masses agree with
dynamical mass measurements for the few NCs in our sample for which
such measurements are available.

The metallicity $Z$ and dust extinction $A_V$ were varied to optimize
the spectral population fits. In general the NC metallicities are close
to solar, but they tend to be somewhat higher in early- than in
late-type spirals. The average extinction towards the NCs is relatively 
low, ${\bar A}_V = 0.3$--$0.4$, and does not differ much between early- 
and late-type spirals. Tests showed that uncertainties in the assumed values 
of $Z$ and $A_V$ do not significantly affect the inferred population ages.

The best spectral fits generally contain a mixture of populations of
different ages. Approximately 53\% of the NCs contain a population
younger than 1\,Gyr. The NCs show a wide range of luminosity-weighted
ages, ranging from 10 Myrs to 10 Gyrs. On average, NCs in late-type
spirals are younger than in early-type spirals ($\langle \log \tau
\rangle_L = 8.37 \pm 0.25$ vs.~$9.23 \pm 0.21$). For both galaxy types
the NC age is less than the typical age of galactic disks. The stellar 
populations of NCs are generally best fit as a mixture of populations of 
different ages (see Section~\ref{ss:ages}). This indicates that NCs did 
not form in a single event, but instead they had additional star formation 
long after the oldest stars formed. The average {\it mass}-weighted 
ages are considerably older than the luminosity-weighted ages, typically 
by $\sim0.7$ dex. So there often is an underlying older population that does 
not contribute much light, but does contain most of the mass.

The average NC masses are smaller in late-type spirals than in
early-type spirals ($\overline{\log M} = 6.25 \pm 0.21$ vs.~$7.63 \pm
0.24$). For both galaxy types the NCs are much more massive than
typical globular clusters in the Milky Way. However, the lowest NC
masses overlap with the high-mass end of the globular cluster mass
distribution in some other galaxies.

We presented a detailed discussion of various systematic effects that
might have influenced our results, including contamination from AGN
emission, uncertainties in the subtraction of underlying bulge and
disk light, and sample biases. Each of these probably play a role at
some level. However, we argued that none of them can explain the
differences that we find between NCs in early- and late-type spirals.

The primary impact of sample biases is that the average NC ages for
our sample might be lower than they would have been for an unbiased NC
sample. This is because older NCs are fainter and are therefore less
likely to yield sufficient $S/N$ for spectral population fitting. However, 
based on a joint analysis of luminosities and $B-V$ colors (available for 
all the NCs in our STIS sample, even those with low $S/N$ spectra) we found 
no evidence that the inferred NC masses are necessarily biased.

Our findings are broadly consistent with those obtained in other
recent spectroscopic studies of NCs, in particular those of
\citet{wal05,wal06} and \citet{sar05}. The primary new contribution
of the present work is the added insight about the variations in NC
properties as a function of Hubble type. This particular issue was
addressed previously by \citet{car02}. They found that NCs in
early-type spirals have a higher average luminosity than those in
late-type spirals. Our work shows that this is {\it not} because NCs
in early-type spirals are younger. To the contrary, we actually find
them to be older. Therefore, NCs in early-type spirals are not just
more luminous than those in late-type spirals, but also more massive.

The NC mass $M_{\rm NC}$ correlates loosely with total galaxy
luminosity. It correlates more strongly with both the Hubble type of
the galaxy and the luminosity of the host galaxy bulge. But it should
be kept in mind that not all spiral galaxies necessarily have a bulge
or a NC. Also, the correlations do have considerable scatter
($0.6$--$0.7$ dex in $\log M_{\rm NC}$ at a fixed Hubble type or bulge
luminosity). The correlation of $\log M_{\rm NC}$ with bulge
luminosity is particularly intriguing because it has the same slope as
the well-known correlation between supermassive black hole mass
$M_{\rm BH}$ and bulge luminosity. At fixed bulge luminosity the
correlations predict $M_{\rm NC} \approx 3.3 \times M_{\rm BH}$. This
does not imply that NCs and black holes necessarily exist in the same
galaxies, so the implications of this relation remain
unclear. However, it does appear that the properties of both NCs and
black holes in the centers of spiral galaxies are intimately connected
to the properties of the host galaxy, and in particular to its bulge
component.

%%%%%%%%%%%%%%%
% Acknowledgments
%%%%%%%%%%%%%%%

\acknowledgments 
Support for proposals \#9070 and \#9783 was provided by NASA through a grant 
from the Space Telescope Science Institute, which is operated by the 
Association of Universities for Research in Astronomy, Inc., under NASA 
contract NAS 5-26555. We thank the anonymous referee for suggestions that 
helped improving the presentation of the paper. This research has made use of 
the NASA/IPAC Extragalactic Database (NED) which is operated by the Jet 
Propulsion Laboratory, California Institute of Technology, under contract 
with the National Aeronautics and Space Administration. We also made use of 
the Lyon Extragalactic Database (LEDA).

\clearpage

%%%%%%%%%%%%%%%
% Reference List
%%%%%%%%%%%%%%%

\clearpage

%%%%%%%%%%%%%%%
% Figures 
%%%%%%%%%%%%%%%

%%% FIGURE %%%

%\clearpage

\begin{figure}
\psfig{file=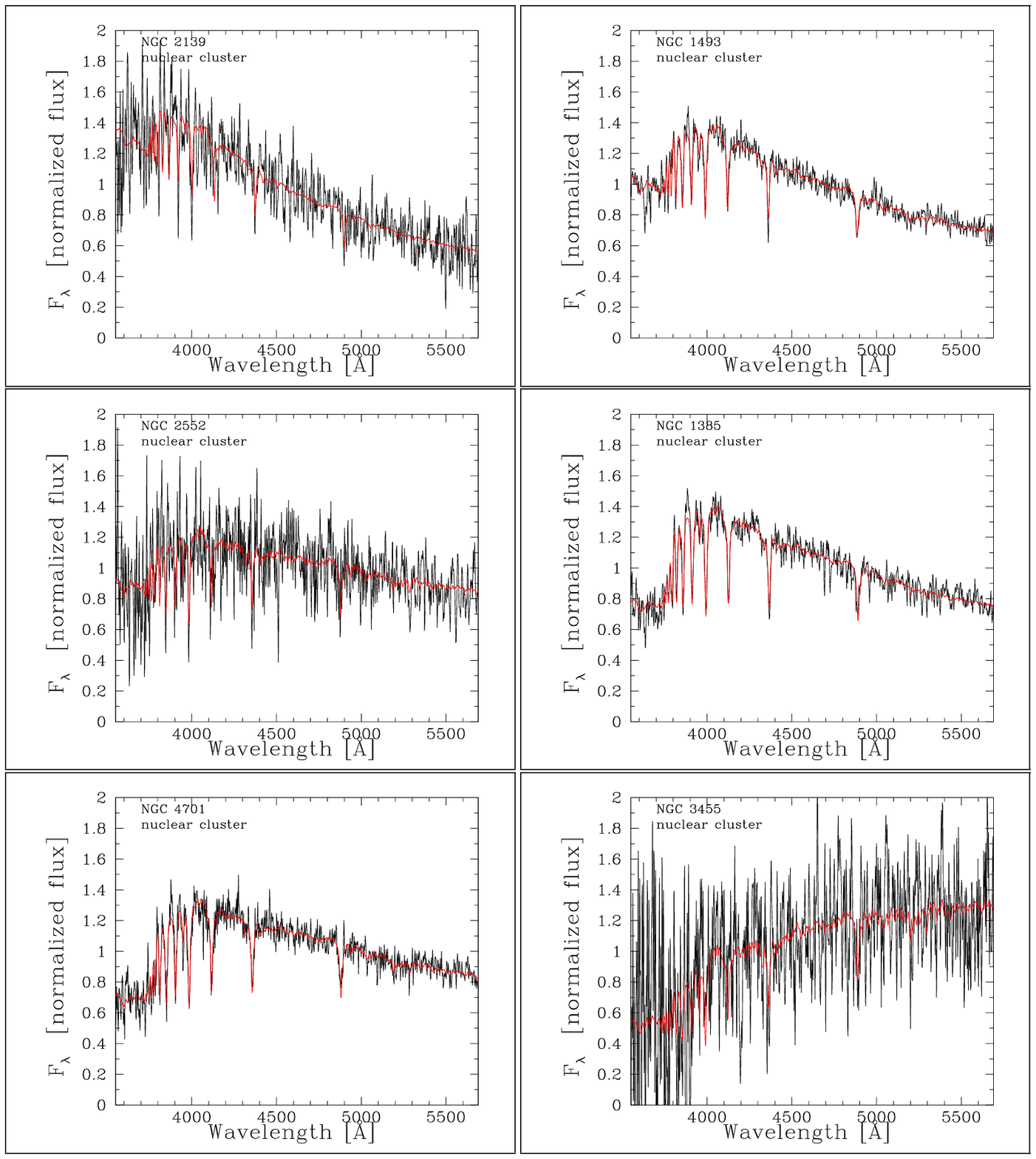,width=16cm,angle=0,clip=t}
\caption{HST/STIS spectra of the NCs 
(black), overlaid with the best spectral population fits
(red) obtained as described in the text. Characteristic properties of
the fits are listed in Table~\ref{t:age}. The spectra are sorted (from
left to right and from top to bottom) in order of increasing
luminosity-weighted age $\langle \log \tau \rangle_L$, with galaxy names 
indicated in the individual sub-panels.\label{f:spectraone}}
\setcounter{figure}{0}
\end{figure} 

%%% FIGURE %%%

\clearpage

\setcounter{figure}{0}
\begin{figure}
\psfig{file=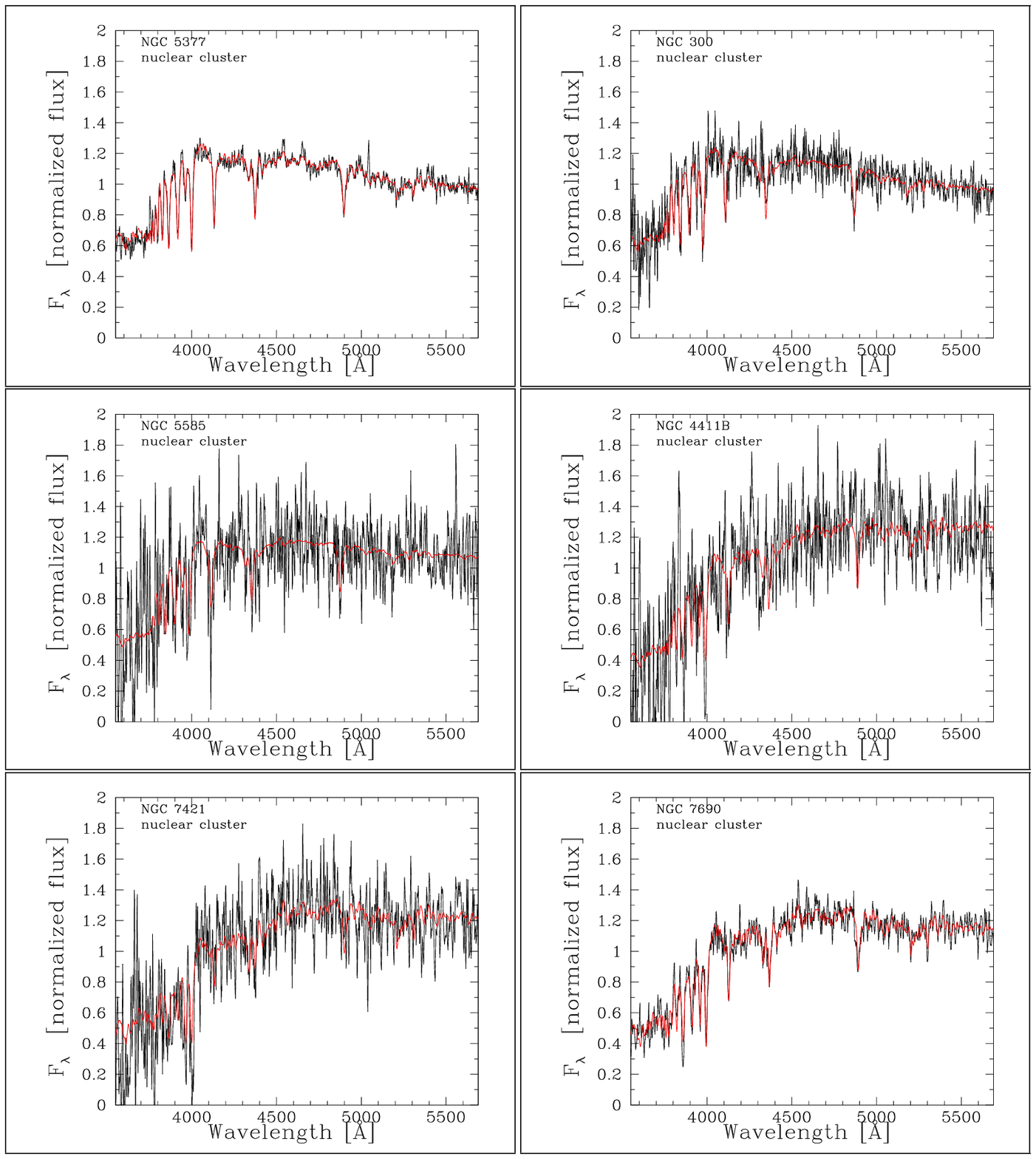,width=16cm,angle=0,clip=t}
\caption{(continued)}
\end{figure} 

%%% FIGURE %%%

\clearpage

\setcounter{figure}{0}
\begin{figure}
\psfig{file=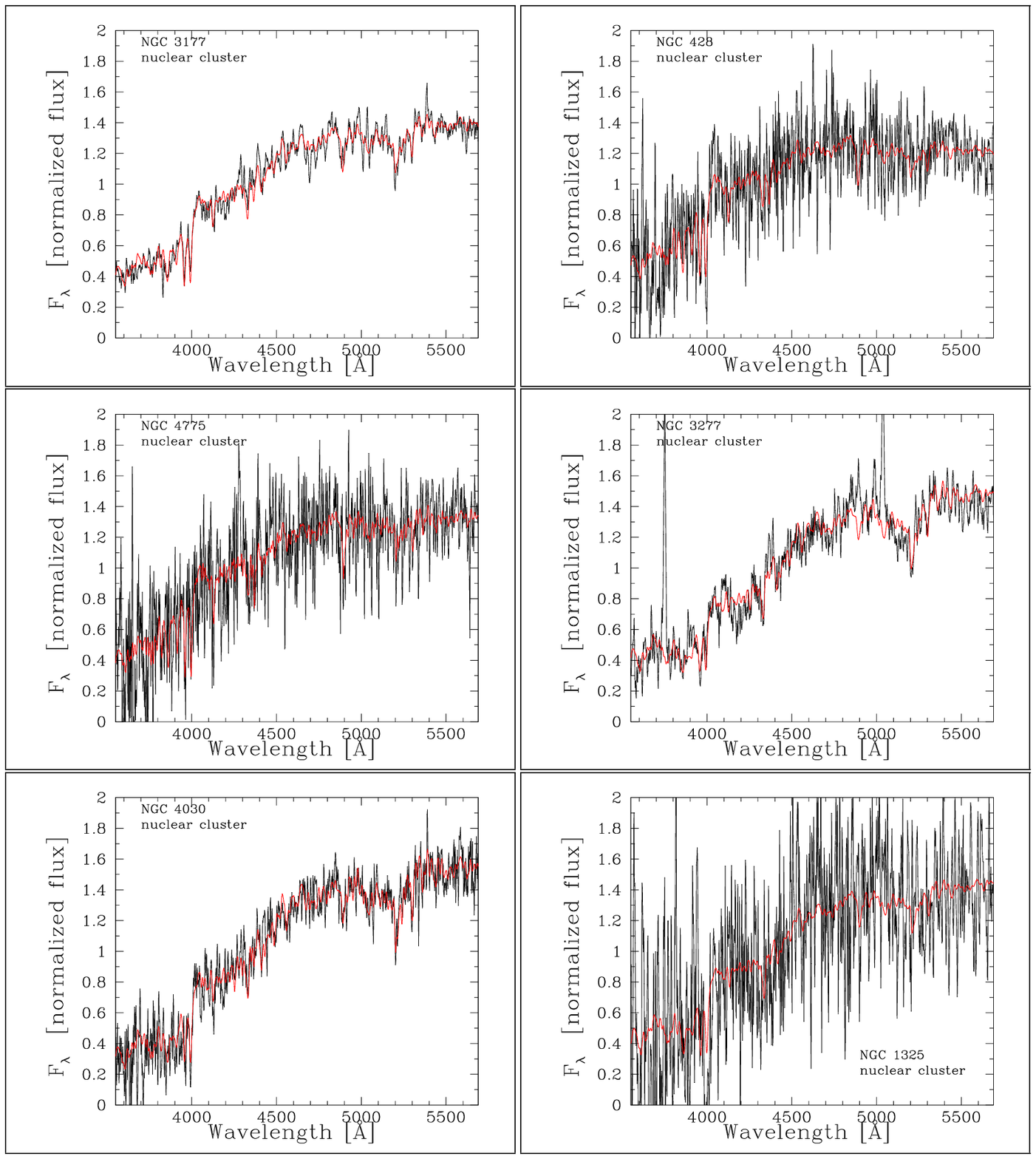,width=16cm,angle=0,clip=t}
\caption{(continued)}
\end{figure} 

%%% FIGURE %%%

\clearpage

\setcounter{figure}{0}
\begin{figure}
\psfig{file=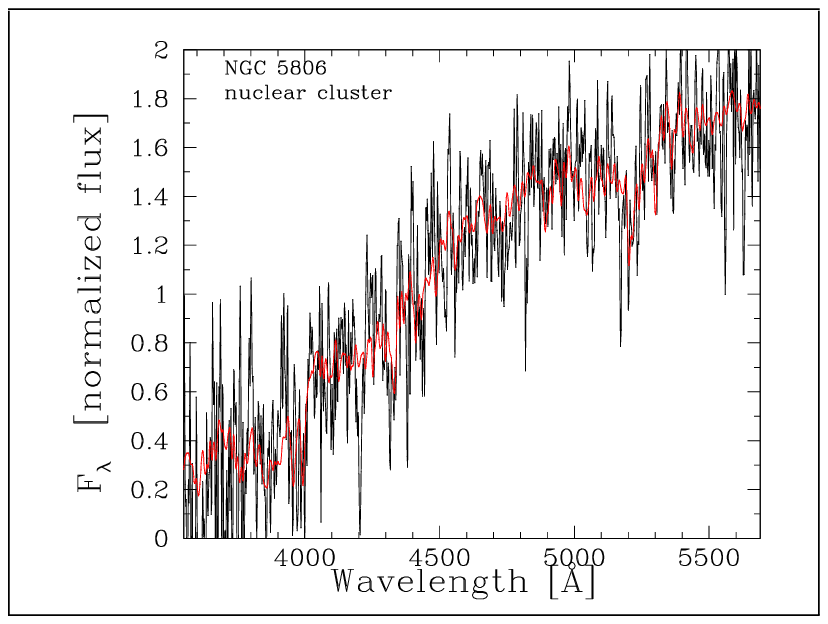,width=16cm,angle=0,clip=t}
\caption{(continued)}
\end{figure} 

%%% FIGURE %%%

\clearpage

\begin{figure}
\psfig{file=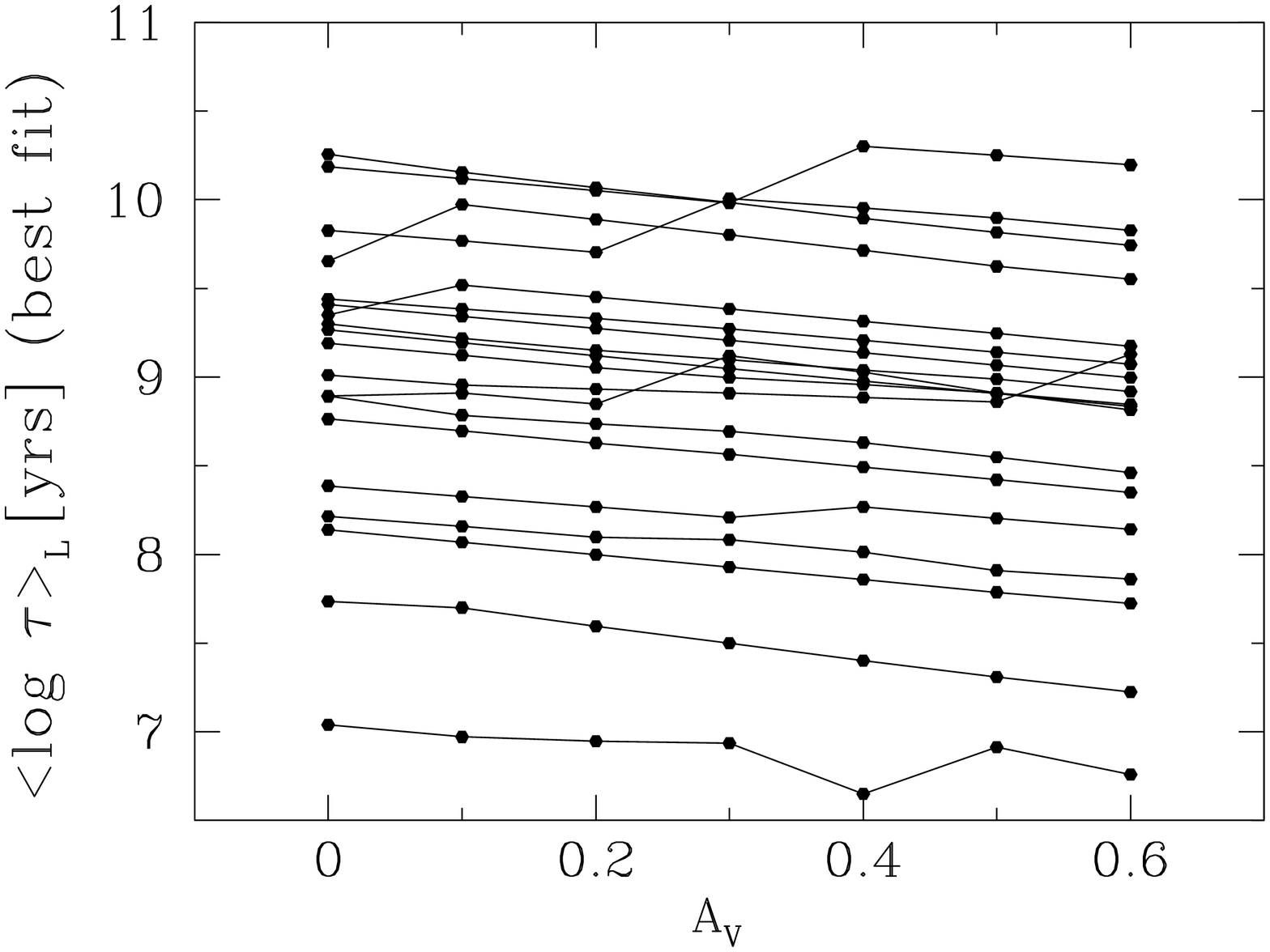,width=16cm,angle=270,clip=t}
\caption{Dependence of the best-fit derived luminosity-weighted ages as 
a function of the extinction $A_V$ for all 19 NCs. Results for the same 
NC at different extinction are connected by a line. Age uncertainties due 
to uncertainties in extinction are relatively small and are much less than 
the variation in age from cluster to cluster. The extinction range shown 
is $0.0$ to $0.6$, which contains the best-fit extinction for most of the 
galaxies in the sample (see Table~\ref{t:age}).\label{f:ageext}}
\end{figure} 

%%% FIGURE %%%

\clearpage

\begin{figure}
\psfig{file=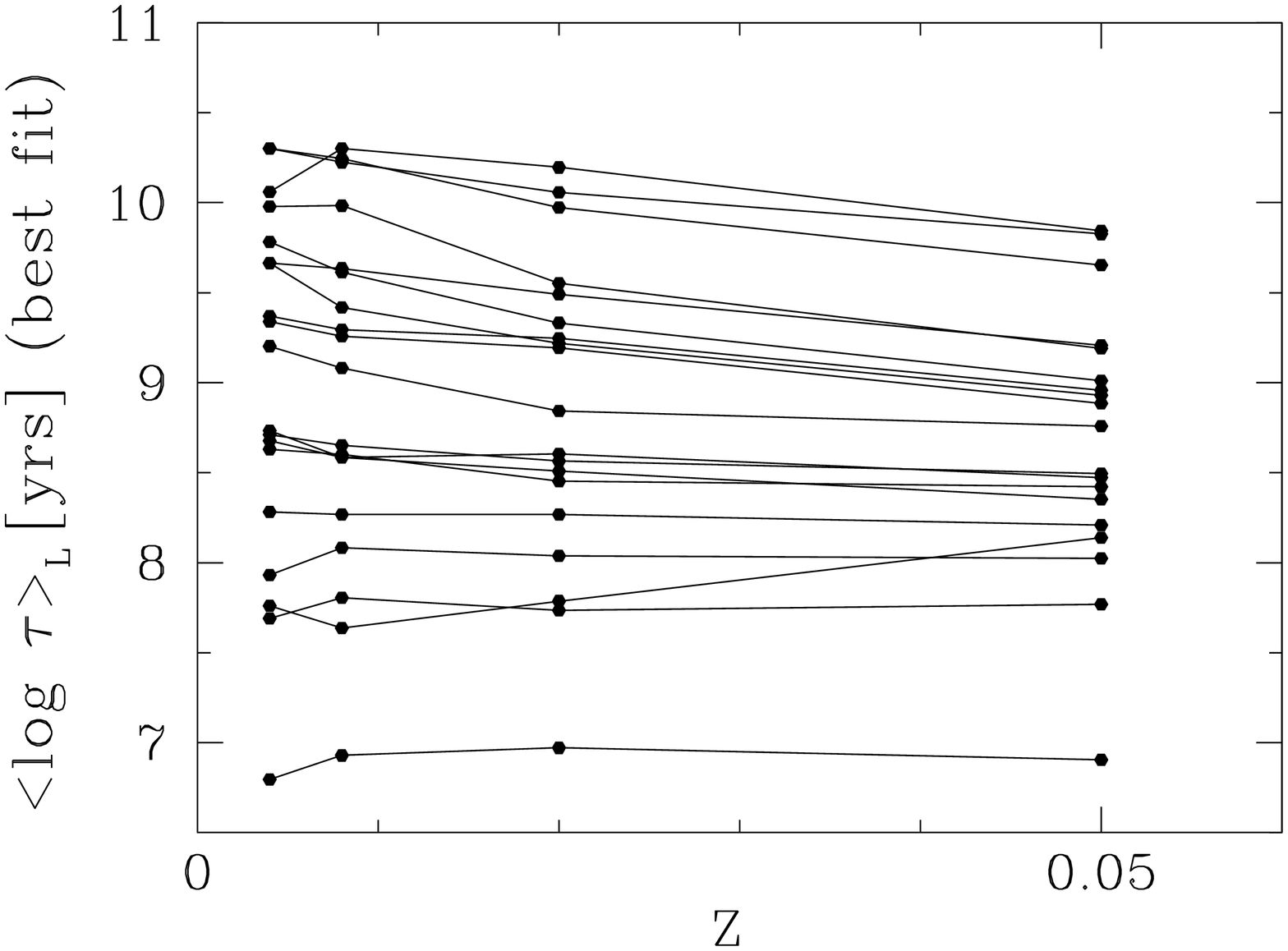,width=16cm,angle=270,clip=t}
\caption{Dependence of the best-fit derived luminosity-weighted ages as a 
function of the metallicity $Z$ for all 19 NCs. Results for the same NC at 
different metallicity are connected by a line.\label{f:agemetal}}
\end{figure} 

%%% FIGURE %%%

\clearpage

\begin{figure}
\psfig{file=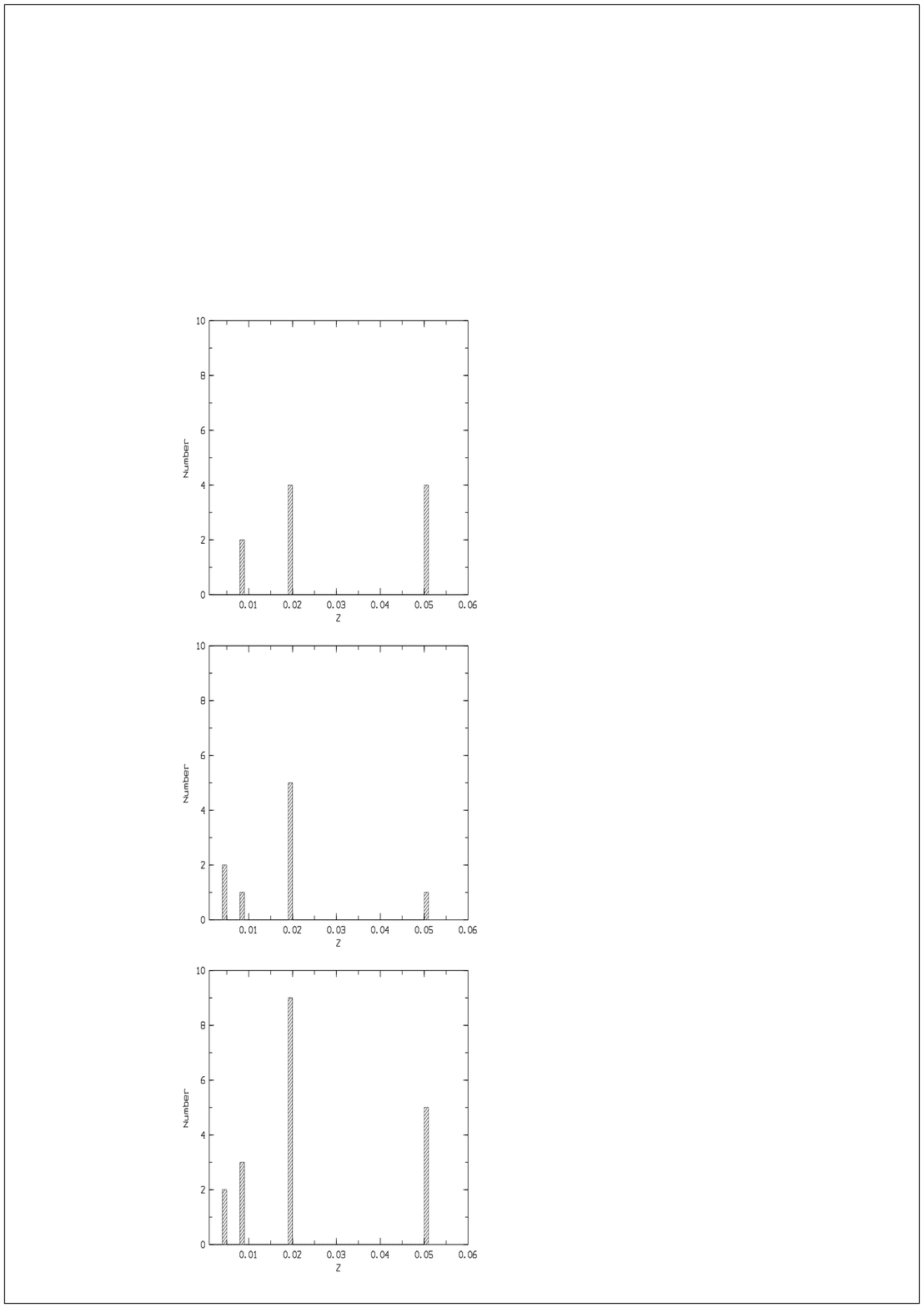,width=6cm,angle=0,clip=t}
\caption{Histogram of the distribution of the 
best-fitting nuclear cluster metallicities $Z$ listed in
Table~\ref{t:age}. The metallicities were obtained from 
spectral population fits to the STIS spectra. The top panel shows the
distribution for the ten early-type spirals, the middle panel shows
the distribution for the nine late-type spirals, and the bottom panel
shows the distribution for the combined sample of all 19 galaxies.
The discrete sampling along the horizontal axis reflects the fact that
Bruzual-Charlot models are available only for a fixed set of
metallicities.\label{f:Zhisto}}
\end{figure} 

%%% FIGURE %%%

\clearpage

\begin{figure}
\psfig{file=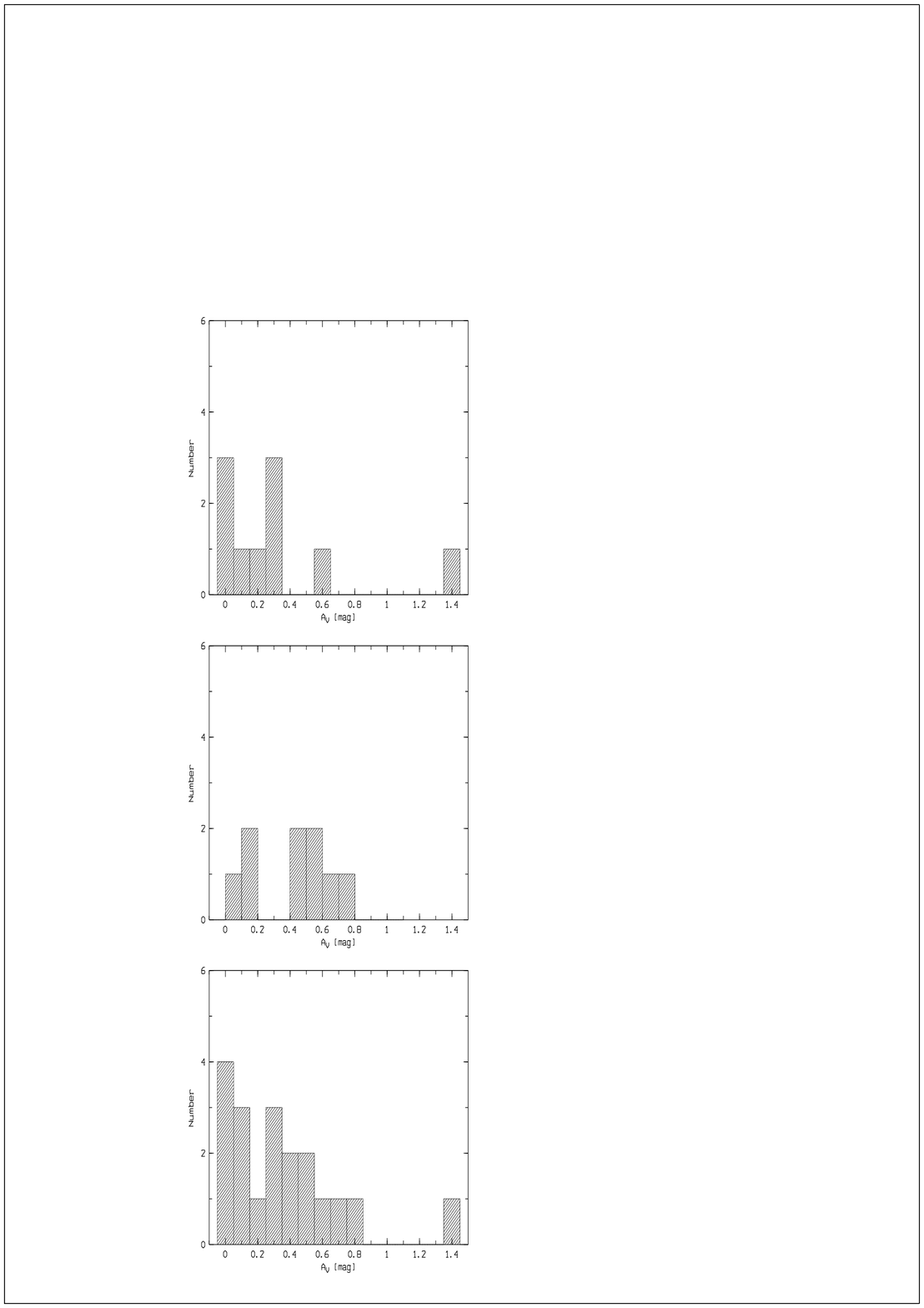,width=6cm,angle=0,clip=t}
\caption{Histogram of the distribution of the best-fitting extinction 
$A_V$ listed in Table~\ref{t:age}. The extinction values were obtained from 
spectral population fits to the STIS spectra. The top panel shows the 
distribution for the ten early-type spirals, the middle panel shows the 
distribution for the nine late-type spirals, and the bottom panel shows the 
distribution for the combined sample of all 19 galaxies.\label{f:avhisto}}
\end{figure} 

%%% FIGURE %%%

\clearpage

\begin{figure}
\psfig{file=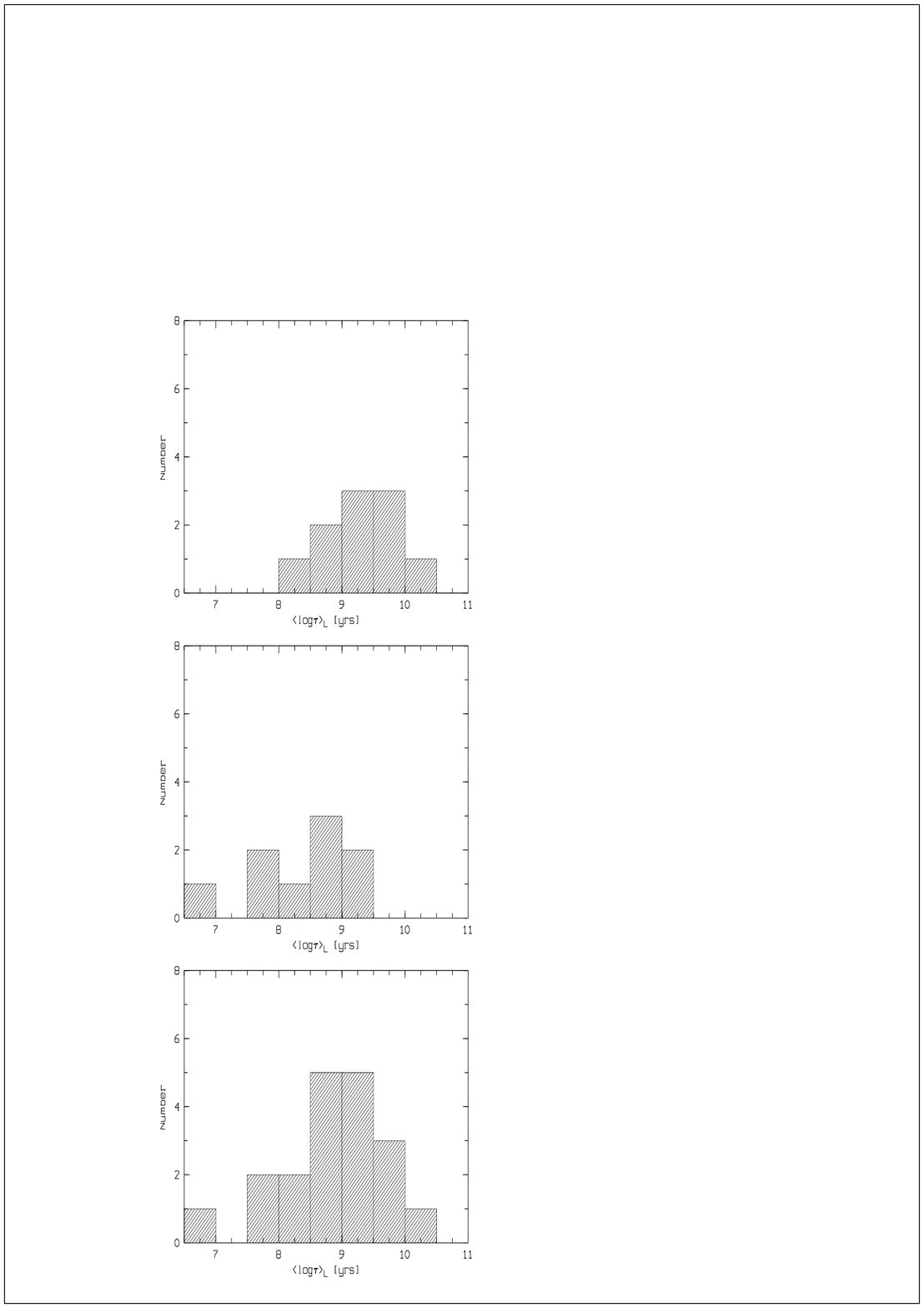,width=6cm,angle=0,clip=t}
\caption{Histogram of the distribution of the luminosity-weighted age 
$\langle\log\,\tau\rangle_L$ listed in Table~\ref{t:age}. The ages were 
obtained from spectral population fits to the STIS spectra. The top panel 
shows the distribution for the ten early-type spirals, the middle panel 
shows the distribution for the nine late-type spirals, and the bottom panel 
shows the distribution for the combined sample of all 19 
galaxies.\label{f:agehisto}}
\end{figure} 

%%% FIGURE %%%

\clearpage
\begin{figure}
\psfig{file=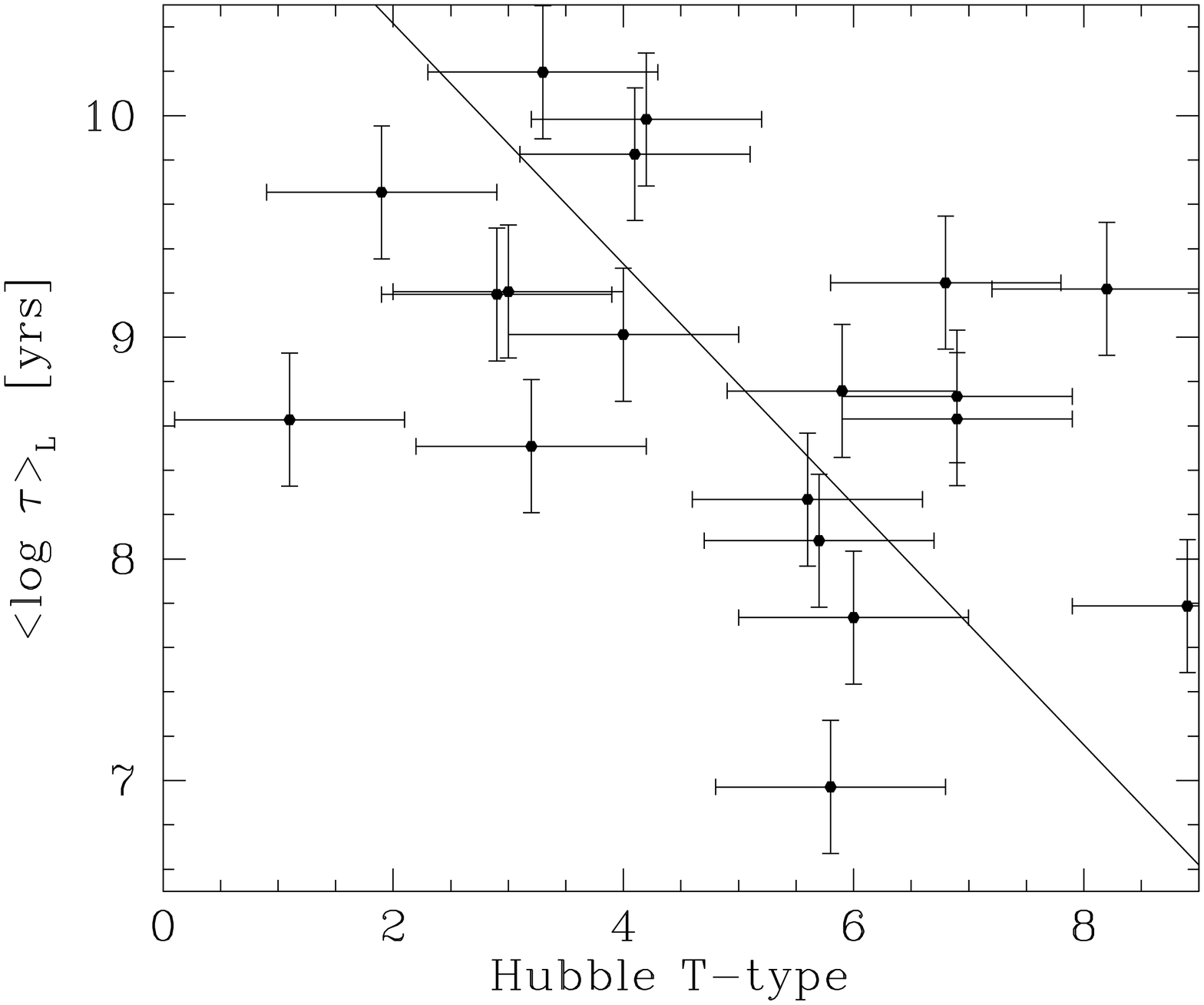,width=16cm,angle=270,clip=t}
\caption{Relation between the luminosity-weighted age 
$\langle\log\,\tau\rangle_L$ listed in Table~\ref{t:age} and the
Hubble T-type of the host galaxy (Tables~\ref{t:obs1}
and~\ref{t:obs2}). The ages were obtained from spectral population
fits to the STIS spectra and were assigned a measurement uncertainty
of $0.3$ dex, based on the discussion in
Section~\ref{ss:robustness}. Hubble T-types were, somewhat
arbitrarily, assigned an uncertainty of 1.0. Early-type spirals tend
to have older NC populations. The solid line indicates the best linear
fit, the parameters of which are listed in
Table~\ref{t:fitcoeff}.\label{f:hubblevsage}}
\end{figure} 

%%% FIGURE %%%

\clearpage

\begin{figure}
\psfig{file=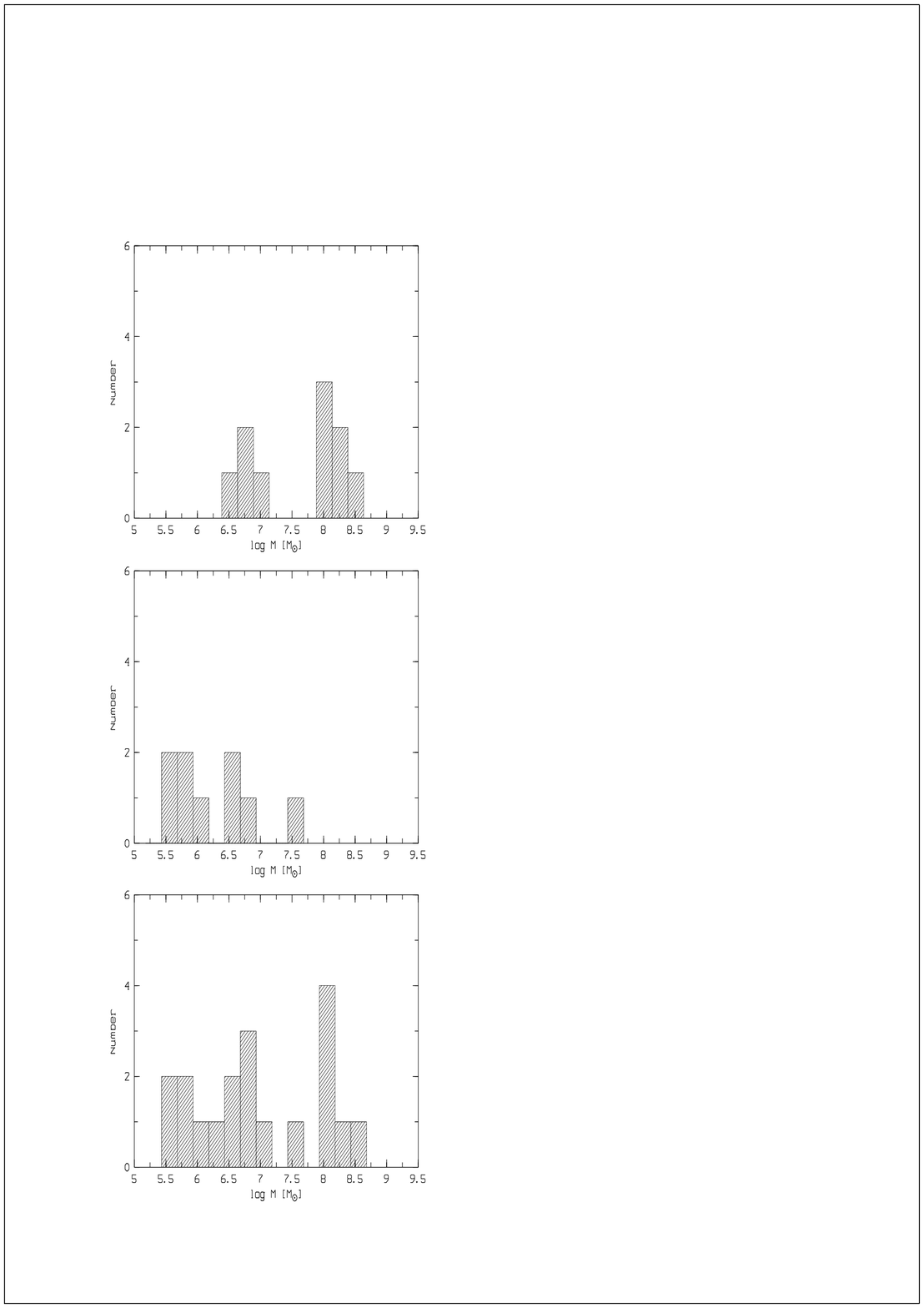,width=7cm,angle=0,clip=t}
\caption{Histogram of the distribution of $\log M$,
the logarithm of the NC mass in solar units, as listed in
Table~\ref{t:age}. The masses were obtained by taking the product of
the mass-to-light ratios inferred from spectral population fits to the
STIS spectra, and the luminosities directly measured from the STIS
spectra (corrected for extinction and aperture losses).  The top panel
shows the distribution for the ten early-type spirals, the middle
panel shows the distribution for the nine late-type spirals, and the
bottom panel shows the distribution for the combined sample of all 19
galaxies.\label{f:masses}}
\end{figure} 

%%% FIGURE %%%

\clearpage

\begin{figure}
\psfig{file=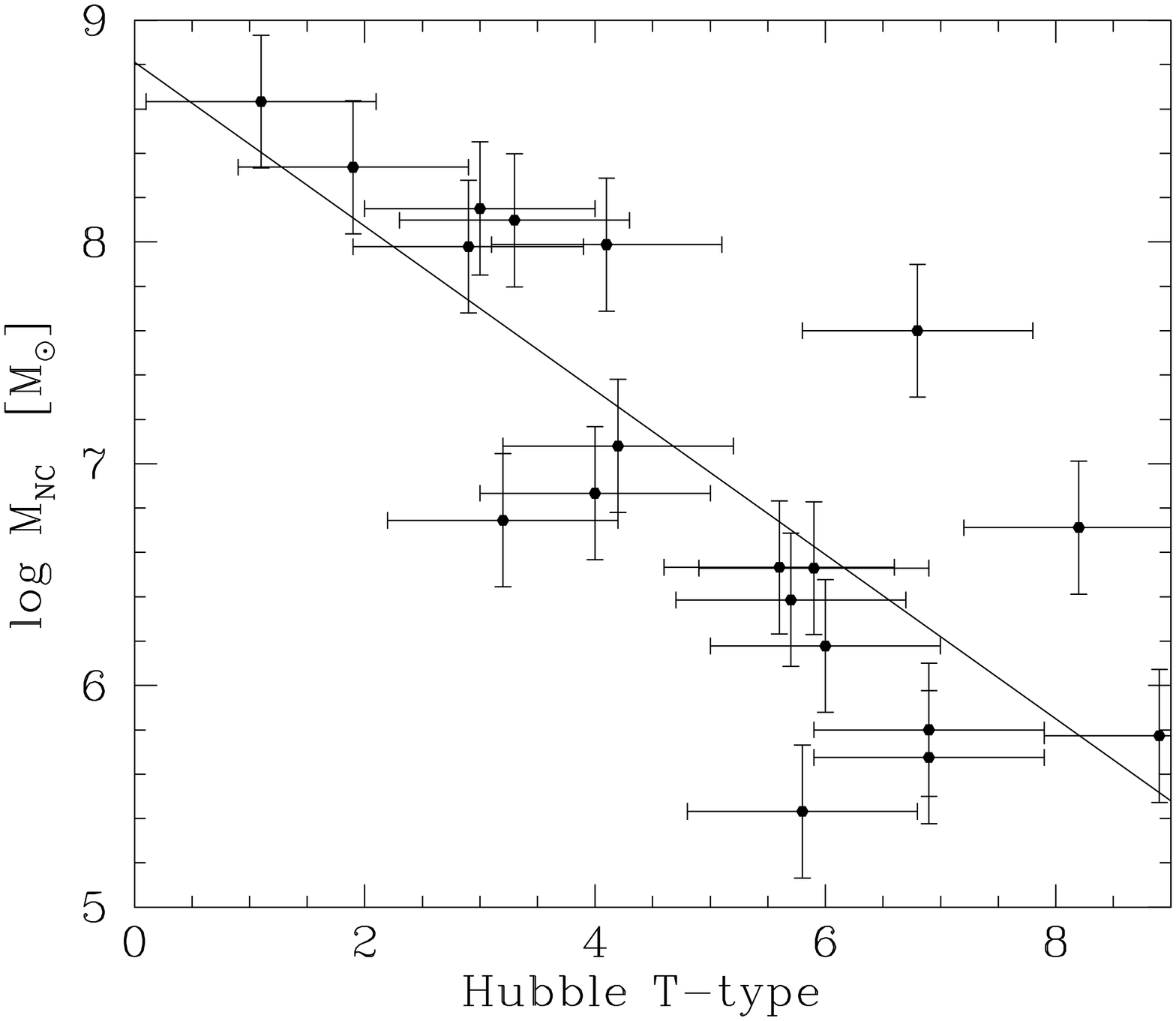,width=16cm,angle=270,clip=t}
\caption{Relation between the NC mass $M$ listed in Table~\ref{t:age} and 
the Hubble T-type of the host galaxy (Tables~\ref{t:obs1} and~\ref{t:obs2}). 
Masses were assigned a measurement uncertainty of $0.3$ dex, based on the 
discussion in Section~\ref{ss:masses}. Hubble T-types were, somewhat 
arbitrarily, assigned an uncertainty of 1.0. Early-type spirals tend to have 
more massive NCs. The solid line indicates the best linear fit, the
parameters of which are listed in
Table~\ref{t:fitcoeff}.\label{f:hubblevsmass}}
\end{figure} 

%%% FIGURE %%%

\clearpage

\begin{figure}
\psfig{file=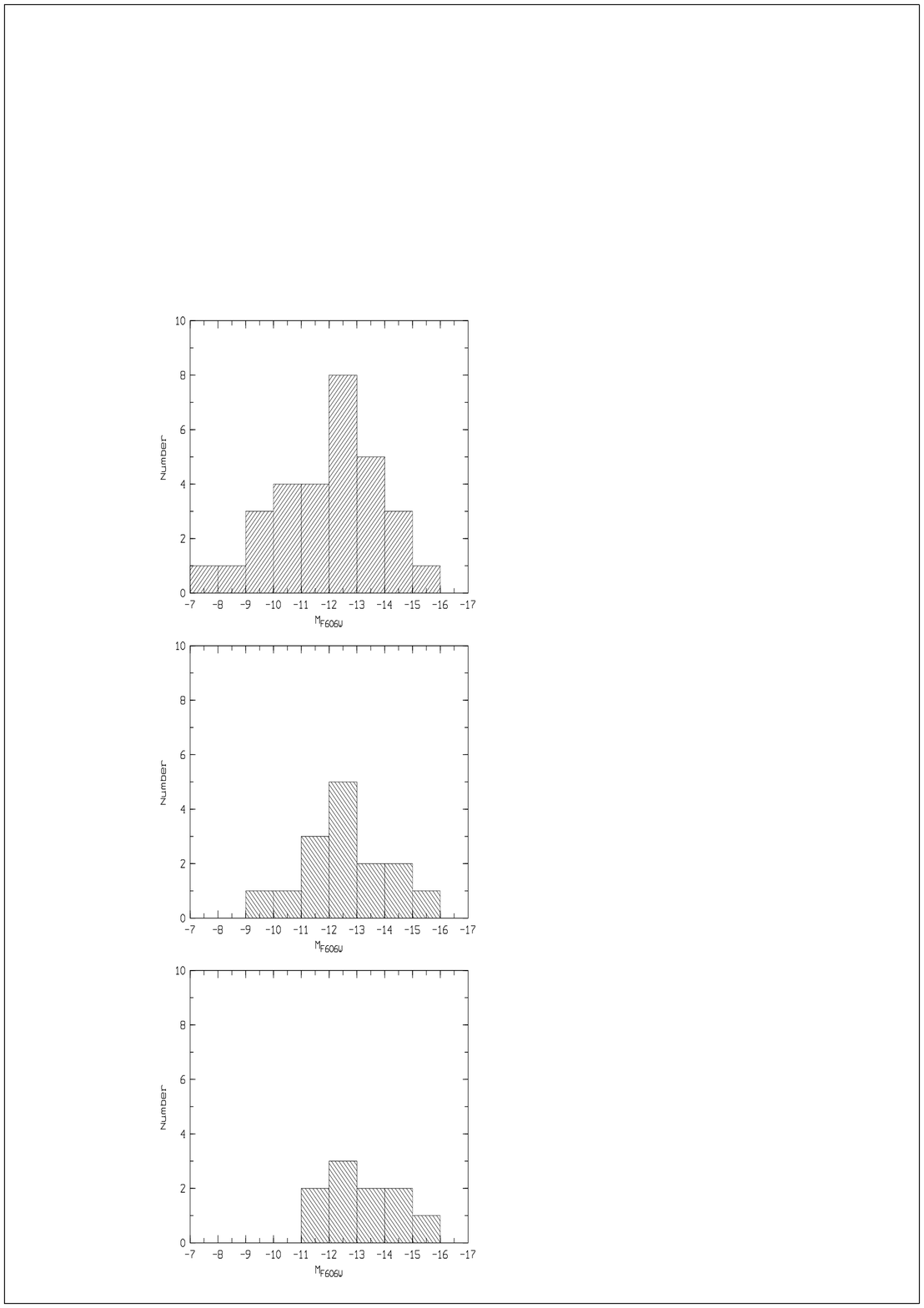,width=6cm,angle=0,clip=t}
\caption{Histogram of the absolute NC magnitudes 
in early-type spirals. The magnitude $M_{\rm{F606W}}$ is measured in
the VEGAMAG system of the WFPC2/F606W filter (similar to the
$V$-band). It was calculated from the published $m_{\rm{F606W}}$
values listed in Table~\ref{t:obs1}. The top panel shows the
distribution for all NCs detected by \citet{car98}. The middle panel
shows the histogram for the NCs which we have observed with
HST/STIS. The bottom panel shows the histogram for those NCs for which
we performed spectral population fits.\label{f:maghistcar}}
\end{figure} 

%%% FIGURE %%%

\clearpage

\begin{figure}
\psfig{file=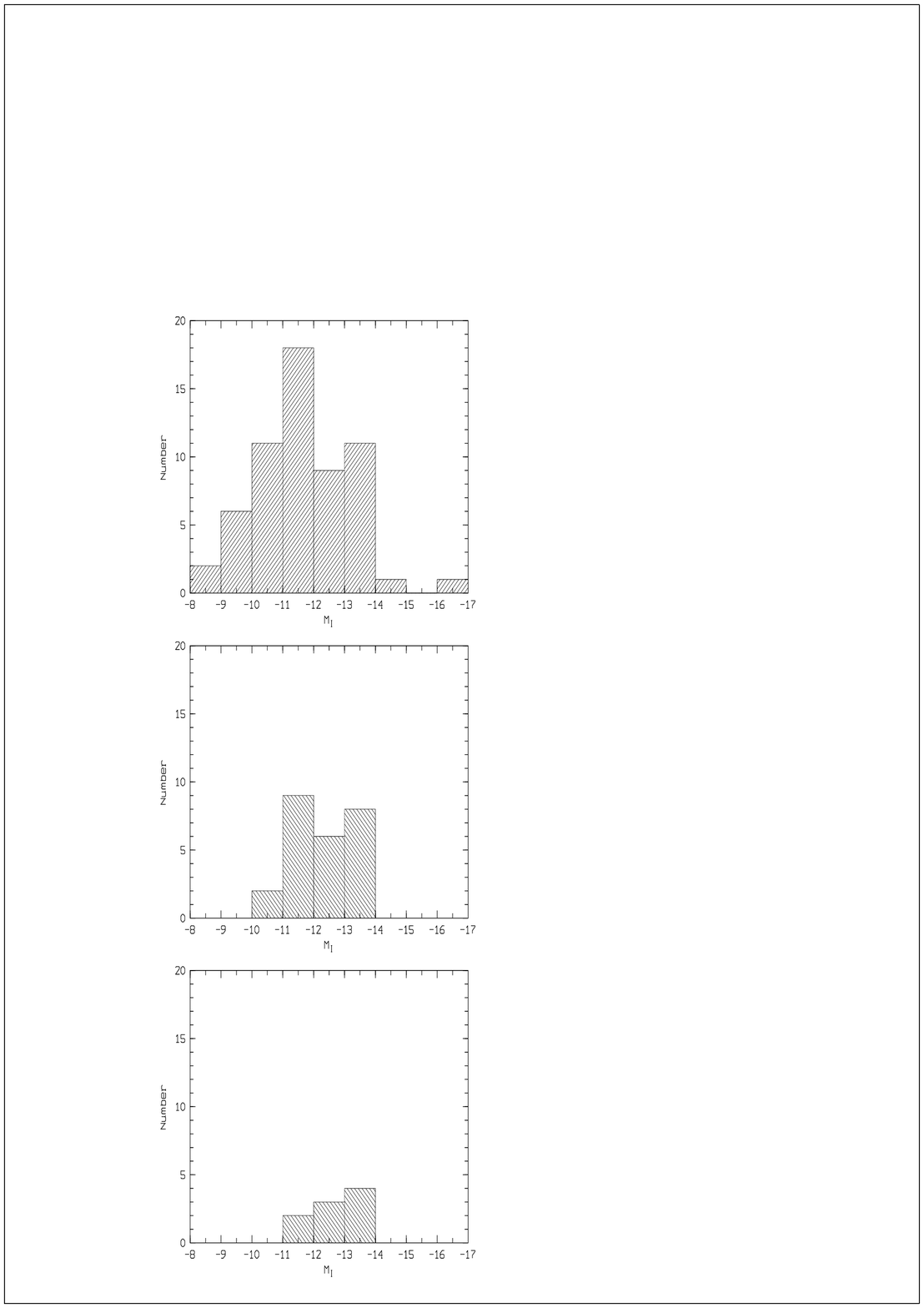,width=6cm,angle=0,clip=t}
\caption{Histogram of the absolute $I$-band NC magnitudes $M_I$ in late-type 
spirals, calculated from the published $m_I$ values listed in 
Table~\ref{t:obs2}. The top panel shows the distribution for all NCs 
detected by \citet{boe02}. The middle panel shows the histogram for the 
NCs which we have observed with HST/STIS. The bottom panel shows the 
histogram for those NCs for which we performed spectral population 
fits.\label{f:maghistboe}}
\end{figure} 

%%% FIGURE %%%

\clearpage

\begin{figure}
\psfig{file=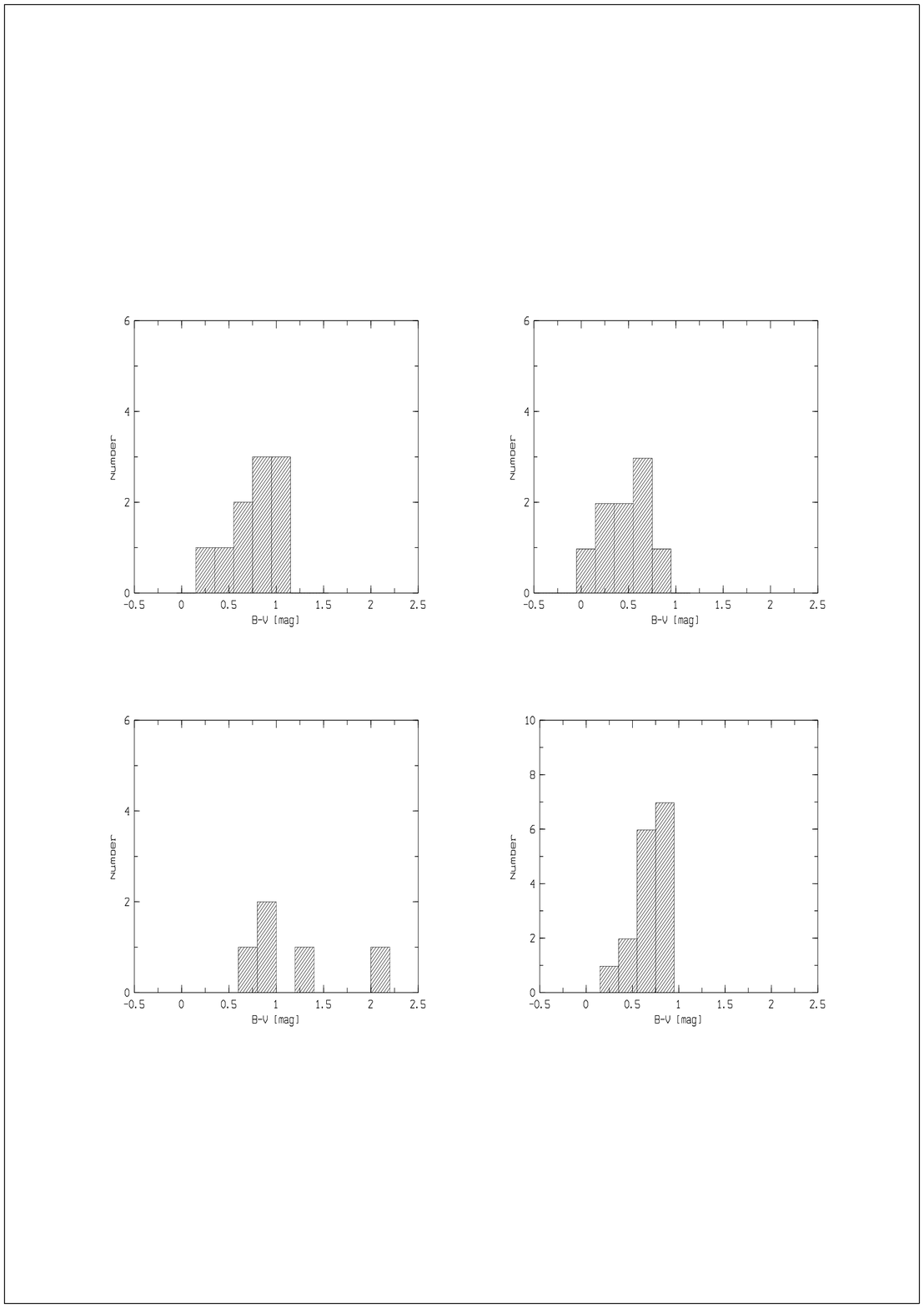,width=16cm,angle=0,clip=t}
\caption{Histograms of the $B-V$ distribution for early- and late-type 
spirals. The left panels present the early-type sample above (top panel) 
and below (bottom panel) our selected magnitude cutoff. The right panels 
depict the late-type sample above (top panel) and below (bottom panel) 
our selected magnitude cutoff.\label{f:bminusv}}
\end{figure} 

%%% FIGURE %%%

\clearpage

\begin{figure}
\psfig{file=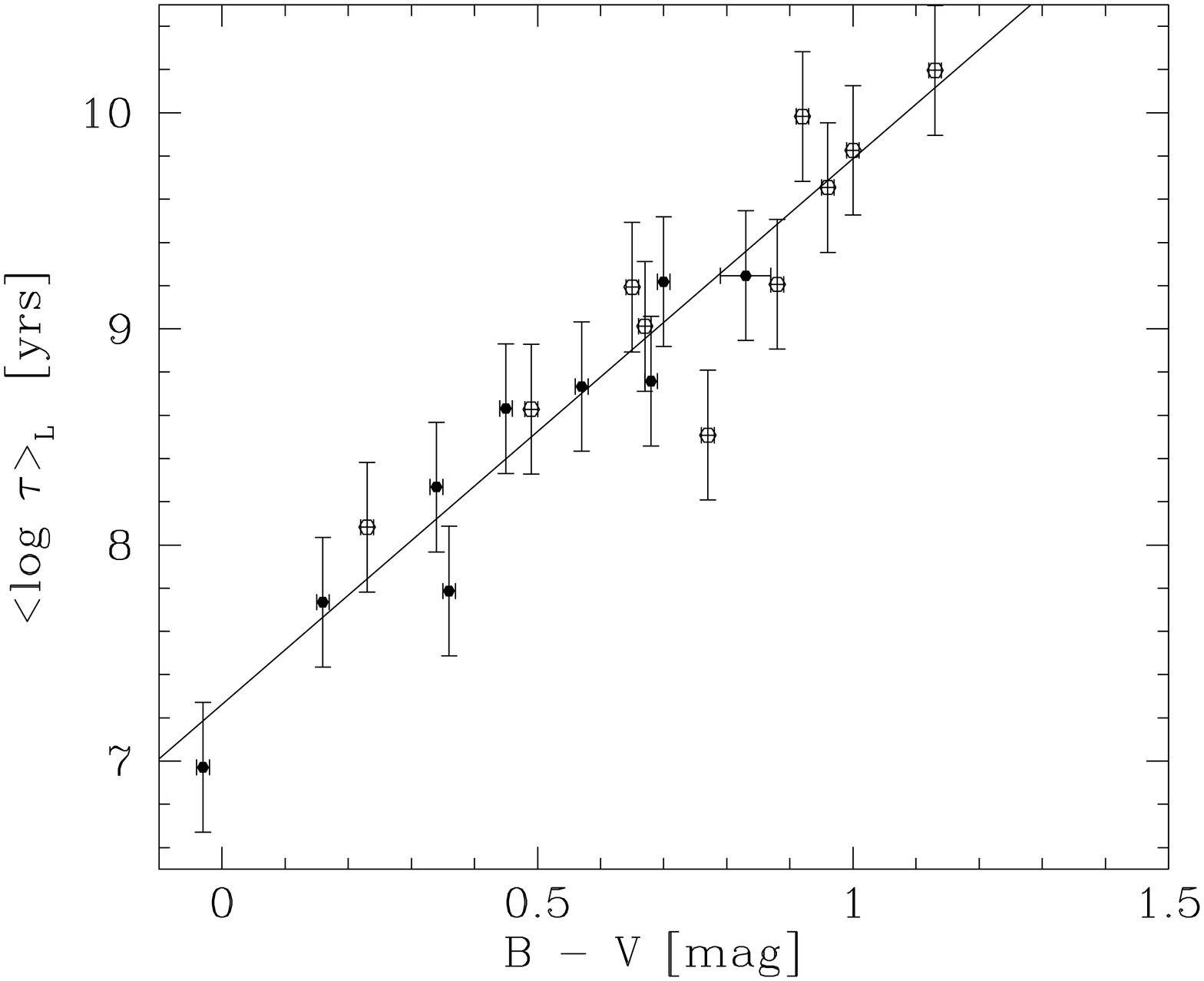,width=16cm,angle=270,clip=t}
\caption{Correlation between the luminosity-weighted age 
$\langle\log\,\tau\rangle_L$, as inferred from spectral population
fits to our STIS spectra, and broadband color $B-V$. Age measurements
were assigned an uncertainty of $0.3$ dex, based on the discussion in
Section~\ref{ss:robustness}. The $B-V$ color increases with increasing
age of the NC. Open symbols denote NCs in early-type spirals and solid
symbols denote NCs in late-type spirals. The solid line indicates the
best linear fit, the parameters of which are listed in
Table~\ref{t:fitcoeff}.\label{f:bvage}}
\end{figure} 

%%% FIGURE %%%

\clearpage

\begin{figure}
\psfig{file=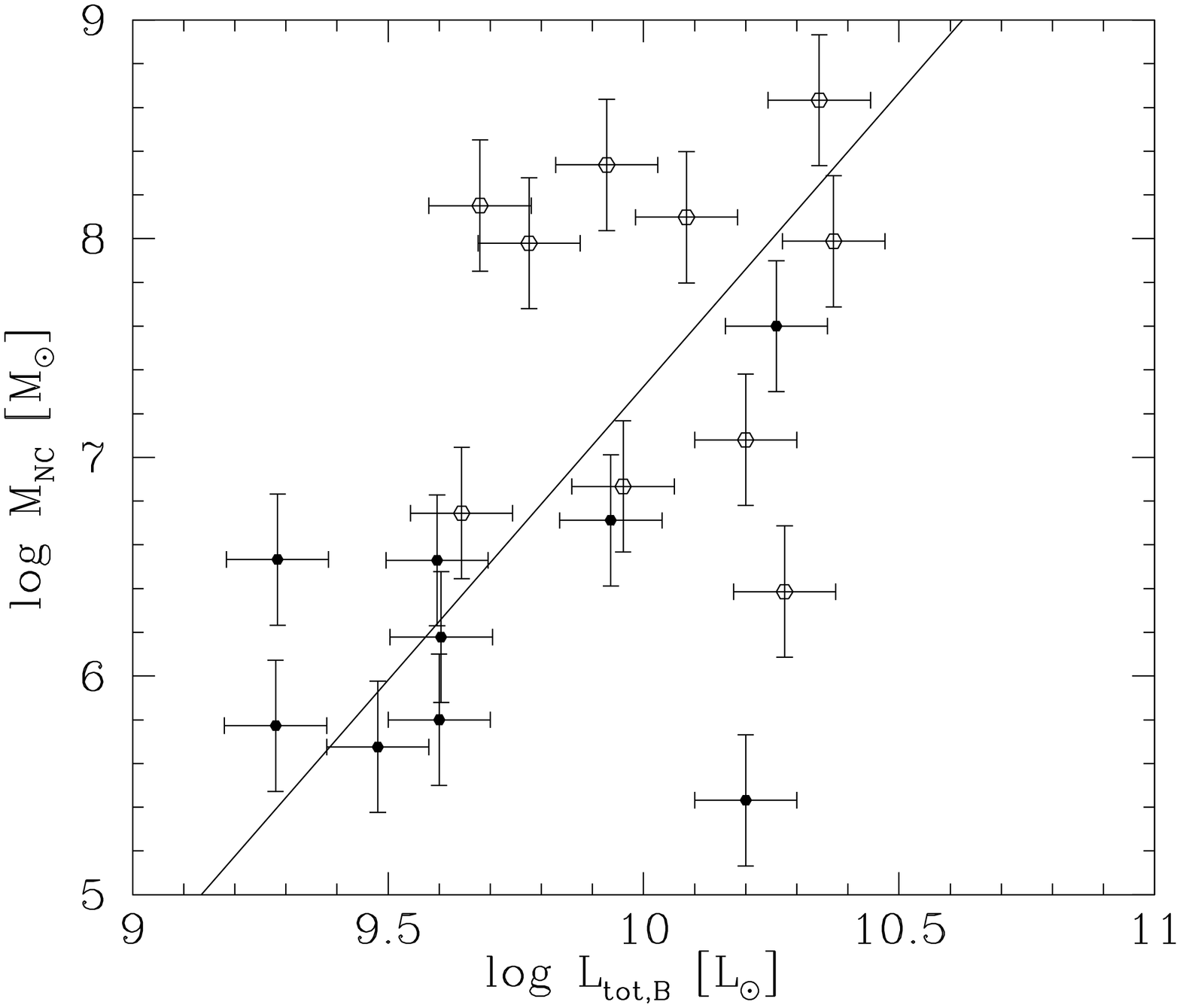,width=16cm,angle=270,clip=t}
\caption{Relation between the NC mass $M$ listed in Table~\ref{t:age} and 
the total $B$-band luminosity $L_B$ of the host galaxy. Masses were assigned 
a measurement uncertainty of $0.3$ dex, based on the discussion in 
Section~\ref{ss:masses}. The measurement uncertainties in $L_B$ are not 
well known, and were somewhat arbitrarily assigned a value of $0.1$ dex. 
There is a weak correlation in the sense that more luminous galaxies have 
more massive NCs. Open symbols denote early-type spirals and solid symbols 
denote late-type spirals. The solid line indicates the best linear fit, the
parameters of which are listed in
Table~\ref{t:fitcoeff}.\label{f:masstotallum}}
\end{figure} 

%%% FIGURE %%%

\clearpage

\begin{figure}
\psfig{file=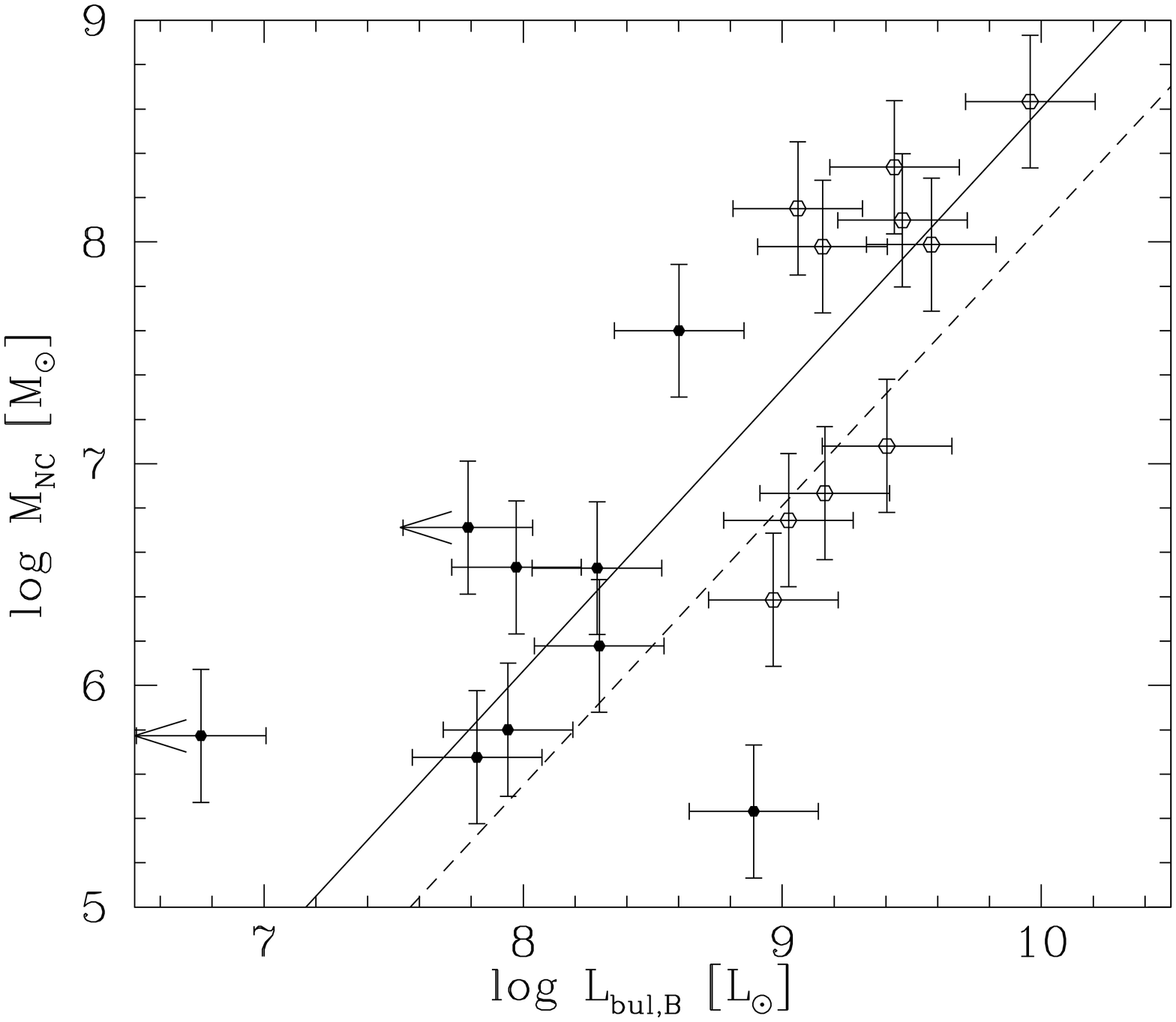,width=16cm,angle=270,clip=t}
\caption{Relation between the NC mass $M$ (listed in Table~\ref{t:age}) 
and the $B$-band luminosity $L_{{\rm bul,}B}$ of the host galaxy bulge. 
Masses were assigned a measurement uncertainty of $0.3$ dex, based on 
the discussion in Section~\ref{ss:masses}. The measurement uncertainties 
in $L_{{\rm bul,}B}$ are dominated by variations in disk-to-bulge ratios 
among galaxies, and were somewhat arbitrarily assigned a value of $0.25$
dex. There is a strong correlation in the sense that galaxies with
more luminous bulges have more massive NCs. Open symbols denote
early-type spirals and solid symbols denote late-type spirals. For the
very late-type spirals NGC\,428 and NGC\,2552 we assigned a
bulge-to-total light ratio based on the extrapolation of the fitting
function of \citet{sim86}.  Therefore, the resulting values of
$L_{{\rm bul,}B}$ for these two cases should be considered as upper
limits (indicated by arrows). The solid line indicates the best linear
fit, the parameters of which are listed in Table~\ref{t:fitcoeff}.
The dashed line indicates the relation between the BH mass
$M_{\rm{BH}}$ and $B$-band luminosity $L_{{\rm bul,}B}$ for all
galaxies studied by \citet{marc03}.\label{f:massbulgelum}}
\end{figure} 

%%% FIGURE %%%

\clearpage

\begin{figure}
\psfig{file=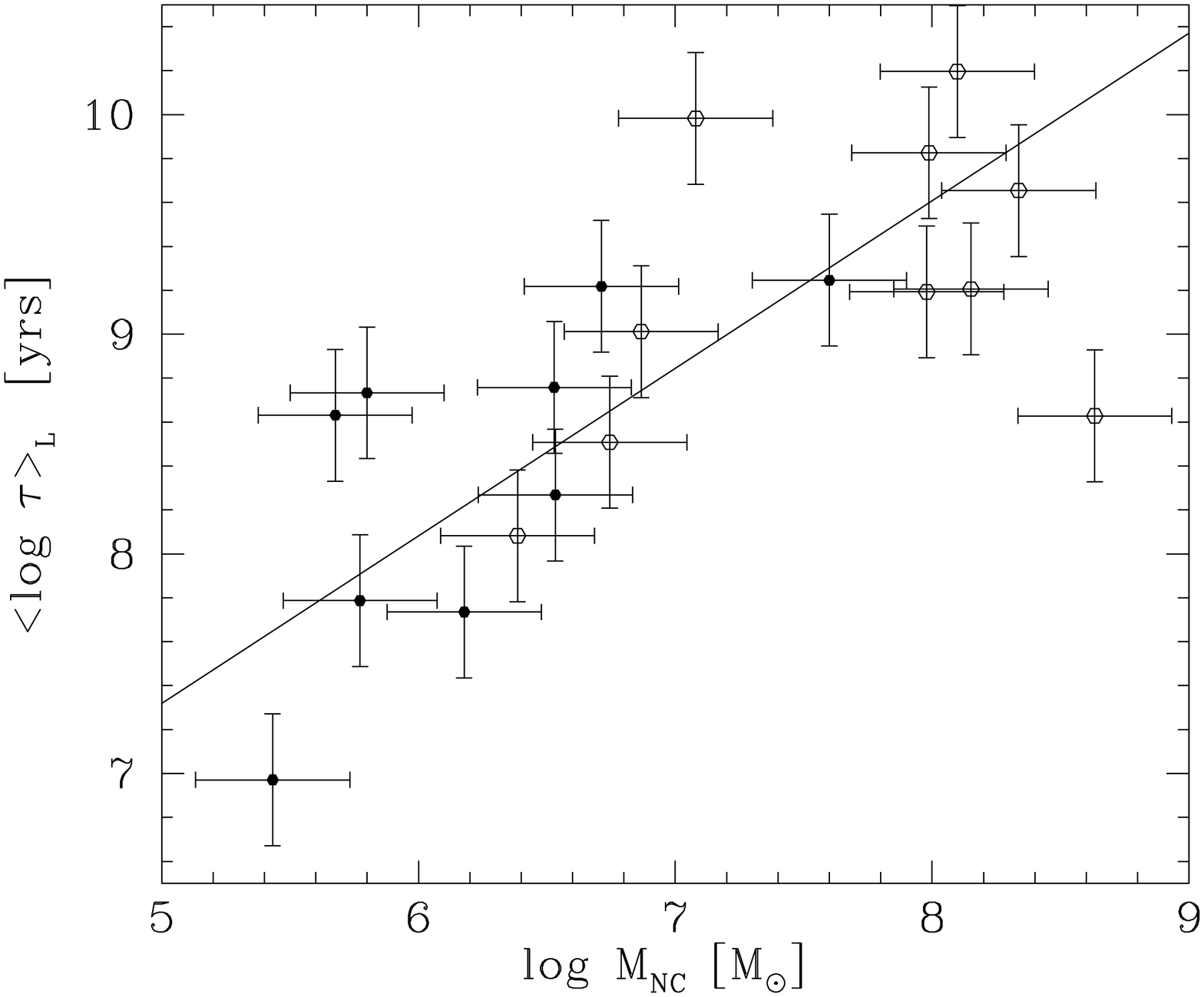,width=16cm,angle=270,clip=t}
\caption{Correlation between the NC mass $M$ and the luminosity-weighted 
age $\langle\log\,\tau\rangle_L$ (listed in Table~\ref{t:age}). Both
quantities were assigned a measurement uncertainty of $0.3$ dex, based
on the discussions in Sections~\ref{ss:robustness}
and~\ref{ss:masses}. More massive clusters tend to have older stellar
populations (and tend to reside in earlier-type spirals) than less
massive clusters. Open symbols denote NCs in early-type spirals and
solid symbols denote NCs in late-type spirals. The solid line
indicates the best linear fit, the parameters of which are listed in
Table~\ref{t:fitcoeff}.\label{f:massvsage}}
\end{figure} 

%%% FIGURE %%%

\clearpage

%%% END OF FIGURES %%%

%%%%%%%%%%%%%%%
% Tables 
%%%%%%%%%%%%%%%

%\clearpage

%%% TABLE %%%

\begin{deluxetable}{lccccccccc}
\tabletypesize{\scriptsize}
\tablecaption{Early-type Spiral Galaxy Sample\label{t:obs1}}
\tablehead{
Galaxy & R.A. & Dec. & Classification & T &
$v_{\rm sys}$ & $D$ & $A_V$ & $m_{\rm{F606W}}$ & $M_{\rm{F606W}}$ \\
 & [hh mm ss.ss] & [dd\degr mm\arcmin ss\farcs s] & & & [km~s$^{-1}$] & 
[Mpc] & [mag] & [mag] & [mag] \\
\colhead{(1)} & \colhead{(2)} & \colhead{(3)} & \colhead{(4)} &
\colhead{(5)} & \colhead{(6)} & \colhead{(7)} & \colhead{(8)} &
\colhead{(9)} & \colhead{(10)} \\
}
\startdata
NGC\,1325 & 03 24 25.20 & $-$21 32 32.8 & SA(s)bc & 4.2 & 1414 & 20.2 & 
0.072 & 20.0 & $-$11.60 \\
NGC\,1385 & 03 37 28.32 & $-$24 30 04.2 & SB(s)cd & 5.7 & 1305 & 18.6 & 
0.067 & 19.0 & $-$12.41 \\
NGC\,2082 & 05 41 51.21 & $-$64 18 04.0 & SAB(rs+)c & 3.2 & 942 & 13.5 & 
0.192 & 20.9 & $-$09.94 \\
NGC\,3177 & 10 16 34.21 & +21 07 24.8 & SA(rs)b & 3.0 & 1392 & 19.9 & 0.074 & 
17.1 & $-$14.47  \\
NGC\,3277 & 10 32 55.50 & +28 30 42.0 & SA(r)ab & 1.9 & 1539 & 22.0 & 0.088 & 
17.2 & $-$14.60 \\
NGC\,3455 & 10 54 31.07 & +17 17 06.2 & (R')SAB(rs)b & 3.2 & 1185 & 16.9 & 
0.110 & 20.1 & $-$11.15 \\
NGC\,4030 & 12 00 23.65 & +01 06 03.3 & SA(s)bc & 4.1 & 1476 & 21.1 & 0.087 & 
19.3 & $-$12.41 \\
NGC\,4806 & 12 56 12.67 & $-$29 30 13.8 & SB(s)c? & 4.9 & 2350 & 33.6 & 0.285 
& 20.8 & $-$12.12 \\
NGC\,4980 & 13 09 10.24 & $-$28 38 30.8 & SAB(rs)a pec? & 1.0 & 1329 & 19.0 
& 0.237 & 20.2 & $-$11.43\\
NGC\,5188 & 13 31 28.46 & $-$34 47 44.0 & (R':)SAB(rs)b & 3.1 & 2301 & 
32.9 & 0.173 & 19.2 & $-$13.56 \\
NGC\,5377 & 13 56 16.79 & +47 14 17.2 & (R)SB(s)a & 1.1 & 2036 & 29.1 & 
0.054 & 16.5 & $-$15.87 \\
NGC\,5806 & 15 00 00.30 & +01 53 29.0 & SAB(s)b & 3.3 & 1440 & 20.6 & 0.169 & 
18.6 & $-$13.14 \\
NGC\,6384 & 17 32 24.30 & +07 03 38.0 & SAB(r)bc & 3.8 & 1780 & 25.4 & 
0.409 & 19.6 & $-$12.83 \\
NGC\,7421 & 22 56 54.00 & $-$37 20 51.0 & SB(r)bc & 4.0 & 1662 & 23.7 & 0.049 
& 19.0 & $-$12.92 \\
NGC\,7690 & 23 33 02.60 & $-$51 41 52.0 & Sb & 2.9 & 1281 & 18.3 & 0.035 
& 18.0 & $-$13.35 \\
\enddata
\tablecomments{Column~(1) lists the galaxy name. Columns~(2) and~(3) list 
the right ascension and declination (J2000.0) from the NASA
Extragalactic Database (NED). Column~(4) lists the morphological
classification given by NED and column~(5) lists the Hubble T-type
taken from the Lyon Extragalactic Database (LEDA). Much of the
information in Columns~(4)+(5) relies on the information given by the
RC3 \citep{dev91} as its primary source. Column~(6) lists the systemic
velocity corrected for Virgocentric infall using the model of
\citet{san90}, as taken from LEDA. Column~(7) lists the corresponding 
distance for a Hubble constant $H_0 = 70 \kms \Mpc^{-1}$. Column~(8) lists 
the Galactic foreground extinction \citep{sch98} in the $V$-band for 
$R_V = 3.1$ from NED. Column~(9) lists observed apparent magnitude
$m_{\rm{F606W}}$ of the nuclear cluster in the WFPC2 F606W filter, as
given by \citet{car98}. Column~(10) lists the corresponding absolute
magnitude corrected for foreground extinction.}
\end{deluxetable}

\clearpage

%%% TABLE %%%

\begin{deluxetable}{lccccccccc}
\tabletypesize{\scriptsize}
\tablecaption{Late-type Spiral Galaxy Sample\label{t:obs2}}
\tablehead{
Galaxy & R.A. & Dec. & Classification & T &
$v_{\rm sys}$ & $D$ & $A_I$ & $m_I$ & $M_I$ \\
 & [hh mm ss.ss] & [dd\degr mm\arcmin ss\farcs s] & & & [km~s$^{-1}$] & 
[Mpc] & [mag] & [mag] & [mag] \\
\colhead{(1)} & \colhead{(2)} & \colhead{(3)} & \colhead{(4)} &
\colhead{(5)} & \colhead{(6)} & \colhead{(7)} & \colhead{(8)} &
\colhead{(9)} & \colhead{(10)} \\
}
\startdata
NGC\,300 & 00 54 53.72 & $-$37 40 56.9 & SA(s)d & 6.9 & $-$54 & 2.2 & 0.025 & 
15.29 & $-$11.43\\
NGC\,428 & 01 12 55.16 & +00 58 58.7 & SAB(s)m & 8.2 & 1130 & 16.1 & 
0.055 & 17.94 & $-$13.15\\
NGC\,450 & 01 15 30.29 & $-$00 51 40.9 & SBc & 6.0 & 1720 & 24.6 & 0.077 & 
20.07 & $-$11.96\\   
ESO\,358$-$5 & 03 27 16.47 & $-33$ 29 06.1 & SAB(s)m pec:& 8.7 & 1409 & 20.1
& 0.022 & 20.09 & $-$11.45\\
NGC\,1493 & 03 57 27.73 & $-$46 12 38.1 & SB(rs)cd & 6.0 & 796 & 11.4 & 0.020 
& 17.17 & $-$13.13\\
NGC\,2139 & 06 01 08.44 & $-$23 40 24.4 & SAB(rs)cd & 5.8 & 1649 & 23.6 & 
0.065 & 19.09 & $-$12.83\\
UGC\,3574 & 06 53 10.57 & +57 10 44.6 & SA(s)cd & 5.9 & 1635 & 23.4 & 0.103 & 
19.97 & $-$11.98\\
NGC\,2552 & 08 19 20.25 & +50 00 27.2 & SA(s)m? & 8.9 & 695 & 9.9 & 0.090 & 
18.04 & $-$12.04 \\
NGC\,2805 & 09 20 20.38 & +64 06 12.2 & SAB(rs)d & 7.0 & 1968 & 28.1 & 0.100 
& 19.02 & $-$13.33 \\
NGC\,3346 & 10 43 38.70 & +14 52 17.9 & SB(rs)cd & 5.9 & 1315 & 18.8 & 0.054 
& 19.64 & $-$11.78 \\
NGC\,3423 & 10 51 13.25 & +05 50 31.3 & SA(s)cd & 6.0 & 1025 & 14.6 & 0.058 & 
19.03 & $-$11.85 \\
NGC\,3445 & 10 54 35.87 & +56 59 24.4 & SAB(s)m & 8.7 & 2245 & 32.1 & 0.015 & 
19.10 & $-$13.45 \\
A\,1156+52 & 11 59 08.57 & +52 42 26.1 & SB(rs)cd & 6.0 & 1307 & 18.7 & 0.053 
& 20.43 & $-$10.98 \\
NGC\,4411B & 12 26 47.28 & +08 53 02.0 & SAB(s)cd & 5.9 & 1334 & 19.1 & 0.058 
& 18.88 & $-$12.58 \\
NGC\,4540 & 12 34 50.92 & +15 33 09.6 & SAB(rs)cd & 6.0 & 1383 & 
19.8 & 0.065 & 19.25 & $-$12.29 \\
NGC\,4618 & 12 41 32.50 & +41 09 03.0 & SB(rs)m & 8.4 & 748 & 10.7 & 
0.041 & 18.73 & $-$11.46 \\
NGC\,4625 & 12 41 52.95 & +41 16 26.8 & SAB(rs)m pec & 8.7 & 816 & 11.7 & 
0.035 & 19.74 & $-$10.63 \\
NGC\,4701 & 12 49 11.60 & +03 23 19.0 & SA(s)cd & 5.6 & 768 & 11.0 & 0.057 & 
16.80 & $-$13.46 \\
NGC\,4775 & 12 53 46.29 & $-$06 37 27.1 & SA(s)d & 6.8 & 1565 & 22.4 & 0.067 
& 18.04 & $-$13.78 \\
NGC\,5068 & 13 18 55.24 & $-$21 02 21.5 & SB(s)d & 6.2 & 607 & 8.7 & 0.197 & 
17.54 & $-$12.34 \\
NGC\,5585 & 14 19 47.90 & +56 43 46.0 & SAB(s)d & 6.9 & 571 & 8.2 & 0.030 
& 18.24 & $-$11.35 \\
NGC\,5668 & 14 33 24.89 & +04 27 01.7 & SA(s)d & 6.7 & 1665 & 23.8 & 0.071 & 
18.85 & $-$13.10 \\
NGC\,5964 & 15 37 36.33 & +05 58 28.2 & SB(rs)d & 6.7 & 1552 & 22.2 & 0.113 & 
19.21 & $-$12.63 \\
NGC\,6509 & 17 59 24.92 & +06 17 12.1 & SBcd & 7.0 & 1926 & 27.5 & 0.375 & 
19.48 & $-$13.09 \\
UGC\,12732 & 23 40 39.80 & +26 14 10.0 & Sm: & 8.8 & 870 & 12.4 & 0.172 & 
19.35 & $-$11.29 \\
\enddata
\tablecomments{Column~(1) lists the galaxy name. The galaxy A\,1156+52 is 
also known as UGC\,6983. Columns~(2) and~(3) list the right ascension
and declination (J2000.0) from NED. Column~(4) lists the morphological
classification given by NED, except for NGC\,450 which was taken from
LEDA. Column~(5) lists the Hubble T-type taken from LEDA. Much of the
information in Columns~(4)+(5) relies on the information given by the
RC3 \citep{dev91} as its primary source. Column~(6) lists the systemic
velocity corrected for Virgocentric infall using the model of
\citet{san90}, as taken from LEDA. Column~(7) lists the corresponding
distance for a Hubble constant $H_0 = 70 \kms \Mpc^{-1}$.  The value
for NGC\,300 has been taken from \citet{fre92}.  Column~(8) lists the
Galactic foreground extinction \citep{sch98} in the $I$-band using the
\citet{car89} extinction law and $R_V = 3.1$, from NED. Column~(9)
lists observed apparent magnitude $m_I$ of the nuclear cluster in the
$I$-band from \citet{boe04}. Column~(10) lists the corresponding
absolute magnitude corrected for foreground extinction.}
\end{deluxetable}

\clearpage

%%% TABLE %%%

\begin{deluxetable}{lcccc}
\tabletypesize{\scriptsize}
\tablecaption{Journal of HST/STIS G430L Observations\label{t:exp}}
\tablewidth{0pt}
\tablehead{
Galaxy & Data set identifier & Observation Date & Integration Time & 
$\quad\quad\quad \langle S/N \rangle$ \quad\quad\quad \\
& & [mm-dd-yyyy] & [s] & \\
\colhead{(1)} & \colhead{(2)} & \colhead{(3)} & \colhead{(4)} & 
\colhead{(5)} \\
}
\startdata
NGC\,300 & o6dz02 & 04-20-2002 & 900 & 24.5 \\
NGC\,428 & o6dz03 & 08-23-2001 & 900 & \,~7.7 \\
NGC\,450 & o6dz38 & 10-31-2001 & 900 & \,~1.3 \\ 
NGC\,1325 & o8nb02 & 06-27-2003 & 2140 & \,~4.9 \\
ESO\,358$-$5 & o6dz37 & 02-06-2002 & 900 & \,~1.2 \\
NGC\,1385 & o8nb03 & 08-02-2003 & 2180 & 28.3 \\
NGC\,1493 & o6dz10 & 12-12-2001 & 900 & 31.4 \\
NGC\,2082 & o8nb04 & 07-30-2003 & 2300 & \,~2.4 \\
NGC\,2139 & o6dz35 & 04-10-2002 & 900 & 15.5 \\
UGC\,3574 & o6dz05 & 04-15-2002 & 900 & \,~1.2 \\
NGC\,2552 & o6dz06 & 09-26-2001 & 900 & \,~9.6 \\
NGC\,2805 & o6dz08 & 02-28-2002 & 900 & \,~4.1 \\
NGC\,3177 & o8nb05 & 06-24-2003 & 2200 & 51.0 \\ 
NGC\,3277 & o8nb06 & 06-23-2003 & 2200 & 30.8 \\
NGC\,3346 & o6dz07 & 04-10-2002 & 900 & \,~1.9 \\
NGC\,3423 & o6dz12 & 02-28-2002 & 900 & \,~3.7 \\
NGC\,3455 & o8nb07 & 06-27-2003 & 2140 & \,~5.1 \\
NGC\,3445 & o6dz13 & 11-02-2001 & 900 & \,~2.9 \\
A\,1156+52 & o6dz14 & 12-20-2001 & 900 & \,~1.6 \\
NGC\,4030 & o8nb58 & 04-15-2004 & 2180 & 21.0 \\
NGC\,4411B & o8nb01 & 06-26-2003 & 2180 & \,~7.5 \\
NGC\,4540 & o6dz19 & 05-03-2002 & 900 & \,~1.7 \\
NGC\,4618 & o6dz20 & 04-11-2002 & 900 & \,~1.2 \\
NGC\,4625 & o6dz36 & 11-26-2001 & 900 & \,~1.8 \\
NGC\,4701 & o6dz21 & 12-11-2001 & 900 & 26.7 \\
NGC\,4775 & o6dz22 & 04-19-2002 & 900 & \,~8.3 \\
NGC\,4806 & o8nb09 & 06-25-2003 & 2100 & \,~2.0 \\
NGC\,4980 & o8nb10 & 07-30-2003 & 2140 & \,~4.1 \\
NGC\,5068 & o6dz23 & 04-19-2002 & 900 & \,~4.4 \\
NGC\,5188 & o8nb11 & 07-31-2003 & 2200 & \,~3.7 \\
NGC\,5377 & o8nb12 & 06-28-2003 & 2300 & 57.2 \\
NGC\,5585 & o6dz25 & 08-29-2001 & 900 & \,~8.9 \\
NGC\,5668 & o6dz26 & 02-27-2002 & 900 & \,~4.3 \\
NGC\,5806 & o8nb13 & 06-27-2003 & 2180 & 11.9 \\
NGC\,5964 & o6dz27 & 02-21-2002 & 900 & \,~2.8 \\
NGC\,6384 & o8nb14 & 07-13-2003 & 2140 & \,~5.2 \\
NGC\,6509 & o6dz29 & 08-30-2001 & 900 & \,~2.9 \\
NGC\,7421 & o8nb15 & 06-26-2003 & 2200 & 12.1 \\
NGC\,7690 & o8nb16 & 06-29-2003 & 2340 & 36.6 \\
UGC\,12732 & o6dz32 & 08-30-2001 & 900 & \,~1.8 \\
\enddata
\tablecomments{Col.~(1) lists the host galaxy name. Col.~(2) lists the data 
set identifier in the HST Data Archive. Col.~(3) and (4) list the date of the 
observations and the total exposure time. Col.~(5) lists the signal to noise 
$(S/N)$ per pixel in the logarithmically rebinned disk-subtracted spectrum.}
\end{deluxetable}

\clearpage

%%% TABLE %%%

\begin{deluxetable}{lccccccc}
\tabletypesize{\scriptsize}
\tablecaption{NC Properties from Spectral Population Fits
and Derived Masses\label{t:age}}
\tablewidth{0pt}
\tablehead{
Galaxy & $Z$ & $A_V$ & $\langle \log \tau \rangle_L$ &
$\langle \log \tau \rangle_M$ & $M/L_B$ & $\chi^2_{\rm red}$ & $\log M$ \\
& & [mag] & & & [M$_{\odot}$/L$_{B,\odot}$] & & [M$_{\sun}$] \\ 
\colhead{(1)} & \colhead{(2)} & \colhead{(3)} & \colhead{(4)} &
\colhead{(5)} & \colhead{(6)} & \colhead{(7)} & \colhead{(8)} \\
}
\startdata
NGC\,1325 & 0.008 & 0.3 & \,~9.98 & 10.13 & 3.65 & 2.19 & 7.08 \\
NGC\,1385 & 0.008 & 0.3 & \,~8.08 & \,~9.01 & 0.12 & 3.02 & 6.39 \\
NGC\,3177 & 0.05 & 0.3 & \,~9.21 & 10.05 & 2.72 & 9.37 & 8.15 \\
NGC\,3277 & 0.05 & 0.0 & \,~9.65 & 10.24 & 8.18 & 5.89 & 8.34 \\
NGC\,3455 & 0.02 & 1.4 & \,~8.51 & \,~9.48 & 0.57 & 2.26 & 6.75 \\
NGC\,4030 & 0.05 & 0.0 & \,~9.83 & 10.28 & 8.96 & 3.66 & 7.99 \\
NGC\,5377 & 0.02 & 0.2 & \,~8.63 & 10.13 & 1.39 & 3.28 & 8.63 \\
NGC\,5806 & 0.02 & 0.6 & 10.20 & 10.30 & 7.97 & 2.42 & 8.11 \\
NGC\,7421 & 0.05 & 0.0 & \,~9.01 & \,~9.14 & 0.74 & 2.80 & 6.87 \\
NGC\,7690 & 0.02 & 0.1 & \,~9.19 & 10.13 & 2.12 & 2.58 & 7.98 \\
\tableline
NGC\,300 & 0.004 & 0.4 & \,~8.63 & \,~9.62 & 0.51 & 3.46 & 5.68 \\
NGC\,428 & 0.02 & 0.1 & \,~9.22 & \,~9.65 & 1.18 & 2.42 & 6.71 \\
NGC\,1493 & 0.02 & 0.0 & \,~7.74 & \,~8.72 & 0.12 & 2.95 & 6.18 \\
NGC\,2139 & 0.02 & 0.1 & \,~6.97 & \,~7.74 & 0.02 & 3.03 & 5.43 \\
NGC\,2552 & 0.02 & 0.5 & \,~7.79 & \,~8.87 & 0.17 & 2.84 & 5.77 \\
NGC\,4411B & 0.05 & 0.8 & \,~8.76 & \,~8.85 & 0.36 & 2.35 & 6.53 \\
NGC\,4701 & 0.008 & 0.4 & \,~8.27 & \,~9.00 & 0.17 & 3.49 & 6.53 \\
NGC\,4775 & 0.02 & 0.5 & \,~9.25 & 10.04 & 1.87 & 3.56 & 7.60 \\
NGC\,5585 & 0.004 & 0.7 & \,~8.73 & \,~9.14 & 0.26 & 3.06 & 5.80 \\
\enddata
\tablecomments{Column~(1) lists the galaxy name. Columns~(2) and~(3) list
the metallicity and extinction that give the best fit to the observed
spectrum. Columns~(4) and~(5) list the luminosity-weighted and
mass-weighted average value of $\log \tau$ for the best spectral
population fit, where $\tau$ is the population age in
years. Column~(6) lists the $B$-band mass-to-light ratio for the NCs in
solar units. Column~(7) gives the $\chi^2$ per degree of freedom of
the fit. Column~(8) lists the NC masses, which were calculated from
the $M/L$ ratio, the $B$-magnitude, and from the best-fitted
extinction. The horizontal line divides the NCs in early-type spirals
(above the line) from those in late-type spirals (below the line).}
\end{deluxetable}

\clearpage

%%% TABLE %%%

\begin{deluxetable}{lccc}
\tabletypesize{\scriptsize}
\tablecaption{Photometric Properties of the NCs in Early- and Late-type 
Spirals\label{t:color}}
\tablewidth{0pt}
\tablehead{
Galaxy & $m_B$ & $m_V$ & $B-V$ \\
& [mag] & [mag] & [mag] \\
\colhead{(1)} & \colhead{(2)} & \colhead{(3)} & \colhead{(4)} \\
}
\startdata
NGC\,1325 & 21.11$\pm$0.01 & 20.19$\pm$0.01 & \,~~0.92$\pm$0.01 \\
NGC\,1385 & 18.99$\pm$0.01 & 18.76$\pm$0.01 & \,~~0.23$\pm$0.01 \\
NGC\,2082 & 21.94$\pm$0.01 & 20.99$\pm$0.01 & \,~~0.95$\pm$0.02 \\ 
NGC\,3177 & 18.08$\pm$0.01 & 17.20$\pm$0.01 & \,~~0.88$\pm$0.01 \\ 
NGC\,3277 & 18.63$\pm$0.01 & 17.67$\pm$0.01 & \,~~0.96$\pm$0.01 \\ 
NGC\,3455 & 21.01$\pm$0.01 & 20.24$\pm$0.01 & \,~~0.77$\pm$0.01 \\ 
NGC\,4030 & 19.51$\pm$0.01 & 18.51$\pm$0.01 & \,~~1.00$\pm$0.01 \\ 
NGC\,4806 & 22.12$\pm$0.02 & 21.13$\pm$0.01 & \,~~0.99$\pm$0.02 \\ 
NGC\,4980 & 21.20$\pm$0.01 & 20.44$\pm$0.01 & \,~~0.76$\pm$0.01 \\ 
NGC\,5188 & 21.26$\pm$0.01 & 19.26$\pm$0.01 & \,~~2.00$\pm$0.01 \\ 
NGC\,5377 & 16.84$\pm$0.01 & 16.35$\pm$0.01 & \,~~0.49$\pm$0.01 \\ 
NGC\,5806 & 19.85$\pm$0.01 & 18.72$\pm$0.01 & \,~~1.13$\pm$0.01 \\ 
NGC\,6384 & 21.64$\pm$0.02 & 20.31$\pm$0.01 & \,~~1.33$\pm$0.02 \\ 
NGC\,7421 & 19.86$\pm$0.01 & 19.19$\pm$0.01 & \,~~0.67$\pm$0.01 \\ 
NGC\,7690 & 17.79$\pm$0.01 & 17.14$\pm$0.01 & \,~~0.65$\pm$0.01 \\ 
\tableline
NGC\,300 & 17.79$\pm$0.01 & 17.34$\pm$0.01 & \,~~0.45$\pm$0.01 \\
NGC\,428 & 20.05$\pm$0.01 & 19.35$\pm$0.01 & \,~~0.70$\pm$0.01 \\
NGC\,450 & 20.61$\pm$0.04 & 20.08$\pm$0.03 & \,~~0.53$\pm$0.05 \\
ESO\,358-5 & 22.01$\pm$0.02 & 21.41$\pm$0.02 & \,~~0.60$\pm$0.03 \\
NGC\,1493 & 18.03$\pm$0.01 & 17.87$\pm$0.01 & \,~~0.16$\pm$0.01 \\
NGC\,2139 & 19.47$\pm$0.01 & 19.50$\pm$0.01 & $-$0.03$\pm$0.01 \\
UGC\,3574 & 21.85$\pm$0.02 & 21.10$\pm$0.02 & \,~~0.75$\pm$0.03 \\
NGC\,2552 & 19.75$\pm$0.01 & 19.39$\pm$0.01 & \,~~0.36$\pm$0.01 \\
NGC\,2805 & 20.61$\pm$0.01 & 20.00$\pm$0.01 & \,~~0.61$\pm$0.01 \\
NGC\,3346 & 21.83$\pm$0.02 & 21.01$\pm$0.01 & \,~~0.82$\pm$0.02 \\
NGC\,3423 & 20.97$\pm$0.01 & 20.10$\pm$0.01 & \,~~0.87$\pm$0.01 \\
NGC\,3445 & 21.34$\pm$0.01 & 20.52$\pm$0.01 & \,~~0.82$\pm$0.01 \\
A\,1156+52 & 21.96$\pm$0.02 & 21.72$\pm$0.02 & \,~~0.24$\pm$0.03 \\
NGC\,4411B & 20.52$\pm$0.01 & 19.84$\pm$0.01 & \,~~0.68$\pm$0.01 \\
NGC\,4540 & 21.84$\pm$0.02 & 20.92$\pm$0.01 & \,~~0.92$\pm$0.02 \\
NGC\,4618 & 20.24$\pm$0.03 & 19.47$\pm$0.02 & \,~~0.77$\pm$0.04 \\
NGC\,4625 & 20.32$\pm$0.02 & 19.96$\pm$0.02 & \,~~0.36$\pm$0.03 \\
NGC\,4701 & 17.98$\pm$0.01 & 17.64$\pm$0.01 & \,~~0.34$\pm$0.01 \\
NGC\,4775 & 19.57$\pm$0.03 & 18.74$\pm$0.03 & \,~~0.83$\pm$0.04 \\
NGC\,5068 & 20.76$\pm$0.01 & 19.83$\pm$0.01 & \,~~0.93$\pm$0.01 \\
NGC\,5585 & 20.01$\pm$0.01 & 19.44$\pm$0.01 & \,~~0.57$\pm$0.01 \\
NGC\,5668 & 20.74$\pm$0.01 & 20.02$\pm$0.01 & \,~~0.72$\pm$0.01 \\
NGC\,5964 & 21.31$\pm$0.01 & 20.62$\pm$0.01 & \,~~0.69$\pm$0.01 \\
NGC\,6509 & 21.35$\pm$0.01 & 20.64$\pm$0.01 & \,~~0.71$\pm$0.01 \\
UGC\,12732 & 21.45$\pm$0.02 & 20.80$\pm$0.01 & \,~~0.65$\pm$0.02 \\
\enddata
\tablecomments{Column~(1) lists the Galaxy name. Column~(2) and~(3) 
list the total $B$- and $V$-band magnitudes of the NC, including
aperture corrections. Column~(4) lists the color $B-V$ of the NC.  The
listed quantities are not corrected for foreground extinction or
extinction intrinsic to the galaxy. The horizontal line divides the
NCs in early-type spirals (above the line) from those in late-type
spirals (below the line).}
\end{deluxetable}

\clearpage

%%% TABLE %%%

\begin{deluxetable}{ccccccccc}
\tabletypesize{\scriptsize}
\tablecaption{Fit Coefficients and Statistical Significance\label{t:fitcoeff}}
\tablewidth{0pt}
\tablehead{
$x$ & $y$ & Figure & $a$ & $b$ & RMS($x$) & RMS($y$) & $r_s$ & $1-P$\\ 
\colhead{(1)} & \colhead{(2)} & \colhead{(3)} & \colhead{(4)} & 
\colhead{(5)} & \colhead{(6)} & \colhead{(7)} & \colhead{(8)} &
\colhead{(9)}\\
}
\startdata
T & $\langle\log\,\tau\rangle_L$ & \ref{f:hubblevsage} & \,~$+11.54\pm0.68$ & 
$-0.55\pm0.13$ & 1.97 & 1.08 & $-$0.3080 & 0.8005\\
T & $\log M$ & \ref{f:hubblevsmass} & \,~\,~$+8.81\pm0.61$ & $-0.37\pm0.11$ & 
1.66 & 0.61 & $-0.7933$ & 0.9999 \\
$B-V$ & $\langle\log\,\tau\rangle_L$ & \ref{f:bvage} & \,~\,~$+7.26\pm0.16$ & 
$+2.53\pm0.23$ & 0.10 & 0.26 & $+$0.9246 & 0.9999\\
$\log L_{{\rm tot},B}$ & $\log M$ & \ref{f:masstotallum} & \,~$-19.50\pm3.82$ & 
$+2.68\pm0.38$ & 0.37 & 0.98 & $+$0.4906 & 0.9670\\
$\log L_{{\rm bul},B}$ & $\log M$ & \ref{f:massbulgelum} & 
\,~\,~$-4.10\pm1.33$ & $+1.27\pm0.15$ & 0.57 & 0.72 & $+$0.7965 & 0.9999\\
$\log M$ & $\langle\log\,\tau\rangle_L$ & \ref{f:massvsage} & 
\,~\,~$+3.50\pm0.72$ & $+0.76\pm0.10$ & 0.78 & 0.60 & $+$0.6895 & 0.9989\\
\enddata
\tablecomments{Cols.~(1) to (3) list the quantities to be correlated using 
the generic expression $y = a + b \times x$, and the corresponding
figure, in which they are plotted against each other. Cols.~(4) and
(5) are the coefficients $a$ and $b$ of the best linear regression
fit, taking into account the errors in both $x$ and $y$. Col.~(6) and
(7) list the RMS scatter around the best fit in both $x$ and
$y$. Cols.(8) and (9) list the Spearman rank-order correlation
coefficient $r_s$ and its significance ($1-P$).}
\end{deluxetable}

\clearpage

%%% TABLE %%%

\begin{deluxetable}{lccc}
\tabletypesize{\scriptsize}
\tablecaption{Luminosity Fractions of NCs per Age Bin of Best Composite 
Population Fit\label{t:contr}}
\tablewidth{0pt}
\tablehead{
Galaxy & $f(\log \tau)$ & $f(\log \tau)$ & $f(\log \tau)$ \\ 
& ($\tau < 7.8$) & ($\tau$ = 7.8 - 9.2) & ($\tau > 9.2$)\\ 
\colhead{(1)} & \colhead{(2)} & \colhead{(3)} & \colhead{(4)} \\
}
\startdata
NGC\,1325 & ... & 0.088 & 0.912 \\
NGC\,1385 & ... & 0.974 & 0.026 \\
NGC\,3177 & 0.005 & 0.822 & 0.174 \\
NGC\,3277 & 0.077 & 0.133 & 0.791 \\
NGC\,3455 & 0.170 & 0.668 & 0.163 \\
NGC\,4030 & ... & 0.331 & 0.669 \\
NGC\,5377 & ... & 0.842 & 0.158 \\
NGC\,5806 & ... & 0.062 & 0.938 \\
NGC\,7421 & ... & 0.880 & 0.120 \\
NGC\,7690 & ... & 0.765 & 0.235 \\
\tableline
NGC\,300 & ... & 0.698 & 0.302 \\
NGC\,428 & ... & 0.629 & 0.371 \\
NGC\,1493 & 0.380 & 0.620 & ... \\
NGC\,2139 & 0.868 & 0.132 & ... \\
NGC\,2552 & 0.447 & 0.553 & ... \\
NGC\,4411B & ... & 1.000 & ... \\
NGC\,4701 & ... & 0.889 & 0.111 \\
NGC\,4775 & ... & 0.811 & 0.189 \\
NGC\,5585 & ... & 0.805 & 0.195 \\
\enddata
\tablecomments{The luminosity fractions in Columns~(2) to (4) are shown 
in three, somewhat arbitrarily chosen, age-bins. The bins roughly
reflect young, intermediate-age and old populations. The horizontal
line divides the NCs in early-type spirals (above the line) from
those in late-type spirals (below the line).}
\end{deluxetable}

%%% END OF TABLES %%%

%%% END OF DOCUMENT %%%

\end{document}